\documentclass[a4paper,11pt]{article}
\pdfoutput=1
\usepackage{jheppub}
\usepackage[T1]{fontenc}
\usepackage{lmodern}
\usepackage[utf8]{inputenc}
\usepackage{amsmath,amssymb,mathtools}
\usepackage{booktabs,array}
\usepackage{tikz}
\usetikzlibrary{decorations.pathreplacing}
\newcommand{\SO}{\mathrm{SO}}
\newcommand{\Or}{\mathrm{O}}
\newcommand{\SU}{\mathrm{SU}}
\newcommand{\USp}{\mathrm{USp}}
\newcommand{\Spin}{\mathrm{Spin}}
\newcommand{\so}{\mathfrak{so}}
\newcommand{\su}{\mathfrak{su}}
\DeclareMathOperator{\Pf}{Pf}
\newcommand{\tr}{\operatorname{tr}}
\newcommand{\eps}{\epsilon}
\newcommand{\Fp}{\mathbb{F}_p}
\newcommand{\Qp}{Q_{+}}
\newcommand{\sect}[2]{(#1;\,#2)}
\newcommand{\Nfour}{\mathcal{N}=4}
\newcommand{\XN}{\mathcal{X}}
\newcommand{\iv}{\iota_{\rm v}}
\newcommand{\is}{\iota_{\rm s}}

\title{\boldmath The Pfaffian sector of $\SO(2N)$ BPS cohomology:\\ brane excitations, triality and fortuity}

\author[a]{Zheng Miao}
\author[b,c]{and Yuezhang Tang}

\affiliation[a]{Department of Physics, Lehigh University,\\16 Memorial Drive East, Bethlehem, PA 18015, USA}
\affiliation[b]{Interdisciplinary Center for Theoretical Study, University of Science and Technology of China,\\Hefei, Anhui 230026, China}
\affiliation[c]{Peng Huanwu Center for Fundamental Theory,\\Hefei, Anhui 230026, China}

\emailAdd{zhm323@lehigh.edu}
\emailAdd{yuezhangtang@ustc.edu.cn}


\abstract{We study the classical $\frac1{16}$-BPS cohomology of $\Nfour$ super-Yang--Mills theory with gauge group $\SO(2N)$ in the BMN sector, including the operators built with one $\eps$-tensor. These operators are odd under the reflection in $\Or(2N)$, carry the $\mathbb Z_2$ charge of a D3-brane wrapped on $\mathbb{RP}^3\subset\mathbb{RP}^5$, and form the Pfaffian sector. For $\SO(8)$ we determine this sector through level $j=18$, where the level is twice the total charge of an operator. All but one of its $106$ $\SU(3)_R$ multiplets are products of trace gravitons and excited Pfaffians. The exception is a singlet, represented by the singlet projection of the rank-four member of a four-term family of $\eps$-tensor operators. We construct a covering space for the Pfaffian sector in which this family is closed at every rank, and the family defines a non-graviton class for every $N\ge4$. Triality pairs the singlet with a reflection-even class that is fortuitous with respect to the trace covering. The two classes restrict, along the spinor and the vector embedding of $\so(7)$ respectively, to the known non-graviton singlet of $\SO(7)$, which is lifted at first order. We show that the two $\SO(8)$ classes, and the family for every $N\ge4$, are protected to all orders of perturbation theory. Their protection is compatible with the lifting in $\SO(7)$ because restriction to a subalgebra does not commute with the quantum corrections to the supercharge.}

\preprint{USTC-ICTS/PCFT-26-78}

\begin{document}

\maketitle

\section{Introduction}
\label{sec:intro}

The $\frac1{16}$-BPS states of $\Nfour$ super-Yang--Mills (SYM) theory account for the entropy of supersymmetric black holes in $\mathrm{AdS}_5$~\cite{Gutowski:2004ez,Kunduri:2006ek,Hosseini:2017mds,Cabo-Bizet:2018ehj,Choi:2018hmj,Benini:2018ywd}. At weak coupling they are described by the cohomology of a nilpotent supercharge $Q$ acting on gauge-invariant operators~\cite{Kinney:2005ej,Grant:2008sk,Chang:2013fba}. A central question is which cohomology classes describe gravitons and which do not. Chang and Lin found the first classes that are not products of gravitons~\cite{Chang:2022mjp}. They later proposed to classify cohomology classes by comparing ranks: a class is \emph{monotone} if it lifts to a class of a rank-independent covering complex, and \emph{fortuitous} otherwise~\cite{Chang:2024zqi}. At large $N$ monotone classes are expected to describe horizonless objects and fortuitous classes the microstates of black holes~\cite{Chang:2024zqi}.

Fortuitous classes, called non-graviton or black-hole cohomologies in earlier work, have been constructed or counted for $\SU(N)$ at small $N$~\cite{Chang:2022mjp,Choi:2022caq,Choi:2023znd,deMelloKoch:2024pcs,Kim:2026mch}. A threshold class proposed for $\SU(3)$ by Choi, Choi, Kim, Lee and Lee was later found to lie in the graviton span~\cite{Choi:2023vdm,Kim:2026mch,FortuitySummary}. De Mello Koch, Kim and Liu gave an explicit construction for every $N\ge3$~\cite{deMelloKoch:2026sbd}. For gauge group $\SO(7)$ such a class was found by Gadde, Lee, Raj and Tomar~\cite{Gadde:2025yoa}. It has no counterpart in the classical cohomology of the S-dual $\USp(6)$ theory~\cite{Chang:2025mqp,Chang:2026bil}. The public repository of the Fortuity collaboration records examples across several gauge groups~\cite{FortuitySummary}.

While the classical cohomology already gives the one-loop BPS spectrum, Chang, Feng, Lin and Tao, and Budzik, Murali and Vieira, obtained the one-loop BPS representative of the black-hole class of $\SU(2)$ by diagonalizing the one-loop dilatation operator~\cite{Chang:2023zqk,Budzik:2023vtr}. Budzik, Murali and Vieira followed this state to non-integer $N$ and checked that a first supercharge correction preserves its class~\cite{Budzik:2023vtr}. Within perturbation theory, whether a class remains BPS beyond one loop is decided by the corrections to the supercharge~\cite{Choi:2025bhi,Kim:2026rnx,Kibe:2026qid,Kim:2026mch}. Choi and Lee obtained part of the first correction from the generalized Konishi anomaly~\cite{Choi:2025bhi}. Budzik and Kulp derived it in complete form, building on the analysis of semi-chiral operators~\cite{Budzik:2025zvu,Budzik:2023xbr}.

The theory with gauge group $\SO(2N)$ has a feature that these examples lack. Its gauge invariants include operators built with one $\eps$-tensor, the simplest of which is the Pfaffian of an adjoint scalar. These operators are odd under the reflection that extends $\SO(2N)$ to $\Or(2N)$, and they are not polynomials in traces. Holographically the theory is dual to type IIB string theory on $\mathrm{AdS}_5\times\mathbb{RP}^5$, and Witten identified the Pfaffian with a D3-brane wrapped on the non-contractible $\mathbb{RP}^3$~\cite{Witten:1998xy}. Aharony, Antebi, Berkooz and Fishman substantiated the identification of Pfaffian, determinant and subdeterminant operators with wrapped D3-branes~\cite{Aharony:2002nd}. Imamura and Yokoyama identified the leading contribution of this brane to the superconformal index~\cite{Imamura:2016abe}. The corresponding $\frac18$-BPS partition function, the large-$N$ Pfaffian sector of the index and its giant-graviton expansion were obtained by Imamura and collaborators~\cite{Arai:2018utu,Arai:2019xmp,Fujiwara:2023bdc}. Giant-graviton expansions of the index were developed more generally by Imamura, by Gaiotto and Lee and by Murthy~\cite{Imamura:2021ytr,Gaiotto:2021xce,Murthy:2022ien}. Caputa, de Mello Koch and Diaz constructed bases of large half-BPS operators for orthogonal and symplectic gauge groups~\cite{Caputa:2013hr,Caputa:2013vla}. In the half-BPS and free-field quarter-BPS sectors, $\eps$-contracted operators of orthogonal groups were constructed by Lewis-Brown and Ramgoolam~\cite{Lewis-Brown:2018dje}. Holguin and Wang studied half-BPS coherent states of the orthogonal and symplectic theories and the place of the Pfaffian among them~\cite{Holguin:2022drf}. Eleftheriou, Murthy and Rossell\'o obtained the Pfaffian term of the half-BPS index from the topologically stable brane in the bulk~\cite{Eleftheriou:2025lac}. These works concern the index and sectors that preserve more supersymmetry.

In the study of the $\frac1{16}$-BPS cohomology the operators of the even orthogonal groups that are built with the $\eps$-tensor have so far been left aside. Gadde, Lee, Raj and Tomar noted that already the graviton spectrum of $\SO(2N)$ must include them~\cite{Gadde:2025yoa}. Chang and Lin point out that no covering is known for the $\SO(2N)$ theory and attribute this to the Pfaffian operators~\cite{Chang:2024zqi}. Kibe, van Leuven and Mathieson restrict their general analysis of the first quantum correction to the multi-trace sector, which for $\SO(2N)$ is the reflection-even sector~\cite{Kibe:2026qid}. The Fortuity collaboration lists $\SO(2N)$ with its Pfaffian invariants among the open problems~\cite{FortuitySummary}. A possible way forward was indicated by Choi, Choi and Kim. For baryonic operators of the conifold theory they introduced a covering space built from fermionic integrals with an inverse letter, called the resulting classes \emph{generalized monotones}, and anticipated that the idea should apply to other theories with topological baryons~\cite{Choi:2026faq}.

In this paper we study the reflection-odd part of the cohomology of the $\SO(2N)$ theory, which we call the Pfaffian sector, and follow its non-graviton class through triality and through the quantum corrections to the supercharge. We work in the BMN sector, which consists of the operators built without derivatives and gauginos. Unless stated otherwise, cohomology means the cohomology of the classical supercharge in this sector. It describes BPS states at leading order in the coupling, and whether a class survives the quantum corrections is a separate question, to which we turn at the end.

We compute the cohomology on weight spaces of the $\SU(3)_R$ symmetry that rotates the three scalars, from spanning sets of $\eps$-tensor words evaluated over a finite field. The counts are made exact with the dimensions of the spaces of gauge invariants, which we obtain from Weyl integrals over the two components of $\Or(2N)$. The integration formula for the component that does not contain the identity is due to Wendt, and Bourget and Pini applied it to the counting of gauge-invariant operators~\cite{Wendt:2001,Bourget:2017tmt}. The index of the reflection-odd sector, computed separately, tests these dimensions. It combines the ordinary index with the index twisted by the reflection, the analogue of the index twisted by charge conjugation that Zwiebel introduced~\cite{Zwiebel:2011wa}. The algebras $\so(4)$ and $\so(6)$ reduce to unitary algebras with known cohomology and serve as tests~\cite{Chang:2022mjp,Kim:2026mch,FortuitySummary}. The algebra $\so(8)$ is the first of the even orthogonal series without such an isomorphism, and we determine the Pfaffian sector of $\SO(8)$ through level $j=18$, where the level is twice the total charge of an operator. Trace relations of $\Or(8)$ begin at nine letters, so this is the first level at which a closed reflection-even operator can owe its closure to a trace relation. It turns out to be the first level with non-graviton classes in either reflection sector. We use $\SO(10)$ as a control, since at the relevant charges its reflection-even operators obey no trace relations.

Our first result is the cohomology itself. All but one of the $106$ $\SU(3)_R$ multiplets of the Pfaffian sector of $\SO(8)$ in this range are products of trace gravitons with supersymmetry multiplets of excited Pfaffians, in which the entries of the $\eps$-tensor are symmetrized words of odd length. The bosonic operators of this kind are the chiral-ring operators counted by Arai, Fujiwara and Imamura~\cite{Arai:2018utu}. The exception is an $\SU(3)_R$ singlet at level eighteen with one fermion. Its image on the Coulomb branch is an explicit function that is not generated by supersymmetry descendants of bosonic invariants. The class is therefore not a combination of products of supersymmetry multiplets whose bottom components are bosonic, a set that includes all gravitons and excited Pfaffians.

To understand the origin of this class we adapt the covering of Choi, Choi and Kim to Pfaffians~\cite{Choi:2026faq}. When all but a few entries of an $\eps$-tensor are filled with the same scalar, the operator equals the Pfaffian of that scalar times a polynomial in traces of words that contain its inverse. Closure of such polynomials can be tested independently of the rank. The singlet is represented by the $\SU(3)_R$-singlet projection of the rank-four member of an explicit four-term family of such operators, which is displayed in section~\ref{sec:singletcov}. The family is closed in the covering, and hence at every rank, so the singlet is a generalized monotone. We propose to read it as an excitation of the wrapped brane by words that are not symmetrized. A restriction that separates one brane from the stack relates the members of the family at neighbouring ranks and preserves graviton products, and the family therefore defines a non-graviton class of $\SO(2N)$ for every $N\ge4$. A separate result concerns the covering as a whole. At fixed excitation charges above the Pfaffian and for sufficiently large rank, its cohomology coincides with the Pfaffian-sector cohomology, so every class with these charges is then a generalized monotone.

The algebra $\so(8)$ is the only simple Lie algebra with triality, and triality acts on the cohomology because the classical supercharge is built from the Lie bracket alone. Triality relates the singlet to a second class of $\SO(8)$ at the same charges, the only non-graviton class of the reflection-even sector there, and the two classes form a doublet, the two-dimensional irreducible representation of the triality group $S_3$. The reflection-even member is a polynomial in traces that is closed only by virtue of trace relations of $\Or(8)$. It is fortuitous with respect to the trace covering and to Pfaffian coverings in which every index of an $\eps$-tensor is carried by a word. Our argument does not exclude rank-uniform families with direct contractions between two $\eps$-tensors. The non-graviton property is thus triality covariant at these charges, whereas the distinction between generalized monotones and fortuitous classes, which is defined through coverings built on the vector representation, is not.

\begin{figure}[t]
\centering
\begin{tikzpicture}[x=1cm,y=1cm,every node/.style={font=\small},box/.style={draw,rounded corners=2pt,align=center,inner sep=5pt,text width=5.0cm}]
\node[box] (odd) at (-4.2,0) {reflection-odd singlet of $\SO(8)$\\[1pt] singlet projection of\\ the four-term family;\\ generalized monotone};
\node[box] (even) at (4.2,0) {reflection-even singlet of $\SO(8)$\\[1pt] closed by trace relations\\ of $\Or(8)$; fortuitous for\\ the trace covering};
\node[box] (seven) at (0,-3.4) {non-graviton singlet of $\SO(7)$\\[1pt] fortuitous; lifted at first order};
\draw[->,thick] (odd) -- node[above] {triality image,} node[below] {even part} (even);
\draw[->,thick] (odd) -- node[left,xshift=-2pt] {spinor restriction} (seven);
\draw[->,thick] (even) -- node[right,xshift=2pt] {vector restriction} (seven);
\draw[decorate,decoration={brace,amplitude=5pt,raise=6pt}] (odd.north west) -- node[above=12pt] {doublet of $S_3$, protected to all perturbative orders} (even.north east);
\end{tikzpicture}
\caption{The non-graviton classes at level eighteen and the maps that relate them. Triality mixes the two reflection sectors of $\SO(8)$. The partner of the reflection-odd singlet is represented by the reflection-even part of the triality image of its representative, and section~\ref{sec:O8} gives an explicit polynomial in traces in the same class, up to normalization. The two members differ in their covering status. Each restricts to the non-graviton singlet of $\SO(7)$, modulo graviton classes and up to normalization, along a different embedding of $\so(7)$.}
\label{fig:map}
\end{figure}
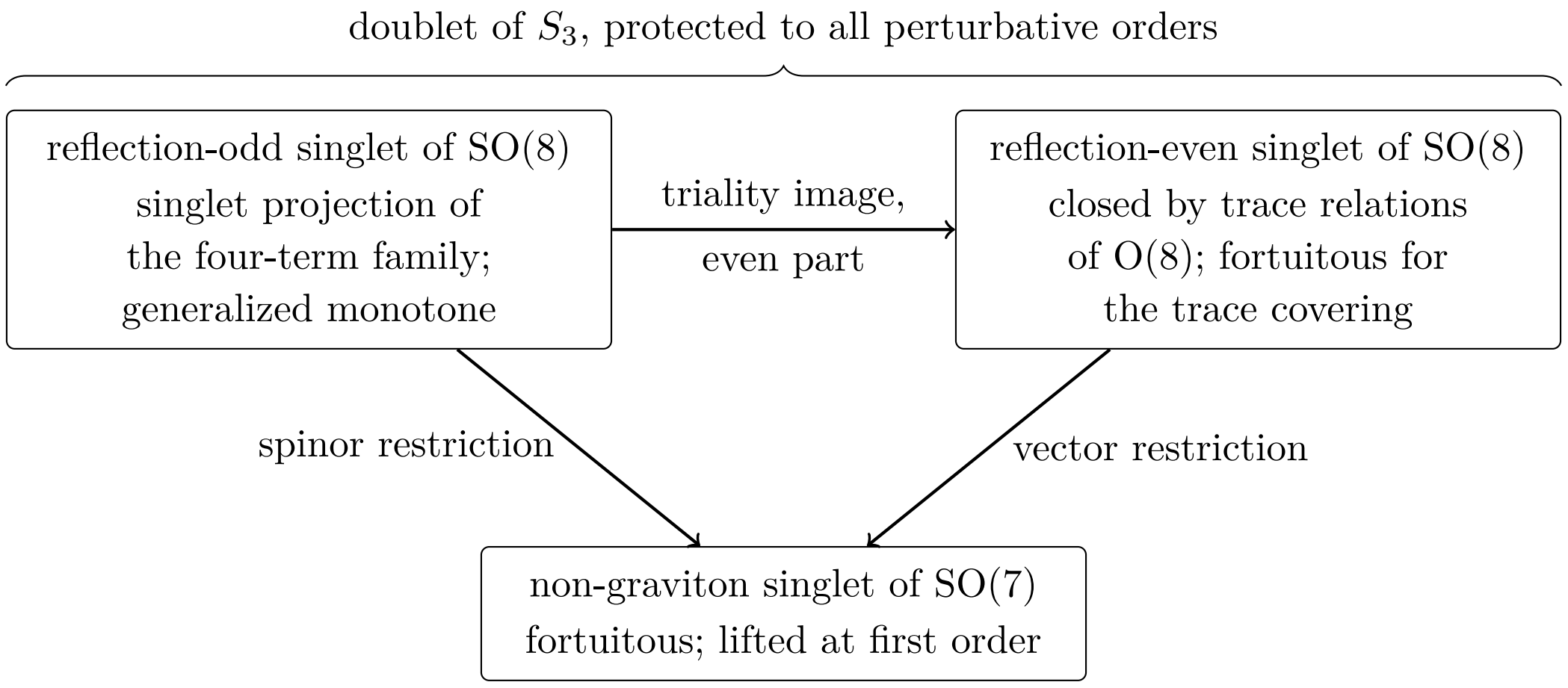

Both members restrict to the non-graviton singlet of $\SO(7)$ found by Gadde, Lee, Raj and Tomar~\cite{Gadde:2025yoa}. The reflection-odd member restricts along the spinor embedding of $\so(7)$ and the reflection-even member along the vector embedding, under which every reflection-odd operator vanishes. Figure~\ref{fig:map} summarizes these relations. Earlier works had argued, or had observed in the examples then known in $\Nfour$ SYM, that fortuitous classes vanish on the Coulomb branch once graviton classes are subtracted~\cite{Chang:2025mqp,Choi:2025bhi,Choi:2025pqr}. Kibe, van Leuven and Mathieson noted that the $\SO(7)$ class is visible on the Coulomb branch modulo monotone classes~\cite{Kibe:2026qid}. The doublet provides an explicit witness for this visibility, a function on the Coulomb branch that is orthogonal to the images of all graviton classes and along which the $\SO(7)$ class has a non-zero component.

Finally we determine the quantum fate of these classes. For the corrections to the supercharge we use the pair operation of Budzik and Kulp in the form used by the Fortuity collaboration, which we extend from odd to even orthogonal algebras~\cite{Budzik:2025zvu,FortuitySummary}. For a BMN class built from scalars and a single fermion, the first-order variation consists of scalars and gauginos only, and thus lies outside the BMN sector. In $\SO(2N)$ the cohomology of such operators is represented faithfully on the Cartan subalgebra, for reflection-odd operators without restriction and for reflection-even ones with at most $2N$ letters, below the first trace relation. Kibe, van Leuven and Mathieson showed that first-order variations vanish on the Cartan subalgebra~\cite{Kibe:2026qid}. We extend their result to the Pfaffian sector and conclude that the doublet survives at first order, and with it the four-term family at every rank. For the doublet we also construct the corrected operators explicitly to first order.

In the scheme of Budzik and Kulp, in which the correction of order $r$ changes the number of letters by $1-r$~\cite{Budzik:2025zvu,Kim:2026rnx}, a grading of the perturbative complex then leaves only two further orders at which an obstruction could arise. For the reflection-odd classes the cohomology in both obstruction sectors vanishes. For the reflection-even member it vanishes at second order, and triality excludes the obstruction at third order. The doublet and the family are therefore protected to all orders of perturbation theory. The same grading protects the two multiplets of $\SO(8)$ at level eighteen that consist of products of two Pfaffian-type factors and are fortuitous with respect to the trace covering, although they are products of brane states.

The $\SO(7)$ singlet to which the doublet restricts is lifted at first order, as Choi and Lee found~\cite{Choi:2025bhi}. The two facts are reconciled by a restriction defect. The quantum corrections contract pairs of letters through all generators of the ambient algebra, including those outside the subalgebra, so restriction commutes with the classical differential and not with its corrections. In $\SO(7)$ the argument that protects the doublet is not available, because the Cartan subalgebra is too small to detect the variation.

Several of these ingredients are known. The $\SO(7)$ singlet, its first-order lifting and its Coulomb-branch visibility are known from the works cited above. Restriction maps between gauge-algebra complexes were used previously for block embeddings~\cite{deMelloKoch:2024pcs,deMelloKoch:2026sbd,FortuitySummary,deMelloKoch:2026cvn}. Chang used restriction from the stable algebra, and the Fortuity collaboration used restriction to $\mathfrak g_2\subset\so(7)$ and noted that block restriction is not a chain map for the first quantum correction, although on the classes they tested it commutes with that correction in cohomology~\cite{Chang:2026bil,FortuitySummary}. New here are the Pfaffian-sector cohomology of $\SO(8)$, the Pfaffian covering, the triality doublet, the cohomologically non-trivial restriction defect for embeddings that do not separate branes, and the all-orders perturbative protection.

The paper is organized as follows. Section~\ref{sec:setup} sets up the BMN complex of $\SO(2N)$, the Pfaffian sector, the graviton spectrum and the notions of monotone and generalized monotone classes. Section~\ref{sec:method} describes how the cohomology is computed and how the counts are made exact. Section~\ref{sec:results} presents the Pfaffian sector through level eighteen, the parity information carried by its index, the bounds that locate the onset of trace relations and the reflection-even sector of $\SO(8)$. Section~\ref{sec:singlet} analyses the non-graviton singlet and its image on the Coulomb branch. Section~\ref{sec:covering} constructs the covering of the Pfaffian sector and the four-term family. Section~\ref{sec:triality} uses triality to relate the singlet to the $\SO(7)$ class and to its reflection-even partner. Section~\ref{sec:quantum} proves the perturbative protection of the doublet, of the family and of the trace-fortuitous products of two Pfaffian-type factors, and analyses the failure of restriction to commute with the quantum corrections. We conclude in section~\ref{sec:conclusions}. The appendices give our conventions and the Fischer inner product, the exact counting of gauge invariants, the properties of the Pfaffian covering, and the quantum corrections and explicit identities.
\section{Pfaffian operators and the classical supercharge}
\label{sec:setup}

Gauge-invariant operators of $\SO(2N)$ are built with traces and with the invariant $\eps$-tensor, and the reflection in $\Or(2N)\setminus\SO(2N)$ separates the two constructions. This section fixes the classical complex, its decomposition under the reflection, the graviton generators with which cohomology classes are compared, and the coverings that define monotone and generalized monotone classes.

\subsection{The BMN complex}
\label{sec:bmn}

At leading order in the coupling, a $\frac1{16}$-BPS state of $\Nfour$ super-Yang--Mills theory is represented by a gauge-invariant operator annihilated by a nilpotent supercharge $Q$, and two such operators represent the same state if they differ by a $Q$-exact operator. We study this cohomology in the BMN sector~\cite{Berenstein:2002jq,Kim:2003rza,Grant:2008sk,Chang:2013fba,Chang:2022mjp,Gadde:2025yoa}. Our conventions are those of the Fortuity collaboration~\cite{FortuitySummary}. The letters of this sector are three scalars $\phi_m$, three fermions $\psi_m$ ($m=1,2,3$) and one bosonic letter $f$, a component of the self-dual field strength, all in the adjoint representation; gauginos and derivatives are absent. When words are spelled out we also write $X,Y,Z$ for $\phi_1,\phi_2,\phi_3$. The supercharge acts as an odd derivation, with
\begin{equation}\label{eq:Q}
Q\phi_m=0,\qquad Q\psi_m=\eps_{mnp}\,\phi_n\phi_p=\tfrac12\eps_{mnp}[\phi_n,\phi_p],\qquad Qf=\sum_{m}[\phi_m,\psi_m],
\end{equation}
where products are matrix products, and it satisfies $Q^2=0$ on gauge invariants. In sections~\ref{sec:setup}--\ref{sec:triality} the symbol $Q$ denotes this classical supercharge; its quantum deformation is the subject of section~\ref{sec:quantum}.

The $\SU(3)_R$ symmetry, which we also call the flavour symmetry, rotates $\phi_m$ in the fundamental and $\psi_m$ in the antifundamental representation. Together with $\SU(3)_R$, the supercharges
\begin{equation}\label{eq:Qplus}
\Qp^m\phi_n=\eps_{mnp}\psi_p,\qquad \Qp^m\psi_n=\delta_{mn}f,\qquad \Qp^mf=0
\end{equation}
generate the $\SU(1|3)$ supermultiplet structure. On a single letter $\{Q,\Qp^m\}$ acts as the gauge rotation $[\phi_m,\cdot\,]$, so $Q$ and $\Qp^m$ anticommute on gauge invariants. Every $\Qp$-descendant of a $Q$-closed gauge invariant is therefore $Q$-closed.

An operator with $n_{\phi_m}$, $n_{\psi_m}$ and $n_f$ letters of each kind has charges
\begin{equation}
q_m=n_{\phi_m}+(n_\psi-n_{\psi_m})+n_f,\qquad 2J=n_\psi+2n_f,\qquad j=2(q_1+q_2+q_3),
\end{equation}
with $n_\psi=\sum_mn_{\psi_m}$. With $|q|=q_1+q_2+q_3$, the number of letters is $L=|q|-2J$ and the classical scaling dimension is $\Delta=|q|-J$; for instance $\Delta=17/2$ for the $\SO(8)$ singlet of section~\ref{sec:singlet}. We call $j$ the \emph{level} and write a sector of fixed charges as $\sect{2J}{q_1,q_2,q_3}$. The supercharge $Q$ maps $\sect{2J}{q}$ to $\sect{2J-1}{q}$. Weight spaces of $\SU(3)_R$ are labelled by $q$, and the highest-weight multiplicities of an irreducible representation $\lambda=(\lambda_1\ge\lambda_2\ge\lambda_3)$ follow from weight-space data by Weyl alternation. We denote its Dynkin labels by $[\lambda_1-\lambda_2,\lambda_2-\lambda_3]$ and refer to its highest-weight components at all values of $2J$ as the \emph{column} $\lambda$.

The BMN index
\begin{equation}
\mathcal I=\sum_{\text{sectors}}(-1)^{2J}\dim H_{\sect{2J}{q}}\;x^q=\sum_{\text{sectors}}(-1)^{2J}\dim V_{\sect{2J}{q}}\;x^q
\end{equation}
is the Euler characteristic of the classical complex, where $V$ denotes the space of gauge invariants, $H$ the cohomology and $x^q=x_1^{q_1}x_2^{q_2}x_3^{q_3}$. It is not a specialization of the full superconformal index. S-duality requires the full indices of $\SO(2N+1)$ and $\USp(2N)$ to agree, which Gadde, Pomoni, Rastelli and Razamat confirmed to the first orders of a series expansion and Spiridonov and Vartanov proved in special cases~\cite{Gadde:2009kb,Spiridonov:2010qv}. The BMN indices of $\SO(7)$ and $\USp(6)$ differ, as found by Gadde, Lee, Raj and Tomar~\cite{Gadde:2025yoa}.

\subsection{Reflection parity and Pfaffian operators}
\label{sec:invariants}

For gauge group $\SO(2N)$ the letters are antisymmetric $2N\times2N$ matrices. By the first fundamental theorem of invariant theory for $\Or(2N)$, the $\Or(2N)$-invariants are spanned by multi-traces~\cite{Procesi:1976}. Aslaksen, Tan and Zhu showed that the $\SO(2N)$-invariants are generated by traces and polarized Pfaffians~\cite{Aslaksen:1995}. Since products of two $\eps$-tensors reduce to traces through $\eps\eps=\det(\delta)$, the $\SO(2N)$-invariants are spanned by multi-traces and by words with a single $\eps$-tensor multiplied by multi-traces. These $\eps$-words have the form
\begin{equation}\label{eq:epsword}
\begin{gathered}
\eps(W_1,\dots,W_N)\;\tr M_1\cdots\tr M_r,\\
\eps(W_1,\dots,W_N)=\eps_{a_1b_1\cdots a_Nb_N}(W_1)_{a_1b_1}\cdots(W_N)_{a_Nb_N},
\end{gathered}
\end{equation}
where the \emph{slot words} $W_s$ and the \emph{necklaces} $M_i$ are words in the letters. Inside $\eps(\cdots)$ we write $W^{\times n}$ for $n$ consecutive slots equal to $W$, whereas a power without $\times$ is a matrix power and hence a single word. The Grassmann order of an $\eps$-word is the slot order followed by the traces, and appendix~\ref{app:conventions} collects the remaining sign conventions. The theorems of Procesi and of Aslaksen, Tan and Zhu, and the second fundamental theorem used in section~\ref{sec:onset}, concern matrices with commuting entries. They extend to Grassmann-odd letters by polarizing to multilinear expressions in the odd letters and writing each odd factor as $\theta_s\widetilde B_s$ with an independent odd parameter $\theta_s$ and a bosonic matrix $\widetilde B_s$, as in section~\ref{sec:fermionic} and appendix~\ref{app:faithful}.

The reflection $\sigma\in\Or(2N)\setminus\SO(2N)$ acts as $+1$ on traces and as $-1$ on $\eps$-words. It commutes with $Q$, so the space of gauge invariants and the cohomology decompose as $V=V^+\oplus V^-$ and $H=H^+\oplus H^-$. We call the reflection-odd sector, with complex $V^-$ and cohomology $H^-$, the \emph{Pfaffian sector}.

Every letter is antisymmetric. Let $\ell$ be the length of a word $W$, $\operatorname{rev}(W)$ the word read backwards and $s_{\rm rev}(W)$ the sign of reversing the order of its fermions. Then $W^T=(-1)^\ell s_{\rm rev}(W)\,\operatorname{rev}(W)$, which gives the relations
\begin{equation}\label{eq:slotid}
\begin{gathered}
\eps(\dots,W,\dots)=c(W)\,\eps(\dots,\operatorname{rev}W,\dots),\qquad c(W)=-(-1)^\ell s_{\rm rev}(W),\\
\tr W=(-1)^\ell s_{\rm rev}(W)\tr\operatorname{rev}W.
\end{gathered}
\end{equation}
Slots commute up to the Koszul sign of their fermion numbers. In particular a palindromic slot with $c(W)=-1$ vanishes, for instance any symmetrized bosonic word of even length. These relations bring every $\eps$-word to a canonical form.

The lowest Pfaffian-sector operator is the Pfaffian of a single scalar, $\Pf\phi=\eps(\phi,\dots,\phi)/(2^NN!)$, with charges $\sect{0}{N,0,0}$ for $\phi=\phi_1$, level $j=2N$ and dimension $\Delta=N$. For $\so(4)=\su(2)\oplus\su(2)$ it is proportional to $\tr\phi_+^2-\tr\phi_-^2$, where $\phi_\pm$ are the components of $\phi$ in the two factors. For $\so(6)=\su(4)$, where $\sigma$ acts as charge conjugation, the multiplet of the Pfaffian is the $\tr\Phi^3$ multiplet. Here $\tr\Phi^n$ denotes the $\SU(1|3)$ multiplet generated by $\Qp$ from $\tr\phi_{(m_1}\cdots\phi_{m_n)}$.

Holographically, $\Nfour$ SYM with gauge group $\SO(2N)$ is dual to type IIB string theory on the $\mathrm{AdS}_5\times\mathbb{RP}^5$ orientifold without discrete torsion, the near-horizon limit of D3-branes on an O3$^-$-plane, and Witten identified the Pfaffian with a D3-brane wrapped on $\mathbb{RP}^3\subset\mathbb{RP}^5$~\cite{Witten:1998xy}. The wrapping number is a $\mathbb Z_2$ charge, since $H_3(\mathbb{RP}^5,\mathbb Z)=\mathbb Z_2$, and it is identified with the $\sigma$-parity. Two wrapped branes carry trivial $\mathbb Z_2$ charge and can annihilate. Correspondingly $\Pf(\phi)^2=\det\phi$ is a polynomial in traces, the operator dual to the maximal giant graviton~\cite{McGreevy:2000cw,Balasubramanian:2001nh,Corley:2001zk,Aharony:2002nd,Holguin:2022drf}. Subdeterminants describe its non-maximal deformations, as argued by Balasubramanian, Berkooz, Naqvi and Strassler~\cite{Balasubramanian:2001nh}. This dictionary is semiclassical: the wrapped brane has mass of order $N$ in AdS units and is a heavy object only at large $N$. At the ranks studied here, ``brane'' and ``graviton'' are labels that depend on the description.

\subsection{Graviton multiplets and the parity constraint}
\label{sec:gravitons}

Traces of odd length of a single antisymmetric matrix vanish. The single-trace gravitons are therefore the multiplets $\tr\Phi^n$ with $n$ even~\cite{Gadde:2025yoa,Kibe:2026qid}. In the Pfaffian sector we use three nested sets of reflection-odd operators, each closed under $\Qp$. The first is the \emph{Pfaffian multiplet}, generated by $\eps(\phi_{m_1},\dots,\phi_{m_N})$. The second consists of the \emph{excited Pfaffians}, generated by $\eps(\mathrm{Sym}\,\phi^{\ell_1},\dots,\mathrm{Sym}\,\phi^{\ell_N})$, where each slot is a symmetrized word in the scalars of odd length $\ell_s$; even lengths vanish by \eqref{eq:slotid}. The third consists of the products of these operators with trace gravitons. All of them are $Q$-closed, because their bottom components contain only scalars and $\{Q,\Qp^m\}=0$ on gauge invariants.

A \emph{graviton product} in $H^-$ is a product of trace gravitons with an odd number of \emph{Pfaffian-type factors}, which are components of the Pfaffian multiplet or excited Pfaffians, and a \emph{graviton combination} is a linear combination of graviton products; in $H^+$ a graviton product has an even number of Pfaffian-type factors, and for $\SO(7)$ none. In the bulk the Pfaffian-type factors are wrapped-brane states rather than supergravity modes, but we keep the name graviton product, following Gadde, Lee, Raj and Tomar~\cite{Gadde:2025yoa}. Accordingly, a \emph{non-graviton} class is one that is not represented by a graviton combination, and the term carries no bulk implication by itself. The bulk-motivated distinction is the one between generalized monotones and strongly fortuitous classes, introduced in section~\ref{sec:monotone}. A class is \emph{minimal} if it is a graviton product with only the Pfaffian multiplet among its $\eps$-factors.

The bosonic excited Pfaffians are the $\frac18$-BPS twisted-sector operators counted by Arai, Fujiwara and Imamura~\cite{Arai:2018utu}. They are close relatives of the subdeterminant operators~\cite{Balasubramanian:2001nh,Aharony:2002nd}. They are also close to the excited dibaryons of Berenstein, Herzog and Klebanov, in which one entry of an $\eps\eps$ contraction is replaced by a composite word~\cite{Berenstein:2002ke}.

Graviton products obey a parity selection rule. For an operator of definite $\sigma$-parity we write $\sigma=\pm1$ also for its eigenvalue and define
\begin{equation}\label{eq:parity}
P_N=(-1)^L\sigma^N.
\end{equation}
Every graviton product has $P_N=+1$, and every gauge invariant with $P_N=-1$ vanishes when all letters take values in a Cartan subalgebra of $\so(2N)$. This is the $\SO(2N)$ analogue of the charge-conjugation parity of $\SU(N)$ gravitons~\cite{FortuitySummary,Kim:2026mch,Kibe:2026qid}. For $N=3$ it reduces to that parity through $\so(6)=\su(4)$. It is also the analogue of the $\Or(4)/\SO(4)$ parity argument of Choi, Choi and Kim~\cite{Choi:2026faq}.

Trace gravitons have even length and $\sigma=+1$. An excited Pfaffian has $L=\sum_s\ell_s\equiv N\pmod2$ and $\sigma=-1$. Since $\Qp$ preserves $L$ and $P_N$ is multiplicative, every graviton product has $P_N=+1$. On the Cartan subalgebra every letter is block diagonal with $2\times2$ blocks $x^{(i)}\mathbb J$, where $\mathbb J=\left(\begin{smallmatrix}0&1\\-1&0\end{smallmatrix}\right)$. A word of length $\ell$ is then block diagonal with blocks proportional to $\mathbb J^\ell$, which is antisymmetric for odd $\ell$ and proportional to the identity for even $\ell$. Traces of odd length therefore vanish, and so does every $\eps$-word with a slot of even length. A non-vanishing term has $L\equiv N\pmod2$ if it carries an $\eps$-tensor, so that $\sigma=-1$, and even $L$ otherwise; in both cases $P_N=+1$.

The index can detect classes with $P_N=-1$ through the sign of its columns, as explained in section~\ref{sec:index}. Such classes are neither graviton products nor visible under restriction to the Cartan subalgebra.

\subsection{Monotone, fortuitous and generalized monotone classes}
\label{sec:monotone}

For gauge invariants that are polynomials in traces, the covering complex of Chang and Lin consists of formal multi-traces~\cite{Chang:2024zqi}. The rank-$N$ complex is its quotient by the trace relations. A class is \emph{monotone} if it lies in the image of the covering cohomology and \emph{fortuitous} otherwise. We use the standard expectation that, in the trace sector, the monotone classes are the graviton polynomials~\cite{Chang:2013fba,Chang:2024zqi}. The stable homology underlying this expectation is discussed by Chang and, for the orthogonal case, by Loday and Procesi~\cite{Chang:2026bil,Loday:1988}.

The Pfaffian sector has no such covering, because the number of slots of the $\eps$-tensor changes with the rank. Choi, Choi and Kim attach to a baryonic ground state a covering space with an inverse letter~\cite{Choi:2026faq}. We adapt their construction to a Pfaffian ground state in section~\ref{sec:covering}. Fix the ground state $\Pf(\phi_1)$. A \emph{formal monomial} $(\vec W,M)$ consists of $k$ slot words $\vec W=(W_1,\dots,W_k)$, none equal to the single letter $\phi_1$, and a list of necklaces $M=(M_1,\dots,M_r)$. It defines the rank-uniform family
\begin{equation}\label{eq:family}
X_N(\vec W,M)=\frac{1}{(N-k)!}\,\eps\big(W_1,\dots,W_k,\phi_1^{\times(N-k)}\big)\,\tr M,
\end{equation}
where $\tr M=\tr M_1\cdots\tr M_r$ and $X_N(\vec W,M)=0$ for $k>N$. The $N-k$ slots equal to $\phi_1$ are the \emph{filler slots}. A linear combination $a$ of formal monomials is \emph{covering-closed} if $\iota(\delta a)=0$ in the covering algebra $\mathcal C$, where the covering map $\iota$ and the differential $\delta$ induced by $Q$ on formal monomials are constructed in section~\ref{sec:covering}. For a covering-closed combination the operators $X_N(a)$ are $Q$-closed at every rank.

The space $GM_N^-$ of \emph{generalized monotone} classes in $H^-$ is spanned by three kinds of classes. The first kind consists of the classes of $X_N(a)$ for covering-closed combinations $a$, together with their $\SU(3)_R$ images. The second consists of the products of these classes with trace monotones. The third arises for odd $B\ge3$ and consists of the classes of $B$-Pfaffian families, defined in appendix~\ref{app:multipf}, that are $Q$-closed at every rank; it includes the products of $B$ classes of the first two kinds. A class in $H^-$ outside $GM_N^-$ is \emph{strongly fortuitous} with respect to $GM_N^-$.

This notion is weaker than strong fortuity in the sense of Choi, Choi and Kim, which excludes every rank-uniform $Q$-closed construction~\cite{Choi:2026faq}. A class outside $GM_N^-$ could still extend to a family of another type, for instance one with Kronecker contractions between $\eps$-tensors; we return to such families in section~\ref{sec:interpretation}. The first two kinds of classes are the single-Pfaffian version of the baryonic monotones of Choi, Choi and Kim, whose covering space already includes multiplication by mesons, and the third is the analogue of their multi-baryonic sectors.

The definition is $\SU(3)_R$ invariant by construction. The classes of covering-closed families with the fixed ground state $\Pf(\phi_1)$ already form a space that is stable under $\mathfrak{gl}(3)$ and $\Qp$ (appendix~\ref{app:gl3}), so the $\SU(3)_R$ images add nothing. A member of a $B$-Pfaffian family at rank $N$ has at least $BN$ letters and hence level $j\ge2BN$ (appendix~\ref{app:multipf}). For $\SO(8)$ the families with $B\ge3$ start at $j=24$ and play no role for $j\le18$. For $\SO(6)$ and $\SO(4)$ they enter from $j=18$ and $j=12$, respectively.

The parity $P_N$ does not separate these notions. Classes with $P_N=-1$ vanish on the Cartan subalgebra, but they need not be fortuitous. Indeed, monotone classes that vanish there are known~\cite{Chang:2025mqp,Choi:2025bhi,Choi:2025pqr,Kibe:2026qid,Kim:2026mch}. Moreover, rank-uniform families with even-length non-palindromic slot words can have $P_N=-1$. In the conifold theory at $N=2$, written with $\SO(4)$ letters, every generalized monotone built from the generators of Choi, Choi and Kim is even under $\Or(4)/\SO(4)$, so the reflection-odd classes there are strongly fortuitous~\cite{Choi:2026faq}. In $\Nfour$ SYM with gauge group $\SO(2N)$ the lowest reflection-odd operator is instead the Pfaffian, a wrapped-brane state, and the reflection-odd sector contains generalized monotones.

Two other conventions are in use. Chang proposes to define fortuity as the cokernel of restriction from the stable algebra~\cite{Chang:2026bil}. With this definition every class of the Pfaffian sector would count as fortuitous. For baryons in a quark model, Fliss, Jejjala and Parrikar likewise call the baryon states fortuitous~\cite{Fliss:2026rad}. We follow Choi, Choi and Kim, for whom baryonic ground states and their rank-uniform excitations are generalized monotones~\cite{Choi:2026faq}.
\section{Computing the Pfaffian-sector cohomology}
\label{sec:method}

The cohomology is computed sector by sector from three ingredients. Exact dimensions of the spaces of gauge invariants fix the size of each cochain space. Ranks of $Q$ computed over a finite field bound the true ranks from below and therefore bound the cohomology from above. Restriction to the Cartan subalgebra supplies lower bounds and separates graviton combinations from other classes. Where the bounds meet, the count is exact.

\subsection{Exact dimensions}
\label{sec:dims}

Let $V^\pm_s$ be the spaces of reflection-even and reflection-odd gauge invariants in a sector $s=\sect{2J}{q}$. Their dimensions follow from the Molien--Weyl integral over $\SO(2N)$ and from the twisted integral over the component $\Or(2N)^-$~\cite{Wendt:2001,Bourget:2017tmt,Fujiwara:2023bdc}. With measures of unit mass,
\begin{equation}\label{eq:dimprojection}
\dim V^\pm_s=\tfrac12\Big(\int_{\SO(2N)}d\mu(g)\,\chi_s(g)\;\pm\;\int_{\Or(2N)^-}d\mu(g)\,\chi_s(g)\Big),
\end{equation}
where $\chi_s$ is the ungraded character of the free Fock space of the letters restricted to the sector $s$. The first integral counts all $\SO(2N)$ invariants and the second is the trace of $\sigma$ on them. Both reduce to integrals over tori, of rank $N$ for the identity component and of rank $N-1$ for the component of the reflection, and we evaluate them exactly, without floating-point arithmetic. Appendix~\ref{app:dims} gives the two measures, the character, the method of evaluation and the tests of the result. The resulting tables cover $\SO(4)$ to $j=28$, $\SO(6)$ to $j=26$, $\SO(8)$ to $j=40$ and $\SO(10)$ to $j=22$; the $\SO(7)$ dimensions used in section~\ref{sec:triality} are obtained in the same way.

For the reflection-even sector there is an independent count. Below the first trace relation of $\Or(2N)$, which occurs at $L=2N+1$ letters (section~\ref{sec:onset}), $\dim V^+$ equals the number of formal multi-traces. Beyond the threshold the difference locates the trace relations. For $\SO(8)$ with $j\le18$ the formal count exceeds $\dim V^+$ only at $2J=0$ and $L=9$, by exactly one highest weight in each of the columns $(3,3,3)$, $(5,3,1)$ and $(7,1,1)$. For $\SO(10)$ at $j=22$, $2J=0$ and $L=11$, it exceeds $\dim V^+$ by exactly one highest weight in each of the columns $(5,3,3)$, $(5,5,1)$, $(7,3,1)$ and $(9,1,1)$. These three multiplets of trace relations of $\Or(8)$ are the origin of the trace-fortuitous classes of section~\ref{sec:evenscan}. The $\SO(10)$ dimensions, for instance $\dim V^+=237$ and $936$ at $\sect{0}{3,3,3}$ and $\sect{1}{3,3,3}$, enter \eqref{eq:h10h8} and the comparisons of sections~\ref{sec:partner} and~\ref{sec:O8}.

\subsection{Spanning sets and finite-field evaluation}
\label{sec:eval}

In each sector we use as spanning set all $\eps$-words \eqref{eq:epsword} with slot words of length at most $3$ and necklaces of length at most $6$, brought to canonical form with \eqref{eq:slotid}. Whether these cut-offs suffice is decided sector by sector: the set spans $V^-_s$ when the rank of its evaluations equals the exact dimension $\dim V^-_s$, and it does so in every sector used below (section~\ref{sec:validation}).

Operators are evaluated at random points over $\Fp$ with $p=2^{26}-5$. Bosonic letters are random antisymmetric matrices, and fermionic letters are antisymmetric matrices with entries in a Grassmann algebra on $n_\theta=\max(2J,1)$ generators. The contraction $\eps(W_1,\dots,W_N)$ is the top-degree coefficient of $\omega_1\wedge\dots\wedge\omega_N$, where $\omega_s=\sum_{a<b}\big((W_s)_{ab}-(W_s)_{ba}\big)\,e_a\wedge e_b$ and the Grassmann-valued coefficients are multiplied in slot order. Each evaluated vector collects the Grassmann components of the values at all points. The number of points is chosen for every fermion number so that the rank of the spanning set reaches the exact dimension; the evaluation map is then injective on that sector. Ranks of $Q$ are obtained by evaluating the $Q$-images of the spanning set in the target sector. The enumeration of multi-traces, the formal differential and the modular linear algebra are taken from the public library of the Fortuity collaboration~\cite{FortuitySummary}.

The cohomology of a weight space is then
\begin{equation}\label{eq:hcomp}
h_s=\dim V^-_s-\operatorname{rank}(Q:V^-_s\to V^-_{s-})-\operatorname{rank}(Q:V^-_{s+}\to V^-_s),
\end{equation}
where $s\pm=\sect{2J\pm1}{q}$. The rank over $\Fp$ of an evaluated integer matrix is a lower bound for its rank over $\mathbb Q$, because a minor that is non-zero modulo $p$ is a non-zero integer. A single prime thus gives a rigorous lower bound, and it suffices wherever an independent upper bound meets it. Since the dimensions are exact, the computed value $h^{\rm comp}_s$ is an upper bound for $h_s$, with no probabilistic assumption. An excess $h^{\rm comp}_s>h_s$ can arise only from a rank drop modulo $p$, which requires $p$ to divide every minor of an integer matrix whose size equals its rank over $\mathbb Q$, or from a non-generic evaluation point, whose probability is bounded by the Schwartz--Zippel lemma~\cite{Schwartz:1980,Zippel:1979}. Both are excluded wherever a matching lower bound is available. The same formula and the same bounds apply to the reflection-even sector, with $V^+$ in place of $V^-$ and multi-traces as spanning set.

\subsection{The Coulomb map and the graviton test}
\label{sec:coulombtest}

Let $C(\mathcal O)$ denote the restriction of a gauge-invariant operator $\mathcal O$ to letters valued in a Cartan subalgebra, that is, to block-diagonal matrices with blocks $x^{(i)}\mathbb J$. We call $C(\mathcal O)$ the \emph{Coulomb image} of $\mathcal O$ and $C$ the \emph{Coulomb map}; the symbol is not to be confused with the covering algebra $\mathcal C$ of section~\ref{sec:covering}. The Cartan subalgebra serves here as an algebraic test configuration: $C(\mathcal O)$ is a polynomial in the block variables attached to the operator itself, not a vacuum expectation value.

Since $Q$ is a derivation and $Q\psi_m$ and $Qf$ are commutators, which vanish for commuting matrices, every $Q$-exact operator has vanishing Coulomb image. The Coulomb map therefore descends to cohomology. A closed operator with $C(\mathcal O)\neq0$ is not $Q$-exact, and we call its class \emph{Coulomb visible}. The converse fails, because $C$ need not be injective on cohomology. This is the Coulomb-branch criterion of Choi, Choi, Kim, Lee and Lee, used later by Chang and Lin, by Choi and Lee and by Choi and Kim~\cite{Choi:2023vdm,Chang:2025mqp,Choi:2025bhi,Choi:2025pqr}. Restriction to the Cartan subalgebra as a test of non-exactness goes back to Cachazo, Douglas, Seiberg and Witten and to Witten~\cite{Cachazo:2002ry,Witten:2003ye}.

The graviton test compares three ranks in each sector $s$. We denote by $D_s$ the rank of the Coulomb images of the closed subspace and by $G^C_s$ the rank of the Coulomb images of the graviton products of section~\ref{sec:gravitons}. The rank of the graviton products in $H_s$ is $G^H_s$; the same symbol denotes their span, the graviton subspace of $H_s$. The candidate graviton products are the products of trace gravitons $\tr\Phi^n$, with $n$ even and $n\le2N$, with an odd number of Pfaffian-type factors, minimal or excited. With the superscript ``comp'' for computed values,
\begin{equation}\label{eq:chain}
G^{C,\rm comp}_s\;\le\;G^C_s\;\le\;\min(G^H_s,D_s)\;\le\;\max(G^H_s,D_s)\;\le\;h_s\;\le\;h^{\rm comp}_s.
\end{equation}
The first inequality holds because computed ranks over $\Fp$ are lower bounds, and the second because graviton products are closed and the Coulomb map factors through $H_s$. The fourth holds because $G^H_s$ is a subspace of $H_s$ and the image of $C$ is a quotient of $H_s$, and the last is the bound of section~\ref{sec:eval}. A computed Coulomb rank $D^{\rm comp}_s$ of the closed subspace bounds $D_s$ from below only when the computed rank of $Q$ out of the sector is exact, for instance equal to the dimension of its target; otherwise the computed closed subspace can be too large.

By \eqref{eq:chain}, if $G^{C,\rm comp}_s=h^{\rm comp}_s$, all entries of the chain coincide: the count $h_s$ is exact and every class in the sector is a graviton product. A non-graviton class exists if and only if $G^H_s<h_s$. Since the kernel of $C$ on $H_s$ has dimension $h_s-D_s$, one has $G^H_s\le G^C_s+(h_s-D_s)$. It therefore suffices to show $D_s>G^C_s$, or to exhibit a closed operator whose Coulomb image lies outside the span of the Coulomb images of the graviton products. We do the latter exactly in section~\ref{sec:exactproof}.

The same logic determines when a computed count is exact: the upper bound $h^{\rm comp}_s$ must be met by a lower bound. This is automatic when $h^{\rm comp}_s=0$ and when $G^{C,\rm comp}_s=h^{\rm comp}_s$. In the few sectors with a non-graviton class the lower bound comes from exact ranks, from the Coulomb image of an explicit closed operator, or from the exact index of a column. A computed rank of $Q$ that equals the dimension of its target is exact. By nilpotence, the rank of $Q$ into a sector is at most the dimension of its source minus the rank of the differential entering that source, so a computed rank that attains this bound is exact as well. The Coulomb image of an explicit closed operator outside the span of the graviton images gives a class directly, and the exact index of a column forces a class when the cohomology at the other values of $2J$ in the column is known. Unless stated otherwise, every count quoted below is exact in this sense, and we state the counts without further qualification.

\subsection{Comparison with low rank and with a direct construction}
\label{sec:validation}

Two isomorphisms give independent predictions at low rank. For $\so(4)=\su(2)\oplus\su(2)$ the cohomology factorizes as $H_{\SO(4)}=H_{\SU(2)}\otimes H_{\SU(2)}$, and $\sigma$ acts as the graded exchange of the two factors, $\sigma(x\otimes y)=(-1)^{|x||y|}\,y\otimes x$, so $H^-$ is the graded-antisymmetric part of the tensor product. For $\so(6)=\su(4)$ the reflection is charge conjugation, and $H^-$ is the charge-conjugation-odd part of the $\SU(4)$ cohomology. Both predictions follow from the known cohomologies of $\SU(2)$ and $\SU(4)$~\cite{Chang:2022mjp,Kim:2026mch,FortuitySummary}. The spanning sets reach the exact dimension in all $545$ dominant sectors of $\SO(4)$ with $4\le j\le20$, in all $385$ dominant sectors of $\SO(6)$ with $6\le j\le18$ and in all $373$ dominant sectors of $\SO(8)$ with $8\le j\le18$; of these, $204$, $142$ and $124$ have non-zero $V^-$. The computed cohomology agrees with the $\SU(2)\otimes\SU(2)$ prediction in all $545$ such sectors of $\SO(4)$, and its highest-weight multiplicities agree with the $\SU(4)$ prediction in all $385$ such sectors of $\SO(6)$.

A second computation avoids $\eps$-words and random evaluation altogether. It constructs the invariants as the common kernel of the root-vector derivations acting on polynomials in the matrix entries, using the split form $\left(\begin{smallmatrix}A&B\\C&-A^T\end{smallmatrix}\right)$ of $\so(2N)$ and the signed-permutation group $B_N$, and it applies $Q$ directly. Its kernel dimensions and ranks over $\Fp$ bound the rational ones from above and from below, respectively, so agreement with the exact dimensions also excludes a rank drop. This construction is limited to small letter numbers. It reproduces $\dim V^-$ and $h^-$ in all $161$ dominant sectors of $\SO(4)$ with non-empty letter content, $1\le L\le7$ and $4\le j\le20$, in all $106$ such sectors of $\SO(6)$ with $1\le L\le6$ and $6\le j\le18$, and in a selection of $63$ such sectors of $\SO(8)$ with $1\le L\le6$ and $8\le j\le18$, which contain $57$ classes. The selection consists of the sectors with at most four letters, the six-letter sectors with reflection-odd classes and $q_3=0$ or $q_2=q_3=1$, and the five-letter sectors from which $Q$ maps into the latter.

The alternating sum $\sum_{2J}(-1)^{2J}\dim V^-$ reproduces the separately computed reflection-odd index for $N=2,3,4$ through $j=20,18,18$. This tests the dimension tables but not the ranks of $Q$, since each computed rank enters the alternating sum of $h$ at $2J$ and at $2J+1$ with opposite signs.
\section{The Pfaffian sector through level eighteen}
\label{sec:results}

We now apply the method of section~\ref{sec:method} to $\SO(4)$, $\SO(6)$ and $\SO(8)$. Comparing the cohomology counts with the graviton products of section~\ref{sec:gravitons} separates the classes built from the Pfaffian multiplet, those that require excited Pfaffians, and those outside the graviton span. The index gives a complementary probe of the parity $P_N$, and a bound on the letter number determines where trace relations first affect the reflection-even complex. That bound identifies $j=18$ as the first level of $\SO(8)$ at which trace relations enter, and it prepares the comparison with $\SO(10)$ used in sections~\ref{sec:singlet} and~\ref{sec:triality}.

\subsection{Class counts}
\label{sec:classcounts}

We computed the Pfaffian sector $H^-$ on all dominant $\SU(3)_R$ weight spaces with $4\le j\le20$ for $\SO(4)$ and with $2N\le j\le18$ for $\SO(6)$ and $\SO(8)$, for all values of $2J$. For $\SO(8)$ this range contains $46$ dominant weights, and $124$ dominant sectors in it have non-zero $V^-$. Table~\ref{tab:counts} lists the resulting numbers of $\SU(3)_R$ highest-weight classes, summed over $2J$. For $\SO(4)$ and $\SO(6)$ they agree with the predictions that follow from $\so(4)=\su(2)\oplus\su(2)$ and $\so(6)=\su(4)$, described in section~\ref{sec:validation}. The lowest class, at $j=2N$, is the Pfaffian $\Pf(v\cdot\phi)$ with $v\in\mathbb C^3$ and $v\cdot\phi=v_m\phi_m$, which transforms in the representation $[N,0]$.

\begin{table}[t]
\centering
\begin{tabular}{lrrrrrrrrr}
\toprule
$j$ & 4 & 6 & 8 & 10 & 12 & 14 & 16 & 18 & 20\\
\midrule
$\SO(4)$ & 1 & 1 & 3 & 6 & 12 & 19 & 34 & 54 & 85\\
$\SO(6)$ & 0 & 1 & 1 & 4 & 9 & 20 & 40 & 83 & --\\
$\SO(8)$ & 0 & 0 & 1 & 1 & 5 & 11 & 27 & 61 & --\\
\bottomrule
\end{tabular}
\caption{Number of $\SU(3)_R$ highest-weight classes in the Pfaffian sector $H^-$ at level $j$, summed over $2J$. A zero marks a level below that of the Pfaffian, $j<2N$, where the sector is empty, and a dash a level that was not computed.}
\label{tab:counts}
\end{table}

\subsection{Graviton content and excited Pfaffians}
\label{sec:gravcontent}

For every sector that contains classes we computed the Coulomb rank $G^C$ of the graviton products of section~\ref{sec:gravitons} and applied the criterion of section~\ref{sec:coulombtest}: if $G^C=h$, every class in the sector is a graviton product.

For $\SO(4)$ with $j\le20$ and $\SO(6)$ with $j\le18$, all $104$ and $71$ weight sectors with classes satisfy $G^C=h$. In both cases the minimal products, whose $\eps$-factors belong to the Pfaffian multiplet, suffice. For $\SO(6)$ this agrees with the $\SU(4)$ description, in which the Pfaffian multiplet is the $\tr\Phi^3$ multiplet. For $\SO(4)$, seven weight sectors at $j=18$ and $20$ require products with three Pfaffian-type factors. With at most one such factor the graviton Coulomb rank falls short of $h$ by one to three. The shortfall is an exact rational rank in five of these sectors, including the three highest weights, and it follows in the other two by $\SU(3)_R$ covariance. The seven sectors form three $\SU(3)_R$ multiplets with one class each: $[3,0]$ at $2J=3$ and $j=18$, and $[2,1]$ and $[4,0]$ at $2J=4$ and $j=20$, all with $L=6=3N$ letters. A product of three Pfaffian-type factors can lie outside the span of the products with one such factor only if the product of two of them is not cohomologous to a polynomial in trace gravitons. Some products of two components of the Pfaffian multiplet are therefore reflection-even classes outside the trace-graviton polynomials. The same phenomenon occurs for $\SO(8)$ in section~\ref{sec:evenscan}.

For $\SO(8)$ with $j\le18$, $59$ of the $60$ sectors with classes satisfy $G^C=h$. The $106$ highest-weight classes consist of $92$ minimal classes, $13$ classes that require excited Pfaffians, and one class that is not a graviton product. Table~\ref{tab:so8} lists the highest-weight components $(\lambda,2J)$ that are not entirely minimal. The decomposition is exact. In all $23$ sectors where the minimal products do not span the cohomology, their Coulomb rank was computed over $\mathbb Q$ from the closed formula \eqref{eq:epsCartan}, and the Coulomb map is injective on the cohomology of these sectors.

\begin{table}[t]
\centering
\begin{tabular}{rcccrrrr}
\toprule
$j$ & $\lambda$ & $2J$ & $L$ & $h^-_{2J}(\lambda)$ & minimal & excited & non-graviton\\
\midrule
12 & $(3,3,0)$ & 0 & 6 & 1 & 0 & 1 & 0\\
14 & $(3,3,1)$ & 1 & 6 & 1 & 0 & 1 & 0\\
14 & $(4,3,0)$ & 1 & 6 & 2 & 1 & 1 & 0\\
16 & $(4,3,1)$ & 0 & 8 & 1 & 0 & 1 & 0\\
16 & $(4,3,1)$ & 2 & 6 & 3 & 2 & 1 & 0\\
16 & $(4,4,0)$ & 2 & 6 & 1 & 0 & 1 & 0\\
18 & $(3,3,3)$ & 1 & 8 & 1 & 0 & 0 & \textbf{1}\\
18 & $(4,3,2)$ & 1 & 8 & 4 & 2 & 2 & 0\\
18 & $(4,4,1)$ & 1 & 8 & 4 & 3 & 1 & 0\\
18 & $(4,4,1)$ & 3 & 6 & 1 & 0 & 1 & 0\\
18 & $(5,3,1)$ & 1 & 8 & 9 & 7 & 2 & 0\\
18 & $(5,4,0)$ & 1 & 8 & 4 & 3 & 1 & 0\\
\bottomrule
\end{tabular}
\caption{Highest-weight components $(\lambda,2J)$ of the columns of the $\SO(8)$ Pfaffian sector with $j\le18$ that are not entirely minimal gravitons, and their letter numbers $L$. The table gives the highest-weight multiplicity $h^-_{2J}(\lambda)$ of the cohomology and its decomposition into minimal gravitons, gravitons that require excited Pfaffians, and classes that are not graviton products. All other highest-weight components consist of minimal gravitons.}
\label{tab:so8}
\end{table}

The first excited class has highest weight $\lambda=(3,3,0)$, that is $[0,3]$, at $j=12$ and $2J=0$, with $L=6$. It is the $[0,3]$ component of $\eps(\mathrm{Sym}\,\phi^3,\phi,\phi,\phi)$, which transforms in $\mathrm{Sym}^3\otimes\mathrm{Sym}^3$. This component lies in $\Lambda^2(\mathrm{Sym}^3)=[4,1]+[0,3]$, whereas the minimal product ${\Pf}\cdot\tr\phi^2$ contains only $[4,0]\otimes[2,0]=[6,0]+[4,1]+[2,2]$. For $N=3$ the analogous operator $\eps(\mathrm{Sym}\,\phi^3,\phi,\phi)$ transforms in $\mathrm{Sym}^3\otimes\mathrm{Sym}^2=[5,0]+[3,1]+[1,2]$, which are the representations in $[3,0]\otimes[2,0]$, carried by ${\Pf}\cdot\tr\phi^2$. This is consistent with the absence of excited classes for $\SO(6)$ in the computed range.

At $2J=0$ the cohomology is the chiral ring, and the excited Pfaffians account for all of it. On the Cartan subalgebra, a polynomial that is invariant under the Weyl group $W(D_N)$, which permutes the blocks and flips the signs of an even number of them, and odd under $\sigma$ is odd under every single sign flip, and hence odd in the block values $(x_i,y_i,z_i)\in\mathbb C^3$ of the three scalars, for each block separately. Such polynomials form $\mathrm{Sym}^N(\mathbb C[x,y,z]_{\rm odd})$. Its generating function is the coefficient of $t^N$ in the twisted-sector $\frac18$-BPS grand partition function $\mathrm{Pexp}(tI_1^{\mathbb Z_2})$ of Arai, Fujiwara and Imamura, the plethystic exponential of the twisted single-particle partition function $I_1^{\mathbb Z_2}$~\cite{Arai:2018utu}. The space $\mathrm{Sym}^N(\mathbb C[x,y,z]_{\rm odd})$ is the orientifold counterpart of the description of the $\frac18$-BPS chiral ring of the unitary theory by $N$ bosons in a three-dimensional harmonic oscillator, which Biswas, Gaiotto, Lahiri and Minwalla obtained by quantizing the giant gravitons of Mikhailov~\cite{Mikhailov:2000ya,Biswas:2006tj}. On every dominant weight space at $2J=0$ in our range, which comprises $65$, $49$ and $46$ weights for $N=2,3,4$, we find that $h=D$ equals the weight multiplicity of $\mathrm{Sym}^N(\mathbb C[x,y,z]_{\rm odd})$. Here $D$ is computed directly for $N=4$ and follows from $G^C=h$ and \eqref{eq:chain} for $N=2,3$.

This agreement holds for every $N$ by the higher-dimensional Chevalley restriction theorem for orthogonal groups of Song, Xia and Xu~\cite{Song:2023chev}. Since the $Q\psi_m$ are the commutators $[\phi_n,\phi_p]$, the $2J=0$ cohomology is the ring of invariant functions on the scheme of commuting triples, as also discussed by Chang~\cite{Chang:2026bil}. Restriction to the complexified Cartan subalgebra $\mathfrak t$ maps this ring isomorphically onto $\mathbb C[\mathfrak t^3]^{W(D_N)}$. By \eqref{eq:epsCartan}, the Coulomb image of an excited Pfaffian whose slots have block values $\mu_1,\dots,\mu_N$, odd monomials in $(x,y,z)$, is a non-zero multiple of
\begin{equation}
\sum_{\pi\in S_N}\prod_{s=1}^N\mu_s(x_{\pi(s)},y_{\pi(s)},z_{\pi(s)}).
\end{equation}
These functions span $\mathrm{Sym}^N(\mathbb C[x,y,z]_{\rm odd})$. Every reflection-odd class at $2J=0$ is therefore a combination of excited Pfaffians, for every $N$. This is the argument of Song, Xia and Xu, who show that the reflection-odd invariants of commuting tuples are spanned by the coefficients of the Pfaffian of a sum of odd monomials~\cite{Song:2023chev}.

The excited class at $j=12$ also appears in the spectrum of the wrapped brane. In the single-brane partition function of Arai and Imamura, written in their fugacities $q$ and $x,y,z$, it is the $N=4$ instance of the term $(xyz)^3\bar\chi_{(N-4,3)}$ at order $q^{N+2}$, where $\bar\chi_{(a,b)}$ denotes the character with highest weight $(a,0,-b)$~\cite{Arai:2019xmp}. Its $\mathrm U(3)$ highest weight is $(N-1,3,0)$. The term arises from one quantum of the degree-$3$ fluctuation mode of the wrapped brane and has the character $\chi_{[N-1,0]}\chi_{[3,0]}-\chi_{[N,0]}\chi_{[2,0]}$ in Dynkin labels~\cite{Arai:2019xmp}. The subtracted term is the character of ${\Pf}\cdot\tr\phi^2$, so the difference reproduces the representation count above: $[0,3]$ at $N=4$ and zero at $N=3$. More generally, a slot $\mathrm{Sym}\,\phi^n$ with odd $n\ge3$ carries one quantum of the degree-$n$ mode, and the filler slots give the orbital factor of the brane partition function.

\subsection{Index signs and the parity of classes}
\label{sec:index}

The parity $P_N=(-1)^L\sigma^N$ of \eqref{eq:parity} equals $+1$ on every graviton product, and every invariant with $P_N=-1$ vanishes on the Cartan subalgebra. The index detects classes of the other parity. Since $2J=|q|-L$ and $\sigma=-1$ on $H^-$, one has $(-1)^{2J}=(-1)^{|q|+L}=(-1)^{|q|+N}P_N$ there. For every $N$, in each $\SU(3)_R$ column the reflection-odd index therefore satisfies
\begin{equation}
(-1)^{|q|+N}\sum_{2J}(-1)^{2J}h^-_{2J}(\lambda)=\#\{P_N=+1\}-\#\{P_N=-1\},
\end{equation}
where $h^-_{2J}(\lambda)$ is the highest-weight multiplicity of $\lambda$ in $H^-$ at the given $2J$, and the right-hand side counts the classes of the column in $H^-$ by their parity. For odd $N$, where $(-1)^{2J}\sigma=(-1)^{|q|}P_N$ on both $H^+$ and $H^-$, the quantity $(-1)^{|q|}\sum_{2J}(-1)^{2J}\tr_H\sigma$ computed from the $\sigma$-twisted index equals the same parity difference summed over both reflection sectors. The twisted index requires only the integral over $\Or(2N)^-$, which reduces to a torus of rank $N-1$ (section~\ref{sec:dims}), and can therefore be evaluated to higher levels, although in it classes with $P_N=+1$ in $H^+$ can mask classes with $P_N=-1$ in $H^-$. We say that a column in which this parity difference is negative has the wrong sign. Such a column contains classes with $P_N=-1$, which are neither graviton products nor visible on the Coulomb branch. This is the reflection analogue of the detection of non-graviton classes by the charge-conjugation-twisted index of Zwiebel, as used by the Fortuity collaboration~\cite{Zwiebel:2011wa,FortuitySummary}.

We evaluated the exact reflection-odd index of $\SO(8)$ to $j=40$ and the $\sigma$-twisted index of $\SO(10)$ to $j=36$. No column has the wrong sign in either range. The reflection-odd indices of $\SO(6)$ through $j=26$ and of $\SO(10)$ through $j=22$, obtained from the exact dimensions of section~\ref{sec:dims}, have no wrong-sign column either. Classes with $P_N=-1$ can cancel in the index against classes with $P_N=+1$, so this result constrains them only through the signed count. For $\SO(4)$ there is exactly one wrong-sign column through $j=28$, at $\lambda=(4,4,4)$ and $j=24$, with index $-1$. It contains the graded-antisymmetrized threshold class of $\SU(2)$ found by Chang and Lin~\cite{Chang:2022mjp}. We return to this class in section~\ref{sec:control}.

The Pfaffian-sector index of $\SO(2N)$, shifted by the charges of the Pfaffian, has lowest coefficients that do not depend on $N$. The shifted coefficients of $\SO(8)$ and $\SO(10)$ coincide for $j-2N\le6$ and differ at $j-2N=8$. In this comparison, columns are compared where they are dominant at both ranks. The coefficients of $\SO(6)$ agree with them only for $j-2N\le4$. At $j-2N=6$ the shifted column $(0,3,0)$ has index $-1$ for $\SO(6)$ and $-2$ for $\SO(8)$ and $\SO(10)$. The additional class is the excited class of table~\ref{tab:so8} at $\lambda=(4,3,0)$ and $j=14$, which has no counterpart for $\SO(6)$. This is compatible with the covering construction of section~\ref{sec:algebra}, which makes an excitation sector independent of $N$ only for $N\ge N_0$, with $N_0$ growing with the excitation. It is the BMN-sector counterpart of the large-$N$ structure of the full superconformal index, which Arai and Imamura, and Fujiwara et al., obtained from the fluctuations of the wrapped D3-brane~\cite{Arai:2019xmp,Fujiwara:2023bdc}.

\subsection{Letter bounds and the onset of trace relations}
\label{sec:onset}

The charges of a sector fix its letter number, $L=|q|-2J=j/2-2J$. This bounds the levels at which the reflection-even complex can depend on the rank, because of a standard fact about traces of antisymmetric matrices. For $n\ge L$, formal multi-traces of $L$ antisymmetric letters, taken modulo graded cyclicity, the reflection identity \eqref{eq:slotid} and graded commutativity, are linearly independent as functions of $\so(n)$-valued letters.

The argument is the one used for the covering algebra in appendix~\ref{app:faithful}, with the letter $\phi_1$ absent and in a real basis. After full polarization it suffices to treat multilinear monomials in distinct bosonic letters $B_1,\dots,B_L$. For a basis monomial $T$, assign distinct basis vectors of $\mathbb R^n$ to the $L$ gaps between cyclically consecutive letters, which is possible because $n\ge L$. Let $\mu_T$ be the product, over the letters, of the matrix entry that connects the two gaps adjacent to each letter. Then $\mu_T$ occurs in $T$ with a non-zero coefficient and in no other basis monomial, so the basis monomials are linearly independent. This is the multilinear form of the second fundamental theorem for $\Or(n)$, by which trace identities start in degree $n+1$, as established by Procesi and, in Brauer-algebra form, by Lehrer and Zhang~\cite{Procesi:1976,Lehrer:2012}.

Consider now the reflection-even complex of $\SO(2N)$, that is, the complex of $\Or(2N)$-invariant operators, at fixed charges $q$ with level $j=2|q|$. The map from formal multi-traces to invariants commutes with $Q$ and is surjective by the first fundamental theorem~\cite{Procesi:1976}. By the independence statement it is injective in every degree with at most $2N$ letters. If $j\le4N$, every cochain has at most $2N$ letters, and the complex is isomorphic to the formal multi-trace complex at these charges, which is the trace covering of Chang and Lin~\cite{Chang:2024zqi}. Every reflection-even class is then the image of a covering class, and none is fortuitous with respect to the trace covering. The same holds for $\SO(n)$ with $n\ge j/2$.

At fixed $q$ we write $V^\pm_{2J}$ for $V^\pm_{\sect{2J}{q}}$. If $j=4N+2$, the complex differs from the formal one only at $2J=0$, where the cochains have $2N+1$ letters. The spaces $V^+_{2J}$ with $2J\ge1$ and the map $Q:V^+_2\to V^+_1$ are formal, while $V^+_0$ is the quotient of the formal space by the trace relations. The cohomology can therefore differ only at $2J=0$ and $2J=1$. At $2J=1$ the closed subspace exceeds the formal one by
\begin{equation}
\operatorname{rank}^{\rm formal}(Q:V_1\to V_0)-\operatorname{rank}(Q:V^+_1\to V^+_0),
\end{equation}
which is at most the number of trace relations at $2J=0$. At $2J=0$ the map from the formal cohomology is onto, since $V^+_0$ is a quotient of the formal space. Classes outside the image of the covering cohomology can therefore occur only at $2J=1$.

For $\SO(8)$ these statements apply with $4N=16$. Through $j=16$ the reflection-even complex coincides with the formal multi-trace complex, and hence with the reflection-even complex of $\SO(n)$ for every $n\ge8$. At $j=18$ trace fortuity can occur only at $2J=1$, and section~\ref{sec:evenscan} shows that it does, in three columns. The letter bound concerns the trace covering and says nothing about graviton content. That every reflection-even class with $j\le16$ is a graviton product is a computed result, $G^C=h$ in all $55$ sectors with classes. The reflection-odd sector has no analogous letter bound at $N=4$. The comparison with the Pfaffian covering is faithful only for $N\ge N_0$, and at the charges of the singlet $N_0=8$, a value that is sharp for the cochains (section~\ref{sec:singletcov}). That $j=18$ is also the first reflection-odd level with a non-graviton class follows from the data, $G^C=h$ in all $43$ sectors with classes and $j\le16$ (section~\ref{sec:gravcontent}). Both statements concern $N=4$ only.

The charges $\sect{1}{3,3,3}$ of the non-graviton classes illustrate the bound and provide the comparison with $\SO(10)$ used below. At these charges the cochains at $2J=2$, $1$ and $0$ have $7$, $8$ and $9$ letters. For $\Or(n)$ with $n\ge9$ the spaces $V^+_2$, $V^+_1$, $V^+_0$, the incoming map $Q:V^+_2\to V^+_1$ and the outgoing map $Q:V^+_1\to V^+_0$ are therefore those of the formal multi-trace complex. The $2J=1$ cohomology at these charges is the same for every $n\ge9$; $\SO(9)$ would serve as well as $\SO(10)$, which we use to stay within the even series. For $\SO(8)$ the incoming map is the same formal map, while the outgoing map lands in the quotient by five trace relations, since $\dim V^+_0=232$ against $237$ formal multi-traces. The incoming rank is $559$: the computed value is a lower bound, and nilpotence bounds it above by $918-359$, where $918=\dim V^+_2$ and $359$ is a lower bound for the rank of $Q:V^+_3\to V^+_2$. Both outgoing maps are onto, since their computed ranks equal the dimensions of their targets. With $\dim V^+_1=936$ this gives
\begin{equation}
h^+_{\SO(10)}=936-237-559=140,\qquad h^+_{\SO(8)}=936-232-559=145.
\label{eq:h10h8}
\end{equation}
These values hold at these charges and do not define a stabilization threshold for the whole cohomology: at higher levels the trace relations of $\Or(10)$, which begin at $11$ letters, enter as well.

\subsection{The reflection-even sector of \texorpdfstring{$\SO(8)$}{SO(8)}}
\label{sec:evenscan}

The triality analysis of section~\ref{sec:triality} also requires the reflection-even sector $H^+$ of $\SO(8)$, the cohomology of $\Or(8)$-invariant operators. These operators are polynomials in traces, so complete spanning sets are available. We computed $H^+$ on all $53$ dominant weights with $j\le18$, for all $2J$, by the same method. The numbers of highest-weight classes for $j=0,2,\dots,18$ are
\begin{equation}
1,\;0,\;1,\;1,\;4,\;5,\;13,\;21,\;47,\;88.
\end{equation}
All classes are visible on the Coulomb branch: $D=h$ in all $72$ weight sectors with classes. In the $71$ sectors in which every class is a graviton product this follows from $G^C=h$ and $G^C\le D\le h$ in \eqref{eq:chain}. In the remaining sector, $\sect{1}{3,3,3}$, the computed rank of $Q$ out of the sector equals the dimension of its target and is therefore exact, so the computed Coulomb rank of the closed subspace is a lower bound for $D$ (section~\ref{sec:coulombtest}).

Trace relations among $\Or(8)$ invariants begin at $9$ letters. For $j\le16$ the reflection-even complex therefore coincides with those of $\SO(9)$ and $\SO(10)$ (section~\ref{sec:onset}). In all $40$ columns with $2\le j\le16$ our counts agree with the $\SO(9)$ data of the Fortuity collaboration~\cite{FortuitySummary}. Trace relations among $\Or(9)$ and $\Or(10)$ invariants begin at $10$ and $11$ letters, so the reflection-even complex of $\SO(10)$ coincides with that of $\SO(9)$ through $j=18$. The same computation for $\SO(10)$ agrees with the $\SO(9)$ data in all $52$ columns with $2\le j\le18$.

The first rank dependence appears at $j=18$, where the reflection-even sector of $\SO(8)$ has $88$ classes against $85$ for $\SO(9)$. The exact dimensions locate the difference. The dimensions of $V^+$ for $\SO(8)$ and $\SO(10)$ agree in every sector with $j\le18$ except at $2J=0$ and $L=9$. There $\Or(8)$ has exactly one trace relation in each of the columns $[0,0]$, $[2,2]$ and $[6,0]$, at $\lambda=(3,3,3)$, $(5,3,1)$ and $(7,1,1)$. At these weights $Q:V^+_1\to V^+_0$ is onto for both $\SO(8)$ and $\SO(10)$, while $V^+_1$, $V^+_2$ and the map between them coincide. Each relation therefore creates one additional $Q$-closed operator at $2J=1$, and hence one additional class.

In the columns $[2,2]$ and $[6,0]$ the additional classes are graviton products that lie outside the trace-graviton span: products of two components of the Pfaffian multiplet, which are polynomials in traces but not cohomologous to polynomials in trace gravitons. In these columns the formal multi-trace complex coincides with the reflection-even complex of $\SO(10)$, so trace monotones account for at most $h^+_{\SO(10)}(\lambda)=h^+_{\SO(8)}(\lambda)-1$ classes at $2J=1$. The Pfaffian-pair classes are therefore fortuitous with respect to the trace covering, although they are products of generalized monotones. Choi, Choi and Kim found the same pattern on the conifold, where multi-trace operators that factorize into baryon--antibaryon pairs are generalized monotones without being graviton polynomials~\cite{Choi:2026faq}. Products of two Pfaffian-type factors are needed at the seven weights $(3,3,3)$, $(4,3,2)$, $(4,4,1)$, $(5,2,2)$, $(5,3,1)$, $(6,2,1)$ and $(7,1,1)$ at $2J=1$. At $(3,3,3)$ they raise the graviton rank from $140$ to $144$, against $h=145$.

In the column $[0,0]$ the additional class is not a graviton product. Section~\ref{sec:partner} identifies it as the triality partner of the reflection-odd singlet of section~\ref{sec:singlet}. Apart from these three columns, that is, the non-graviton singlet and the two columns of Pfaffian pairs, every reflection-even class with $j\le18$ is a product of trace gravitons, so that the whole sector except the singlet is spanned by graviton products. In terms of the index, the reflection-even BMN index of $\SO(8)$ and the index of $\SO(9)$ first differ at $j=18$, by $-\chi_{[0,0]}-\chi_{[2,2]}-\chi_{[6,0]}$, the contribution of the three additional classes.
\section{The non-graviton singlet at level eighteen}
\label{sec:singlet}

Table~\ref{tab:so8} contains exactly one highest-weight component $(\lambda,2J)$ with $j\le18$ that is not spanned by graviton products: the $\SU(3)_R$ singlet at $j=18$ and $2J=1$, with charges $\sect{1}{3,3,3}$. In this section we determine the cohomology at these charges exactly and identify the Coulomb image of the additional class. We then show that this image is orthogonal, in the Fischer inner product, to everything generated by supersymmetry descendants of bosonic invariants. The class is therefore a non-graviton class.

\subsection{The cohomology at the charges \texorpdfstring{$\sect{1}{3,3,3}$}{(1;3,3,3)}}
\label{sec:singletcohomology}

The same charges $\sect{1}{3,3,3}$ recur in the reflection-even sector of $\SO(8)$, in $\SO(10)$ and in $\SO(7)$, and table~\ref{tab:seedcomplexes} collects the four complexes. Here we derive its first row. The exact reflection-odd dimensions are $\dim V^-=167$, $633$ and $520$ at $2J=0$, $1$ and $2$, so the relevant part of the complex is
\begin{equation}
V^-_2\xrightarrow{\;Q\;}V^-_1\xrightarrow{\;Q\;}V^-_0.
\end{equation}
The computed ranks are $\operatorname{rank}(Q:V^-_1\to V^-_0)=167$ and $\operatorname{rank}(Q:V^-_2\to V^-_1)=347$, and both are exact. The rank $167$ equals $\dim V^-_0$, so the outgoing map is onto. The rank $347$ is a lower bound that nilpotence turns into an equality. Since $Q^2=0$, the image of $Q:V^-_3\to V^-_2$ lies in the kernel of $Q:V^-_2\to V^-_1$, and therefore
\begin{equation}
\operatorname{rank}(Q:V^-_2\to V^-_1)\le\dim V^-_2-\operatorname{rank}(Q:V^-_3\to V^-_2)\le520-173=347,
\end{equation}
where $173$ is the computed rank of $Q:V^-_3\to V^-_2$, itself a lower bound. Hence
\begin{equation}
h=633-167-347=119.
\end{equation}

\begin{table}[t]
\centering
\begin{tabular}{lrrrrrrr}
\toprule
Complex & $\dim V_2$ & $\dim V_1$ & $\dim V_0$ & $r_{\rm in}$ & $r_{\rm out}$ & $h$ & $G^H$\\
\midrule
$\SO(8)$, reflection-odd & 520 & 633 & 167 & 347 & 167 & 119 & 118\\
$\SO(8)$, reflection-even & 918 & 936 & 232 & 559 & 232 & 145 & 144\\
$\SO(10)$, reflection-even & 918 & 936 & 237 & 559 & 237 & 140 & 140\\
$\SO(7)$ & 918 & 903 & 220 & 559 & 220 & 124 & 123\\
\bottomrule
\end{tabular}
\caption{The complexes at the charges $\sect{1}{3,3,3}$. The entries are the dimensions of the spaces $V_{2J}$ of gauge invariants at $2J=2,1,0$, the ranks $r_{\rm in}$ of $Q:V_2\to V_1$ and $r_{\rm out}$ of $Q:V_1\to V_0$, the dimension $h=\dim V_1-r_{\rm in}-r_{\rm out}$ of the cohomology at $2J=1$, and the rank $G^H$ of the graviton products in that cohomology. In each row exactly one class lies outside the graviton span, except for $\SO(10)$, where there is none. The first row is derived in this section; the reflection-even rows are derived in sections~\ref{sec:onset} and~\ref{sec:partner}, and the $\SO(7)$ row in section~\ref{sec:SO7restriction}.}
\label{tab:seedcomplexes}
\end{table}

All $119$ classes are visible on the Coulomb branch. The computed Coulomb rank of the closed subspace is $119$; it is a lower bound for $D$ because the outgoing rank is exact, and $D\le h$ by \eqref{eq:chain}, so $D=h=119$ and the Coulomb map is injective on the cohomology of this sector. The graviton products have Coulomb rank $G^C=118$, computed over $\mathbb Q$ by fraction-free elimination on their exact Coulomb images. Injectivity then gives $G^H=G^C=118$: the graviton products span $118$ of the $119$ classes, and exactly one class lies outside their span. The same conclusion follows without the nilpotence bound. The computed ranks give $h\le119$ and $G^H\ge G^C\ge118$, and the explicit closed operator of section~\ref{sec:exactproof} lies outside the graviton span, which forces $h=119$ and $G^H=118$.

The additional class is an $\SU(3)_R$ singlet. In a polynomial $\mathrm{GL}(3)$ representation of degree nine with weight multiplicities $m(q)$, Weyl alternation gives the multiplicity of the $\SU(3)_R$ singlet as
\begin{equation}
m(3,3,3)-2\,m(4,3,2)+m(5,2,2)+m(4,4,1)-m(5,3,1).
\label{eq:alternation}
\end{equation}
At the four weights $(4,3,2)$, $(5,2,2)$, $(4,4,1)$ and $(5,3,1)$ every class is a graviton product, $G^C=h$. The quotient of the cohomology by the graviton span therefore vanishes at these weights and is one-dimensional at $(3,3,3)$, so by \eqref{eq:alternation} it is a single $\SU(3)_R$ singlet.

The exact index shows the same class. The reflection-odd index of the singlet column $\lambda=(3,3,3)$ at $j=18$ is
\begin{equation}
\sum_{2J}(-1)^{2J}m^-_{2J}(3,3,3)=9-23+19-10+6-2=-1,
\label{eq:singletindex}
\end{equation}
where $m^-_{2J}$ is the highest-weight multiplicity of $(3,3,3)$ in $V^-$ at $2J=0,\dots,5$. In this column the cohomology vanishes for $2J\ne1$. At $2J=0,2,4,5$ the computed $h$ at weight $(3,3,3)$ is zero, and computed values are upper bounds. At $2J=3$ the cohomology has dimensions $15$, $12$, $8$, $5$ and $4$ at the weights of \eqref{eq:alternation}, so it contains no singlet. With the cohomology concentrated at $2J=1$, the value $-1$ of the index requires exactly one class in the column, independently of the rank of the incoming map. The index does not distinguish graviton from non-graviton classes. That this class is not a graviton product follows from $G^H=118<h$, and also, without any rank of $Q$, from the obstruction derived next applied to the explicit closed family of section~\ref{sec:exactproof}.

\subsection{Coulomb image and the Fischer-pairing obstruction}
\label{sec:exactcert}

On the Cartan subalgebra of $\so(8)$ every letter reduces to four $2\times2$ blocks. We write $\phi^{(i)}_m$ and $\psi^{(i)}_m$, $i=1,\dots,4$, for the block values. The Weyl group $W(D_4)$ permutes the blocks and flips an even number of block signs. The Coulomb image of a reflection-odd invariant is a polynomial that is invariant under $W(D_4)$ and odd under a single sign flip. In the sector $\sect{1}{3,3,3}$ these functions form a space of dimension $147$, and the Coulomb image of $H^-$ is a $119$-dimensional subspace of it. The analogous dimensions at $(4,3,2)$, $(5,2,2)$, $(4,4,1)$ and $(5,3,1)$ are $122$, $89$, $77$ and $68$, so by \eqref{eq:alternation} the function space contains
\begin{equation}
147-2\cdot122+89+77-68=1
\end{equation}
$\SU(3)_R$ singlet. Since the Coulomb map is injective and $\SU(3)_R$-equivariant, this singlet is, up to normalization, the Coulomb image of the singlet class. It is
\begin{equation}
F=\sum_{i<j}\Delta_{ijk}\,\Delta_{ijl}\,\big(s_{ij}+s_{ji}\big),\qquad \Delta_{ijk}=\eps_{mnp}\phi^{(i)}_m\phi^{(j)}_n\phi^{(k)}_p,\qquad s_{ij}=\psi^{(i)}\cdot\phi^{(j)},
\label{eq:F}
\end{equation}
where $\{k,l\}$ is the complement of $\{i,j\}$ in $\{1,2,3,4\}$.

We compare $F$ with the functions that supersymmetry descendants of bosonic invariants can produce. Let $R_{D_4}$ be the ring of $W(D_4)$-invariant polynomials in the bosonic block variables, and let
\begin{equation}
S=\mathrm{span}\{B\,\Qp^m(A)\;:\;A,B\in R_{D_4}\}.
\end{equation}
Below, $S$ denotes the component of this space of weight $(3,3,3)$, and $S_{\rm odd}$ and $S_{\rm even}$ denote its reflection-odd and reflection-even parts. The Coulomb map commutes with $\Qp^m$, and a graviton product at $2J=1$ contains exactly one first descendant of a bosonic generator. Its Coulomb image therefore lies in $S$, and so does the Coulomb image of any sum of products of $\Qp$-descendants of bosonic invariants.

The function $F$ is orthogonal to $S$ in the Fischer inner product $\langle\phi^\alpha,\phi^\beta\rangle=\delta_{\alpha\beta}\,\alpha!$ of monomials in the bosonic block variables, in which the single $\psi$ is treated as a commuting variable (appendix~\ref{app:conventions}). This form is positive definite and $\mathfrak u(3)$ invariant. To establish the orthogonality, let $A$, non-constant, and $B$ run over the bases of $W(D_4)$-orbit sums of $R_{D_4}$ in all multidegrees compatible with the weight $(3,3,3)$, with $A$ and $B$ of opposite reflection parity and $\Qp^m(A)\neq0$. This gives $435$ generators $B\,\Qp^m(A)$ that span $S_{\rm odd}$, and exact computation gives
\begin{equation}
\langle F,B\,\Qp^m(A)\rangle=0\quad\text{for all $435$ generators},\qquad \langle F,F\rangle=4320.
\end{equation}
The reflection $\sigma$ preserves the Fischer form and $F$ is reflection-odd, so $F$ is also orthogonal to $S_{\rm even}$. Since the form is positive definite and $F\ne0$, it follows that $F\notin S$.

The orthogonality can also be derived by hand, from the stronger statement that $S$ contains no $\SU(3)_R$ singlet at this weight. We regard the functions as representations of $\mathrm{GL}(3)$, under which $\phi_m$ and the supercharges $\Qp^m$ transform in the fundamental representation, so that a singlet of weight $(3,3,3)$ is the cube of the determinant. The products $B\,\Qp^m(A)$ with $A$ of degree $d$ and $B$ of degree $8-d$ form the image of an equivariant map. They can contain a singlet only if the product of the invariants of degree $d$ with those of degree $8-d$ contains the representation of highest weight $(3,3,2)$. The characters of the invariants show that it does not when the two factors have the same reflection parity. For factors of opposite parity it occurs only in the product of the quadratic invariants $g_{np}=\sum_i\phi^{(i)}_n\phi^{(i)}_p$, of highest weight $(2,0,0)$, with the reflection-odd invariants of degree six, once for each assignment of the two degrees to $A$ and $B$. In the latter one block occurs cubically and the other three linearly, so they form $\mathrm{Sym}^3\otimes\mathrm{Sym}^3=(6,0,0)+(5,1,0)+(4,2,0)+(3,3,0)$, and by the Pieri rule only $(3,3,0)$ contributes.

This component contains the totally symmetric tensor
\begin{equation}\label{eq:cubiccovariant}
\Theta_{mnp}=\sum_i\sum_{(j,k,l)}\big(\phi^{(i)}\times\phi^{(j)}\big)_m\big(\phi^{(i)}\times\phi^{(k)}\big)_n\big(\phi^{(i)}\times\phi^{(l)}\big)_p.
\end{equation}
Here $(j,k,l)$ runs over the orderings of the three blocks other than $i$, and $(\phi^{(i)}\times\phi^{(j)})_m=\eps_{mab}\phi^{(i)}_a\phi^{(j)}_b$. The cross product takes values in $\Lambda^2\mathbb C^3\cong(\mathbb C^3)^*\otimes\det$, so that its contraction with $g_{np}$ is $\mathrm{GL}(3)$-equivariant. The component in question is $\Theta_{mnp}\,g_{np}$. In this contraction only the term of $g_{np}$ from the block $j$ survives, and
\begin{equation}\label{eq:syzygy}
\Theta_{mnp}\,g_{np}=2\sum_{i\neq j}\big(\phi^{(i)}\times\phi^{(j)}\big)_m\,\Delta_{ijk}\,\Delta_{ijl}=0,
\end{equation}
with $\{k,l\}$ the complement of $\{i,j\}$ as in \eqref{eq:F}, because the summand is odd under the exchange of $i$ and $j$. The two singlets that this component could produce vanish as well. The total symmetry of $\Theta$ gives $\Theta_{mnp}\,\Qp^m(g_{np})=0$, and applying $\Qp^m$ to \eqref{eq:syzygy} then gives $\Qp^m(\Theta_{mnp})\,g_{np}=0$. Hence $S$ contains no singlet, and every singlet function is orthogonal to $S$ because the Fischer form is $\mathfrak u(3)$ invariant.

The singlet class is therefore not cohomologous to any sum of products of $\Qp$-descendants of bosonic gauge invariants. Indeed, $Q$-exact operators vanish on the Cartan subalgebra, so the Coulomb image of such a sum, corrected by any $Q$-exact operator, lies in $S$, whereas the Coulomb image of the singlet class is a non-zero multiple of $F$. The rank of $S_{\rm odd}$ over $\mathbb Q$ is $118$. It equals the graviton Coulomb rank, so at this weight the Coulomb images of the graviton products span all of $S_{\rm odd}$. The obstruction does not depend on a choice of graviton generators, as long as their supersymmetry multiplets have bosonic bottom components, because $S$ contains the descendants of all bosonic invariants. It also excludes products of lower classes of either reflection parity: every lower class in both sectors is a graviton product (sections~\ref{sec:gravcontent} and~\ref{sec:evenscan}), so the Coulomb images of such products lie in $S$.

This obstruction is a statement about cohomology and has no counterpart for arbitrary invariant functions. The naive fermion-linear analogue of the classical polarization theorems, that every $\psi$-linear $W(D_4)$-invariant lies in the corresponding space $S$, fails for elementary reasons. At weight $(3,1,1)$ there are three invariant $\psi$-linear orbit sums, while the reflection-odd analogue of $S$ has rank two, and the Coulomb image of the operator $\sum_m\eps(\psi_m,\phi_m,\phi_1^{\times2})$ lies outside it. This operator is not $Q$-closed, and the cohomology at that weight is spanned by graviton products. The content of the result above is that at $(3,3,3)$ the Coulomb image of the cohomology itself is not contained in $S$.

The bosonic analogue concerns several copies of the reflection representation. For a finite reflection group acting on one copy of its reflection representation, Solomon showed that the invariant differential forms are a free exterior algebra over the invariant polynomials, generated by the differentials $df_i$ of the basic invariants $f_i$~\cite{Solomon:1963}. For the diagonal action of $W(D_N)$ with $N\ge4$ on two or more copies, Wallach observed that polarizations of the basic invariants do not generate the invariants~\cite{Wallach:1993}. For two copies $(x,y)$ one needs in addition Wallach's higher polarizations
\begin{equation}
\mathcal P_{r_1}\cdots\mathcal P_{r_d}(x_1\cdots x_N),\qquad \mathcal P_r=\sum_iy_i^r\partial_{x_i},
\end{equation}
with odd $r_i$, finitely many of which complete a generating set. This is due to Wallach and to Hunziker and is recalled by Losik, Michor and Popov~\cite{Wallach:1993,Hunziker:1997,Losik:2006}. For any number of copies, Song, Xia and Xu give a basis of the invariants that are odd under the reflection, namely the coefficients of the Pfaffian of a sum of odd monomials in the copies~\cite{Song:2023chev}.

Pfaffian operators realize these higher polarizations. On the Coulomb branch $\Pf(\phi_1)$ maps to $x_1\cdots x_N$ up to normalization, and the excited Pfaffians whose non-filler slots do not contain $\phi_1$ map to higher polarizations of this product; more generally, non-filler slots replace distinct factors $x_i$ of $x_1\cdots x_N$ by their block values (section~\ref{sec:exactproof}). For instance,
\begin{equation}
C\big(\eps(YYY,X^{\times3})\big)=-96\,\mathcal P_3(x_1x_2x_3x_4),
\end{equation}
which supplies the excited class at $j=12$. The singlet requires a fermion-linear higher polarization of the same kind that is not in the $R_{D_4}$-module generated by $\Qp(R_{D_4})$. The covering family \eqref{eq:singletfamily} of section~\ref{sec:covering} provides it: the $\SU(3)_R$-singlet projection of its rank-four member is $Q$-closed and has Coulomb image $\tfrac{8}{15}F$.
\section{A covering space for the Pfaffian sector}
\label{sec:covering}

A class found at one rank says something about wrapped branes only once it is placed in a family of operators defined at every rank. In the Pfaffian sector this is not automatic, because the number of slots of the $\eps$-tensor grows with $N$. The generalized monotones of section~\ref{sec:monotone} are defined through a covering algebra in which this growth is absorbed into the ground state $\Pf(\phi_1)$ and an inverse letter. We construct that algebra here. Closure in it implies closure at every rank, and at large rank it reproduces the Pfaffian sector exactly. We then exhibit a covering-closed family whose rank-four member represents the non-graviton singlet of section~\ref{sec:singlet}, and we test the construction on an $\SO(4)$ class that it should not reach.

\subsection{The fermionic-integral identity}
\label{sec:fermionic}

Gaiotto and Lee used a fermionic representation of determinants~\cite{Gaiotto:2021xce}. Following them, Choi, Choi and Kim represent a determinant baryon as an integral over complex Grassmann variables~\cite{Choi:2026faq}. Excitations of the baryon then become multi-traces in an inverse letter, with coefficients that do not depend on the rank. For a Pfaffian the analogous representation uses $2N$ real Grassmann variables, $\Pf A=\int d\chi\,e^{\frac12\chi^TA\chi}$, and Wick contraction with the propagator $A^{-1}$ turns insertions into traces.

The resulting identity is the following. Let $A$ be an invertible antisymmetric $2N\times2N$ matrix with commuting entries, let $u=A^{-1}$, and let $B_1,\dots,B_k$ be antisymmetric matrices, each with entries that are all Grassmann even or all Grassmann odd. For $k\le N$,
\begin{equation}\label{eq:fermionic}
\eps(B_1,\dots,B_k,\underbrace{A,\dots,A}_{N-k})=2^N(N-k)!\,\Pf(A)\,\mathcal F(B_1,\dots,B_k),
\end{equation}
where
\begin{equation}\label{eq:Fdef}
\mathcal F(B_1,\dots,B_k)=\sum_{\pi\in S_k}\kappa(\pi)\prod_{\text{cycles }(c_1\to\cdots\to c_m)\text{ of }\pi}\frac{(-1)^{m+1}}{2}\,\tr\big(uB_{c_1}uB_{c_2}\cdots uB_{c_m}\big).
\end{equation}
Here $\kappa(\pi)$ is the Koszul sign of reordering the matrices from slot order to the order in which they appear when the cycles of $\pi$ are read in sequence. The summand does not depend on where each cycle is started or on the order in which the cycles are read, because graded cyclicity and graded commutativity of the traces compensate the change of $\kappa$. For $k>N$ the function $\mathcal F(B_1,\dots,B_k)$ vanishes identically on $2N\times2N$ matrices.

To derive \eqref{eq:fermionic}, take first bosonic $B_s$ and set $T=\sum_st_sB_s$ with commuting parameters $t_s$. Since $\Pf(A+T)^2=\det(A)\det(1+uT)$, the power series $\Pf(A+T)$ and $\Pf(A)\exp(\frac12\tr\log(1+uT))$ in the $t_s$ have the same square and the same non-zero constant term, so they are equal:
\begin{equation}\label{eq:Pfexp}
\Pf(A+T)=\Pf(A)\exp\Big(\sum_{m\ge1}\frac{(-1)^{m+1}}{2m}\tr(uT)^m\Big).
\end{equation}
The Pfaffian of an antisymmetric matrix is its $\eps$-contraction with $N$ equal slots divided by $2^NN!$, so the coefficient of $t_1\cdots t_k$ on the left-hand side is $\eps(B_1,\dots,B_k,A^{\times(N-k)})/(2^N(N-k)!)$. On the right-hand side, the multilinear part of $\tr(uT)^m$ on $m$ distinct labels is $m$ times the sum over their cyclic orders, which cancels the factor $1/m$. Exponentiation then turns products of sums over cyclic orders into a sum over permutations, in which each cycle of length $m$ contributes $(-1)^{m+1}/2$ times its trace. This is \eqref{eq:Fdef} with $\kappa=1$. The left-hand side of \eqref{eq:Pfexp} is a polynomial of degree $N$ in $T$, so the multilinear coefficients with $k>N$ vanish. For Grassmann-odd matrices we write $B_s=\theta_s\widetilde B_s$ with auxiliary odd parameters $\theta_s$ and bosonic $\widetilde B_s$, and apply the bosonic identity. Moving the $\theta_s$ to the front produces the slot-order sign on the left and $\kappa(\pi)$ on the right.

A slot word $W$ enters an $\eps$-contraction only through its antisymmetric part,
\begin{equation}\label{eq:antisymslot}
\overline W=\tfrac12\big(W-W^T\big)=\tfrac12\big(W+c(W)\,\operatorname{rev}W\big),
\end{equation}
with $c(W)$ as in \eqref{eq:slotid}. The identity therefore assigns to every formal monomial $(\vec W,M)$ of \eqref{eq:family}, with slot words $\vec W=(W_1,\dots,W_k)$, the polynomial
\begin{equation}\label{eq:iota}
\iota(\vec W,M)=\mathcal F(\overline W_1,\dots,\overline W_k)\,\tr M
\end{equation}
in traces of words in the letters and in $u$. Its coefficients do not depend on $N$, and the member of the family at rank $N$ is recovered as
\begin{equation}\label{eq:XNiota}
X_N(\vec W,M)=2^N\,\Pf(\phi_1)\,\iota(\vec W,M)\big|_{u=\phi_1^{-1}}.
\end{equation}
The rank enters only through the prefactor, which is the wrapped-brane ground state, and through the size of the matrices on which the traces are evaluated.

\subsection{The covering algebra and closure at every rank}
\label{sec:algebra}

Let $\mathcal C$ be the graded-commutative algebra of multi-traces of words in the seven letters and in a bosonic antisymmetric letter $u$, subject only to three identities: graded cyclicity of the trace; the reflection identity $\tr W=(-1)^{\ell(W)}s_{\rm rev}(W)\tr\operatorname{rev}W$, in which the length $\ell(W)$ counts $u$ as a letter; and the cancellation $u\phi_1=\phi_1u=1$. The supercharge induces a differential $\delta$ on $\mathcal C$, which acts on the letters as in \eqref{eq:Q} and annihilates $u$. It acts in the same way on formal monomials, slot by slot and on the trace factor. No formal monomial has a slot equal to the single letter $\phi_1$, so the cancellation never produces the rank-dependent trace of the identity, and the coefficients of $\iota(\vec W,M)$ and $\iota(\delta(\vec W,M))$ are rank independent. A combination $a=\sum_ia_i(\vec W_i,M_i)$ of formal monomials is covering-closed, in the sense of section~\ref{sec:monotone}, if
\begin{equation}
\iota(\delta a)=\sum_ia_i\,\iota\big(\delta(\vec W_i,M_i)\big)=0\qquad\text{in }\mathcal C.
\end{equation}
We write $X_N(a)=\sum_ia_iX_N(\vec W_i,M_i)$ for the rank-$N$ member of the family, and $\ker\iota$ for the space of combinations that $\iota$ maps to zero; the latter is not the kernel of $Q$.

Covering closure implies closure at every rank: if $a$ is covering-closed, then $QX_N(a)=0$ for all $N$. Indeed, $Q$ annihilates the filler slots and $\Pf(\phi_1)$, and it commutes with the substitution $u=\phi_1^{-1}$ because $Q\phi_1=0$. On the dense set of invertible $\phi_1$, equation~\eqref{eq:XNiota} therefore gives
\begin{equation}\label{eq:QXN}
QX_N(a)=2^N\,\Pf(\phi_1)\,\iota(\delta a)\big|_{u=\phi_1^{-1}}.
\end{equation}
The identities that define $\mathcal C$ hold for matrices of every size, so the right-hand side vanishes on a dense set and hence identically.

The converse holds at sufficiently large rank, because the evaluation of $\mathcal C$ on matrices is then faithful. Let $c\in\mathcal C$ be homogeneous in the excitation charges and free of the factor $\tr(u\phi_1)=\tr1$, as every $\iota(\vec W,M)$ and $\iota(\delta(\vec W,M))$ is. Let $d$ be the largest number of letters other than $\phi_1$ and $u$ in a monomial of $c$, and let $r$ be the number of distinct non-zero exponents $m$ for which the pure trace $\tr\phi_1^m$, with $\phi_1^{-1}=u$, occurs as a factor of a monomial of $c$. If $c$ vanishes at $u=\phi_1^{-1}$ for all $\so(2N)$-valued letters at a single rank $N\ge d+r$, then $c=0$ in $\mathcal C$. The argument, given in appendix~\ref{app:faithful}, takes $\phi_1$ to be a generic element of the Cartan subalgebra and reads off each basis monomial of $\mathcal C$ from a distinct monomial in the matrix entries of the other letters. It needs $d$ blocks to separate the letters of a monomial and $r$ further blocks to separate the pure traces.

Faithfulness identifies the covering with the Pfaffian sector at large rank. Fix the charges $(e_1,q_2,q_3)$ of an excitation above $\Pf(\phi_1)$, with $e_1=q_1-N$, and a value of $2J$. Let $\mathcal A_{2J}$ be the span of the formal monomials with these charges, and let $N_0=d+r$ be the bound computed for the set of all monomials of $\mathcal C$ that occur in $\iota(\mathcal A_{2J})$ or in $\iota(\delta\mathcal A_{2J})$. For $N\ge N_0$, a combination in $\mathcal A_{2J}$ is $Q$-closed at rank $N$ if and only if it is covering-closed, by \eqref{eq:QXN} and faithfulness applied to $\iota(\delta a)$. For such $N$ the map $a\mapsto X_N(a)$ induces an isomorphism from $\mathcal A_{2J}/\ker\iota$ onto the space of reflection-odd cochains of $\SO(2N)$ in the sector $\sect{2J}{N+e_1,q_2,q_3}$. It is surjective because every $\eps$-word in the sector equals $(N-k)!\,X_N(\vec W,M)$ with $\vec W$ its non-filler slots, and injective by faithfulness applied to $\iota(a)$. For such $N$ it also induces an isomorphism from the covering cohomology at $2J$, defined as the covering-closed combinations modulo $\ker\iota$ and $\delta\mathcal A_{2J+1}$, onto $H^-$ of $\SO(2N)$ in that sector. The reason is that $X_N$ is a chain map, $QX_N(a)=X_N(\delta a)$, which is surjective on cochains at $2J+1$ for every $N$, so that the boundaries at rank $N$ are the images of $\delta\mathcal A_{2J+1}$. The closure criterion also shows that $\delta$ maps $\ker\iota\cap\mathcal A_{2J}$ into $\ker\iota$, so the covering cohomology is well defined.

In particular, every class of the Pfaffian sector with these excitation charges is a generalized monotone once $N\ge N_0$; section~\ref{sec:singletcov} shows that for the charges of the singlet this already holds at $N=4$. The isomorphisms on cochains and on cohomology need the bound for the cochain monomials in $\iota(\mathcal A_{2J})$ as well as for their $\delta$-images; faithfulness on the images alone gives only the closure criterion. At fixed excitation charges, strong fortuity in the Pfaffian sector can therefore occur only for $N<N_0$, as in the trace sector, and black-hole candidates require charges that grow with $N$. Covering-closed combinations are, for the ground state $\Pf(\phi_1)$, the analogue of the baryonic monotones of Choi, Choi and Kim, which are generalized monotones in their sense~\cite{Choi:2026faq}.

\subsection{Restriction in rank}
\label{sec:branesep}

Members of a family at neighbouring ranks are related by a restriction that separates one brane from the stack. Embed $\so(2N)\oplus\so(2)\subset\so(2N+2)$ block-diagonally and restrict the letters by $\phi_1\mapsto\mathrm{diag}(\phi_1,x\mathbb J)$, with an even parameter $x$ and $\mathbb J=\left(\begin{smallmatrix}0&1\\-1&0\end{smallmatrix}\right)$, while every other letter is extended by a vanishing $\so(2)$ block. Let $\rho$ denote the coefficient of $x^1$ after this restriction. The restricted configuration is a point of the Coulomb branch at which one D3-brane and its orientifold image are separated from the stack. On the ground state $\Pf(\mathrm{diag}(A,x\mathbb J))=x\,\Pf A$, so $\rho$ extracts the part linear in the position of the separated brane.

For every combination $a$ of formal monomials, only odd powers of $x$ occur in the restriction of $X_{N+1}(a)$, and
\begin{equation}\label{eq:rho}
\rho\,X_{N+1}(a)=2X_N(a).
\end{equation}
The reason is that the two indices of each slot of the rank-$(N+1)$ $\eps$-tensor must lie in a single block, so exactly one slot occupies the $\so(2)$ block. A filler slot placed there gives $2x$ times the rank-$N$ contraction of the remaining slots, and the $N+1-k$ choices of filler combine with the normalization $1/(N+1-k)!$ into $2x\,X_N(a)$. A slot word that is a pure power $\phi_1^\ell$ with $\ell\ge3$ odd, and the $\so(2)$ parts of pure traces $\tr\phi_1^\ell$, contribute only powers $x^\ell$ with $\ell\ge2$. Even powers are absent because $X_{N+1}(a)$ is odd under the reflection of the $\so(2)$ block, which maps $x\to-x$.

The map $\rho$ commutes with $Q$ and with $\Qp^m$. In $Q\psi_m$ and $Qf$ at most one factor is $\phi_1$, so their $\so(2)$ components vanish, and $\Qp^m\phi_1$ has no $\so(2)$ component. Consequently $\rho$ maps $Q$-exact operators to $Q$-exact operators. It also maps graviton products of $\SO(2N+2)$ to graviton products of $\SO(2N)$. A trace graviton restricts to the same trace graviton plus constant multiples of powers of $x$, and the multiplet $\tr\Phi^{2N+2}$ at rank $N$ is a polynomial in the lower ones. Every $\eps$-factor restricts to odd powers of $x$ only, so the term linear in $x$ of a product with one $\eps$-factor is the linear term of that factor times the trace gravitons, while products with three or more $\eps$-factors have no linear term. Appendix~\ref{app:rankrestriction} gives the restriction formula and these steps in full.

Two consequences follow for every combination $a$ and every $N$. If $X_{N+1}(a)$ is $Q$-closed, so is $X_N(a)$, since $2QX_N(a)=\rho\,QX_{N+1}(a)$. If $X_N(a)$ is not cohomologous to a sum of graviton products, neither is $X_{N+1}(a)$, since $X_{N+1}(a)=\mathcal G+Q\mathcal V$ with a graviton combination $\mathcal G$ would imply $2X_N(a)=\rho\,\mathcal G+Q\rho\mathcal V$. Closure thus propagates downwards in rank and the non-graviton property upwards.

This is an operator-level form of the inclusion $I_{N+1}\subset I_N$ of the rank-$N$ relation spaces in the covering of Choi, Choi and Kim~\cite{Choi:2026faq}. Here the inclusion is realized by a chain map that in addition preserves graviton products. The map $\rho$ is the Pfaffian counterpart of the block restrictions used by the Fortuity collaboration~\cite{FortuitySummary}. The difference is that here the $x^0$ term of the restriction vanishes identically.

\subsection{The singlet sector in the covering}
\label{sec:singletcov}

We now apply the construction to the sector $\sect{1}{3,3,3}$ of $\SO(8)$, that is, to the excitation charges $\sect{1}{-1,3,3}$ above $\Pf(\phi_1)$. A formal monomial with these charges has $k$ non-filler slots of lengths $\ell_s$ and a multi-trace of total length $\ell(M)$ with $\sum_s\ell_s+\ell(M)=k+4$, hence $\sum_s(\ell_s-1)+\ell(M)=4$. Its slot words therefore have length at most $5$ and its traces total length at most $4$. Every non-filler slot either contains one of the at most six letters other than $\phi_1$ or is an odd power $\phi_1^\ell$ with $\ell\ge3$, since even powers have vanishing antisymmetric part. At most two such powers fit, so $k\le8$. With slot words in canonical form and trace words reduced by graded cyclicity, there are $4134$ formal monomials at $2J=1$ and $2504$ at $2J=2$.

The $\iota$-images of the monomials at $2J=1$ involve $7630$ monomials of $\mathcal C$ and their $\delta$-images involve $1735$. Both sets have $d=6$ and pure-trace exponents $\{2,4\}$, so $N_0=8$, and the large-rank identification of section~\ref{sec:algebra} holds for $N\ge8$. The bound is sharp for the cochains. It is also sharp for the equivalence of closure and covering closure. The formal monomial $m_2$ with slots $(X^3,X^3,Y,Y,Y,Z,Z,\psi_2)$ and its analogue $m_3$ with slots $(X^3,X^3,Y,Y,Z,Z,Z,\psi_3)$ have eight non-filler slots, so their members vanish for $N\le7$, although $\iota(m_{2,3})$ and $\iota(\delta m_{2,3})$ are non-zero. Thus $X_7(m_2)=0$ is trivially closed, whereas $m_2$ is not covering-closed. Their $\delta$-images agree, $\iota(\delta m_2)=\iota(\delta m_3)$. For $N\ge8$, the derivation $\mathrm{ad}\,X$ annihilates $\eps(X^3,X^3,Y,Y,Y,Z,Z,Z,X^{\times(N-8)})/(N-8)!$ and acts trivially on its $X^3$ and filler slots. Since $Q\psi_2=[Z,X]$ and $Q\psi_3=[X,Y]$, its action on the $Y$ slots and on the $Z$ slots gives $3QX_N(m_3)$ and $-3QX_N(m_2)$, so $QX_N(m_2)=QX_N(m_3)$ for every $N$, and faithfulness at $N=8$ gives the equality in $\mathcal C$. At $N=7$ the members $X_7(m_{2,3})$ vanish, so the common element $\iota(\delta m_2)$ evaluates to zero and gives one linear relation among the $1735$ monomials of $\iota(\delta\mathcal A_1)$.

Evaluated on $\so(2N)$ over a finite field, these monomials have rank $1323$, $1638$, $1722$, $1734$ and $1735$ at $N=4,\dots,8$, and these values are lower bounds for the ranks over $\mathbb Q$. At $N=8$ the bound is the full rank. At $N=7$ the relation just described gives the matching upper bound, so the rank is exactly $1734$. In the same way the image of the outgoing differential at $N=7$ has rank exactly $310$, one less than in the covering: the relation bounds it from above and a non-zero minor of size $310$ modulo a prime from below. Whether the cohomology itself differs from the covering cohomology at $N=7$ is not settled by these ranks.

The covering complex of this excitation sector can be computed exactly. When we apply the infinitesimal form \eqref{eq:absorbtrace} (section~\ref{sec:lowestdegree}) of $\Pf(g^TAg)=\det g\,\Pf(A)$ to the $\eps$-word with non-filler slots $\vec V=(W_2,\dots,W_k)$ and $N-k+1$ fillers $\phi_1$, each filler contributes $2\phi_1W_0$ by \eqref{eq:slotid}, giving
\begin{equation*}
\begin{aligned}
2X_N\big((\vec V,\phi_1W_0),M\big)={}&(-1)^{|W_0|\sum_{i=2}^k|W_i|}\tr(W_0)\,X_N\big(\vec V,M\big)\\
&-\sum_{s=2}^k(-1)^{|W_0|\sum_{i=s}^k|W_i|}X_N\big(\vec V_s,M\big),
\end{aligned}
\end{equation*}
Here the slot $\phi_1W_0$ stands after those of $\vec V$, and $\vec V_s$ replaces $W_s$ by $\widehat W_s$ of \eqref{eq:absorbtrace}. The symbol $|W|$ denotes the Grassmann parity of a word, and a slot $W_0\phi_1$ is first reversed with \eqref{eq:slotid}. The identity holds at every rank, hence in $\mathcal C$, and its right-hand side has one non-filler slot fewer. Induction on $k$ removes every slot that begins or ends with $\phi_1$, and reducing the trace words in $M$ by reversal \eqref{eq:slotid} as well leaves $311$, $1045$ and $806$ formal monomials at $2J=0,1,2$, which therefore span $\mathcal A_{2J}/\ker\iota$.

Their images under $\iota$ are linearly independent: since no $u$ is adjacent to $\phi_1$, each image is homogeneous of $u$-degree $k$, and its terms with one $u$ in each of $k$ traces are non-zero multiples of $\tr(uW_1)\cdots\tr(uW_k)\tr M$, distinct for distinct monomials. The dimensions $\dim(\mathcal A_{2J}/\ker\iota)=311,1045,806$ are thus exact. The differential from $2J=1$ to $2J=0$ has rank $311$ and is therefore surjective, and the differential from $2J=2$ to $2J=1$ has rank $554$. Both ranks are exact: non-zero minors of these sizes modulo a prime bound them from below, and explicit rational factorizations of the two matrices through $311$ and $554$ of their columns bound them from above. The covering cohomology at $2J=1$ thus has dimension
\begin{equation}\label{eq:stablecount}
1045-311-554=180.
\end{equation}
By the large-rank identification, the weight space $\sect{1}{N-1,3,3}$ of the Pfaffian-sector cohomology of $\SO(2N)$ has dimension $180$ for every $N\ge8$. This is the stable, rank-independent content of the excitation sector.

At $N=4$ the same weight space has dimension $119$ (section~\ref{sec:singletcohomology}), and the map $a\mapsto X_4(a)$ from the covering cohomology to it is surjective: every class of $\SO(8)$ at these charges is represented by the rank-four member of a covering-closed combination. This holds because the families of all graviton products at this weight are covering-closed (appendix~\ref{app:gl3}) with members of Coulomb rank $118$, because the explicit family below supplies a closed operator whose Coulomb image lies outside their span, and because the Coulomb map is injective on the $119$-dimensional cohomology.

The class outside the graviton span is carried by one explicit family. With \mbox{$X,Y,Z=\phi_1,\phi_2,\phi_3$} and $X^{\times n}$ denoting $n$ filler slots, it is
\begin{equation}\label{eq:singletfamily}
\begin{aligned}
\XN_N={}&\frac{2\,\eps(ZYZY\psi_1,X^{\times(N-1)})+\eps(ZY\psi_1YZ,X^{\times(N-1)})}{(N-1)!}\\
&+\frac{\eps(YZY,Z\psi_3Z,X^{\times(N-2)})-2\,\eps(YZY,ZY\psi_2,X^{\times(N-2)})}{(N-2)!}.
\end{aligned}
\end{equation}
This combination is covering-closed, so $\XN_N$ is $Q$-closed at every rank. The cancellation is short. With $Q\psi_1=YZ-ZY$, $Q\psi_2=ZX-XZ$ and $Q\psi_3=XY-YX$, and using only cyclicity, reflection and $uX=Xu=1$, the four terms of \eqref{eq:singletfamily} contribute, in order,
\begin{equation}\label{eq:cancel}
\begin{aligned}
\iota(\delta\XN)={}&\big[\tr(u(YZ)^3)-\tr(uZY^2ZYZ)\big]+\tr(uZY^2ZYZ)\\
&+\big[\tfrac12\tr(uYZY)\tr(uZXYZ)-\tr(uYZY\,uZXYZ)\big]\\
&+\big[-\tr(u(YZ)^3)-\tfrac12\tr(uYZY)\tr(uZXYZ)+\tr(uYZY\,uZXYZ)\big]=0.
\end{aligned}
\end{equation}
The term with a single $u$ in the last bracket arises from the cancellation $Xu=uX=1$, and all products inside the traces are matrix products. None of the four terms is closed on its own, and dropping any one of them spoils closure.

\subsection{The Coulomb image of the family}
\label{sec:exactproof}

Closure of $\XN_N$ is an identity in $\mathcal C$. That its rank-four member represents a class outside the graviton span follows from an equally explicit computation on the Cartan subalgebra, combined with the Fischer orthogonality of section~\ref{sec:exactcert}. On the Cartan subalgebra, a word $W$ of length $\ell$ is block diagonal with blocks $w^{(i)}\mathbb J^\ell$, where $w^{(i)}$ is the ordered product of the block values of its letters and $\mathbb J^2=-1$. Hence $\tr W=2(-1)^{\ell/2}\sum_iw^{(i)}$ for even $\ell$ and $\tr W=0$ for odd $\ell$. In $\eps(W_1,\dots,W_N)$ the index pairs of the slots must occupy distinct blocks, so
\begin{equation}\label{eq:epsCartan}
\eps(W_1,\dots,W_N)\big|_{\rm Cartan}=2^N\prod_{s=1}^N(-1)^{(\ell_s-1)/2}\sum_{\pi\in S_N}\prod_{s=1}^Nw_s^{(\pi(s))}
\end{equation}
if every slot length $\ell_s$ is odd, and zero otherwise; Grassmann-odd block values are multiplied in slot order. We write $(x_i,y_i,z_i)$ for the block values of $(\phi_1,\phi_2,\phi_3)$ and, as in section~\ref{sec:exactcert}, $\psi^{(i)}_m$ for those of $\psi_m$. Applying \eqref{eq:epsCartan} to the four terms of $\XN_4$ gives
\begin{equation}\label{eq:CX4}
C(\XN_4)=16\Big[3\sum_iy_i^2z_i^2\,\psi^{(i)}_1\,x_jx_kx_l+\sum_{i\neq j}y_i^2z_i\,z_j\big(z_j\psi^{(j)}_3-2y_j\psi^{(j)}_2\big)x_kx_l\Big],
\end{equation}
where $\{j,k,l\}$ is the complement of $i$ in the first sum and $\{k,l\}$ the complement of $\{i,j\}$ in the second. The two terms with three fillers combine into the first sum and the two terms with two fillers into the second. This function is not an $\SU(3)_R$ singlet, but its singlet component is non-zero. With the singlet $F$ of \eqref{eq:F} and the descendant space $S$ of section~\ref{sec:exactcert},
\begin{equation}\label{eq:pairing}
\langle F,C(\XN_4)\rangle=2304,\qquad C(\XN_4)-\tfrac{8}{15}\,F\in S.
\end{equation}

It follows that $\XN_4$ is not cohomologous to any sum of products of $\Qp$-descendants of bosonic gauge invariants of either reflection parity; in particular it is not a graviton combination. If $\XN_4$ were $\mathcal G+Q\mathcal V$ with $\mathcal G$ such a sum, then $C(\XN_4)=C(\mathcal G)$ would lie in $S$, because every $Q$-exact operator vanishes on the Cartan subalgebra (section~\ref{sec:coulombtest}), whereas $\langle F,C(\XN_4)\rangle\neq0$ and $F$ is orthogonal to $S$. This argument rests on the cancellation \eqref{eq:cancel} and on exact Fischer pairings, and uses no rank of $Q$. Since the cohomology at this weight is spanned by graviton products and the singlet class (section~\ref{sec:singletcohomology}), the class of $\XN_4$ has a non-zero component along the singlet: modulo graviton products and $Q$-exact operators, $\XN_4$ is a non-zero multiple of the non-graviton singlet.

\subsection{A generalized monotone at every rank \texorpdfstring{$N\ge4$}{N>=4}}
\label{sec:singletgm}

The non-graviton singlet of $H^-_{\SO(8)}$ at $\sect{1}{3,3,3}$ is therefore a generalized monotone. At this weight $H^-/G^H$ is one-dimensional and the graviton span is an $\SU(3)_R$-submodule without singlets, so the singlet component of the class of $\XN_4$ spans the non-graviton singlet. Let $\Pi_0$ be the projector onto $\SU(3)_R$ singlets and define
\begin{equation}\label{eq:X40}
\XN_4^{(0)}=\Pi_0\,\XN_4.
\end{equation}
Since $Q$ commutes with $\SU(3)_R$, the operator $\XN_4^{(0)}$ is $Q$-closed. Its Coulomb image is the singlet component of \eqref{eq:CX4}. The function $F$ is the only singlet among the reflection-odd invariant functions of this weight and the Fischer form is $\SU(3)_R$ invariant, so \eqref{eq:pairing} and $\langle F,F\rangle=4320$ give
\begin{equation}\label{eq:CX40}
C\big(\XN_4^{(0)}\big)=\frac{\langle F,C(\XN_4)\rangle}{\langle F,F\rangle}\,F=\tfrac{8}{15}\,F.
\end{equation}
Thus $\XN_4^{(0)}$ represents the singlet class itself, without a graviton remainder. The classes of covering-closed families with ground state $\Pf(\phi_1)$ form a $\mathfrak{gl}(3)$-stable subspace (appendix~\ref{app:gl3}), so $\XN_4^{(0)}$ is again represented by such a family.

The family gives more than the $\SO(8)$ class. For every $N\ge4$, $\XN_N$ is a $Q$-closed operator in the Pfaffian sector of $\SO(2N)$, with charges $\sect{1}{N-1,3,3}$ and level $j=2N+10$, that is not cohomologous to any sum of graviton products; it is a generalized monotone by construction. The case $N=4$ was shown in section~\ref{sec:exactproof}, and the restriction in rank of section~\ref{sec:branesep} propagates the non-graviton property to all $N>4$. At $N=3$ the operator $\XN_3$ is also closed, but every class of $\SO(6)$ in that sector is a graviton product: the property propagates only upwards.

The Pfaffian sector therefore contains, at every rank $N\ge4$, classical cohomology classes that belong to a rank-uniform family, and are in this sense candidate excitations of the wrapped brane, but are not products of gravitons and excited Pfaffians. Reading them as BPS states of the brane presupposes that they survive the quantum corrections to $Q$. In perturbation theory they do: section~\ref{sec:allorders} shows that the classes of $\XN_N$ are protected to all orders at every rank $N\ge4$. For $N=4$ the singlet class represented by $\XN_4^{(0)}$ forms, together with its triality partner of section~\ref{sec:partner}, a doublet that is protected as a whole.

In the covering, \eqref{eq:singletfamily} is $\Pf(\phi_1)$ times a combination of single and double traces that contain $u$ and the antisymmetric parts \eqref{eq:antisymslot} of the slot words, such as
\begin{equation}
\tr\big(u\,\overline{ZYZY\psi_1}\big)\qquad\text{and}\qquad\tr\big(u\,\overline{YZY}\big)\tr\big(u\,\overline{Z\psi_3Z}\big)-2\,\tr\big(u\,\overline{YZY}\,u\,\overline{Z\psi_3Z}\big).
\end{equation}
These have the form of excitations of a single wrapped brane by words that are not symmetrized in their letters, of the single-determinant type considered by Choi, Choi and Kim~\cite{Choi:2026faq}. We propose this reading for the fermionic family; a match with the spectrum of the wrapped brane is established only for the bosonic excited Pfaffians of section~\ref{sec:gravcontent}. The family can be written as $\XN_N=\eps\big(\Qp^2(ZYZYZ),X^{\times(N-1)}\big)/(N-1)!+\eps\big(YZY,\Qp^1(ZYZ),X^{\times(N-2)}\big)/(N-2)!$, in which a supercharge acts on a single slot rather than on a whole invariant. For determinant operators of unitary gauge groups, a word that replaces an entry of the $\eps\eps$ contraction describes an open string ending on a giant graviton~\cite{Balasubramanian:2002sa,Berenstein:2006qk}. We expect the slot words of a Pfaffian to play the same role, with the antisymmetric projection \eqref{eq:antisymslot} implementing the orientifold projection on the orientation of the open string. None of the excited Pfaffians of section~\ref{sec:gravitons} has this form, and the singlet is the first class in our data that requires such an excitation. It is also not a $\Qp$-descendant, even modulo graviton products: $S$ contains $\Qp^m(C(A))$ for every bosonic invariant $A$, so the bottom component of its $\SU(1|3)$ multiplet is fermionic.

\subsection{A control case: the \texorpdfstring{$\SU(2)$}{SU(2)} threshold class inside \texorpdfstring{$\SO(4)$}{SO(4)}}
\label{sec:control}

A notion of generalized monotone that absorbed every class would carry no information. We therefore test the construction on a class that is expected to be a black-hole state. For $\so(4)=\su(2)\oplus\su(2)$ the reflection $\sigma$ exchanges the two factors, and the reflection-odd sector at $\lambda=(4,4,4)$, $j=24$, contains the graded antisymmetrization over the two factors of the threshold non-graviton class of $\SU(2)$~\cite{Chang:2022mjp,Choi:2022caq}. We write $\mathcal O^-_{24}$ for a representative of this class. The cohomology at weight $(4,4,4)$ has dimensions $h^-=31,0,203,0,150,1,4,0,0$ for $2J=0,\dots,8$, in agreement with the $\SU(2)\otimes\SU(2)$ prediction of section~\ref{sec:validation}, and the same agreement holds at the four other weights that determine the singlet column. This column contains exactly one class, at $2J=5$, as its index $6-15+25-64+110-85+23-2+1=-1$ requires. This class has $L=7$ letters and hence $P_2=-1$, so by the parity constraint of section~\ref{sec:gravitons} it is neither a graviton product nor visible on the Coulomb branch.

To decide whether $\mathcal O^-_{24}$ is a single-Pfaffian generalized monotone, consider the formal monomials with the charges of the class above $\Pf(\phi_1)$ at rank $N=2$. Families whose slot words begin or end with $\phi_1$ are eliminated at every rank as in section~\ref{sec:singletcov}. The class is an $\SU(3)_R$ singlet and the cohomology vanishes at the adjacent weight $(4,5,3)$, so nothing is lost by averaging over the $\SU(2)$ that rotates $\phi_2$ and $\phi_3$ and preserves the ground state.

The invariant cochains at $\sect{5}{4,4,4}$ form a space of dimension $429$, in which the closed subspace has dimension $100$ and the exact subspace dimension $99$. These values follow from the exact ranks $329$ and $99$ of the outgoing and incoming differentials, which we obtain from the factorization of the complex into two $\SU(2)$ complexes. There is a rational linear functional $\omega$ on these cochains that vanishes on the exact subspace, does not vanish on $\mathcal O^-_{24}$, and satisfies
\begin{equation}\label{eq:so4dual}
\omega\big(X_2(a)\big)=y\big(\iota(\delta a)\big)
\end{equation}
for a fixed rational functional $y$ on $\mathcal C$ and every invariant combination $a$ of formal monomials in the sector. The functional $\omega$ therefore annihilates the rank-two member of every covering-closed combination, so each such member is exact. Conversely, the rank-two members of the covering-closed combinations span a space of dimension $99$: a non-zero minor of this size modulo a prime bounds the dimension from below, and the dimension of the exact subspace bounds it from above. The covering-closed families thus fill the exact subspace and miss the class. Since the class is an $\SU(3)_R$ singlet, the same holds for every ground state $\Pf(v\cdot\phi)$.

Families with three Pfaffians are relevant at this level, since for $\SO(4)$ they enter from $j=12$. For these we adopt the analogue of the working assumption of Choi, Choi and Kim that there is no non-factorizable generalized monotone independent of ordinary monotones~\cite{Choi:2026faq}. The families then reduce to products of three reflection-odd classes, which would need letter numbers $2+2+3$; but every reflection-odd $\SO(4)$ class with $j\le20$ has even $L$, so no such product exists. Under this assumption $\mathcal O^-_{24}$ is strongly fortuitous with respect to $GM^-_2$, while the same construction reaches the $\SO(8)$ singlet. This fits a wider pattern. We scanned the excitation sectors $(2J;e_1,q_2,q_3)$ with $P_N=-1$, $2J\le4$, $q_2\ge q_3\ge0$ and $e_1+q_2+q_3\ge0$, with $e_1\le3$ for $1\le q_2+q_3\le3$, $e_1\le1$ for $q_2+q_3=4$, and $2J=0$ and $e_1\le0$ for $(q_2,q_3)=(5,0)$. Of these $124$ sectors, $120$ were evaluated over a finite field (the other four exceed the size limits of the computation), $53$ of them have a non-zero covering cochain space, and the covering cohomology, with cochain and differential ranks both evaluated modulo a prime, vanishes throughout.
\section{Triality and the \texorpdfstring{$\SO(7)$}{SO(7)} non-graviton singlet}
\label{sec:triality}

The algebra $\so(8)$ has an outer automorphism of order three that permutes the vector and the two spinor representations. Because the classical supercharge is built from the Lie bracket alone, this triality acts on the cohomology, although it does not preserve the distinction between traces and Pfaffians. In this section we use triality, together with restriction to $\so(7)$, to relate the singlet of section~\ref{sec:singlet} to two other classes: the known non-graviton singlet of $\SO(7)$ and a reflection-even class of $\SO(8)$ at the same charges. The two $\SO(8)$ classes form a doublet of the triality group, and comparing its members shows which properties of a class are intrinsic and which depend on the covering used to compare ranks.

\subsection{Embeddings of \texorpdfstring{$\so(7)$}{so(7)} and the triality automorphism}
\label{sec:homomorphism}

The supercharge \eqref{eq:Q} acts on gauge invariants through commutators of the letters. Let $\varphi:\mathfrak g\to\mathfrak g'$ be a Lie-algebra homomorphism realized by a linear substitution of matrices, and let $\mathcal O$ be a $\mathfrak g'$-invariant polynomial in $\mathfrak g'$-valued letters. The pullback $\varphi^*\mathcal O$, obtained by substituting $\varphi(\ell)$ for every letter $\ell$, is $\mathfrak g$-invariant and satisfies
\begin{equation}\label{eq:chainmap}
Q\,\varphi^*\mathcal O=\varphi^*\,Q\mathcal O.
\end{equation}
Restriction along $\varphi$ therefore maps $Q$-closed operators to $Q$-closed operators and $Q$-exact operators to $Q$-exact operators, and induces a map on cohomology. For block embeddings this mechanism was used by de Mello Koch and collaborators and by the Fortuity collaboration, the latter also for the embedding $\mathfrak g_2\subset\so(7)$~\cite{deMelloKoch:2024pcs,deMelloKoch:2026sbd,deMelloKoch:2026cvn,FortuitySummary}. The restriction used by Kibe, van Leuven and Mathieson, in which different letters are restricted to different sets of commuting root generators, is a related device but not a pullback of this kind~\cite{Kibe:2026qid}. Restriction from the stable algebra also underlies the definition of fortuity given by Chang~\cite{Chang:2026bil}. Here we apply the same argument to automorphisms of $\so(8)$ and to the spinor embedding of $\so(7)$.

Let $o_1,\dots,o_7$ be the imaginary units of the octonions~\cite{Baez:2001dm}, with the multiplication table fixed by the triples $(123)$, $(145)$, $(176)$, $(246)$, $(257)$, $(347)$ and $(365)$. In the ordered basis $(b_1,\dots,b_8)=(1,o_1,\dots,o_7)$, which we identify with the standard basis of $\mathbb R^8$, let $\gamma_a$ be the real antisymmetric $8\times8$ matrix of left multiplication by $o_a$, so that $o_ab_\beta=\sum_\alpha(\gamma_a)_{\alpha\beta}b_\alpha$. The matrices satisfy $\gamma_i\gamma_j+\gamma_j\gamma_i=-2\delta_{ij}$ and $\gamma_1\cdots\gamma_7=-1$. Let $e_{ij}$ be the matrix unit with entry one in position $(i,j)$, so that the matrices $M_{ij}=e_{ij}-e_{ji}$ with $i,j=1,\dots,8$ generate $\so(8)$. The spinor embedding $\is:\so(7)\to\so(8)$ and a triality automorphism $\tau$ of $\so(8)$ extending it are defined by
\begin{equation}\label{eq:spinorembedding}
\is(M_{ij})=-\tfrac12\gamma_i\gamma_j,\quad \tau(M_{ij})=-\tfrac12\gamma_i\gamma_j,\quad \tau(M_{a8})=\tfrac12\gamma_a,\quad i,j,a=1,\dots,7.
\end{equation}
Both maps preserve the bracket of every pair of generators. We also use the vector embedding $\iv:\so(7)\to\so(7)\oplus0\subset\so(8)$, in terms of which
\begin{equation}\label{eq:trialitydef}
\is=\tau\circ\iv.
\end{equation}

The automorphism $\tau$ is bijective and does not preserve the Pfaffian, so it is outer, and its class in $\mathrm{Out}(\so(8))\cong S_3$ has order three. After composition with an inner automorphism it preserves the Cartan subalgebra, on which it acts as
\begin{equation}\label{eq:tauCartan}
t_\tau=\sigma\cdot(-T),\qquad T=\frac12\begin{pmatrix}1&1&1&1\\1&1&-1&-1\\1&-1&1&-1\\1&-1&-1&1\end{pmatrix},
\end{equation}
where $\sigma$ flips the sign of the fourth block and $T$ acts on the column vectors of block coordinates of each letter, so that $C(\tau^*\mathcal O)(x)=C(\mathcal O)(\sigma(-T)x)$. The matrix $T$ preserves the $D_4$ root system, exchanges the simple roots $\alpha_1$ and $\alpha_4$ in the Bourbaki labelling, and normalizes the Weyl group $W(D_4)$; together with $\sigma$ it generates $\mathrm{Aut}(D_4)/W(D_4)\cong S_3$. One finds $t_\tau^3=-1\in W(D_4)$, while $t_\tau$ itself is not an element of $W(D_4)$.

Triality preserves $Q$-closure, $Q$-exactness and the $\SU(3)_R$ representation content, but it mixes the two reflection sectors. For a single scalar $X$ one has
\begin{equation}\label{eq:Pftriality}
\tau^*\Pf X=-\tfrac12\Pf X-\tfrac1{16}\tr X^4+\tfrac1{64}(\tr X^2)^2,
\end{equation}
which follows by evaluating both sides on the Cartan subalgebra with \eqref{eq:tauCartan}. The Pfaffian is thus mapped to a combination of itself and of trace gravitons. The same mixing distinguishes the two embeddings of $\so(7)$. The reflection $\sigma$ fixes $\so(7)\oplus0$ pointwise, so every reflection-odd operator of $\SO(8)$ vanishes after restriction along $\iv$. Restriction along $\is$ has no such property, and by \eqref{eq:trialitydef} it equals the vector restriction of the triality image.

We use only the action of triality on gauge-invariant local operators. Triality permutes the three $\mathbb Z_2$ subgroups of the centre of $\Spin(8)$, as reviewed by Tong and Turner~\cite{Tong:2019bbk}. It is therefore a symmetry of the $\Spin(8)$ theory, whereas it maps the $\SO(8)$ theory to the theories whose gauge groups are the other two $\mathbb Z_2$ quotients of $\Spin(8)$. Aharony, Seiberg and Tachikawa showed that such global forms differ in their line operators~\cite{Aharony:2013hda}. The local operators do not see this difference: $S^3$ carries no non-trivial bundles or flux sectors, so the Hilbert space of radial quantization, its $Q$-cohomology and the index on $S^3\times S^1$ are the same for every global form of $\so(8)$~\cite{Aharony:2013hda,Razamat:2013opa}. We know of no bulk realization of triality, which would have to mix states of different $\mathbb Z_2$ wrapping charge.

\subsection{Restriction to \texorpdfstring{$\SO(7)$}{SO(7)}}
\label{sec:SO7restriction}

At the charges $\sect{1}{3,3,3}$ the BMN cohomology of $\SO(7)$ consists of $123$ graviton classes and one further class, an $\SU(3)_R$ singlet. This class was found by Gadde, Lee, Raj and Tomar, who showed that it is fortuitous and responsible for a mismatch between the BMN indices of the S-dual pair $\SO(7)$ and $\USp(6)$~\cite{Gadde:2025yoa}. The corresponding mismatch of the BMN cohomologies was established by Chang and Lin and studied further by Chang~\cite{Chang:2025mqp,Chang:2026bil}. Choi and Lee showed that it is lifted by the first quantum correction~\cite{Choi:2025bhi}. The class also appears in the data of the Fortuity collaboration~\cite{FortuitySummary}. The rank-four member $\XN_4$ of the family \eqref{eq:singletfamily} has the same charges, which suggests restricting it to $\so(7)$. The vector restriction vanishes, so we consider $\is^*\XN_4$.

By \eqref{eq:chainmap} the operator $\is^*\XN_4$ is $Q$-closed. To decide whether it is a graviton combination we evaluate it on the Cartan subalgebra of $\so(7)$, which has three blocks and Weyl group $W(B_3)=S_3\ltimes\mathbb Z_2^3$. Since $\is=\tau\circ\iv$, the Coulomb image is
\begin{equation}
f_7=C(\XN_4)\circ t_\tau\big|_{\text{fourth block}=0},
\end{equation}
a $W(B_3)$-invariant polynomial with $786$ monomials. Let $R_{B_3}$ be the ring of $W(B_3)$-invariant bosonic polynomials and $S_{B_3}=\mathrm{span}\{B\,\Qp^m(A):A,B\in R_{B_3}\}$ the space of descendant products, which contains the Coulomb image of every graviton combination of $\SO(7)$. At weight $(3,3,3)$ the $W(B_3)$-invariant functions linear in $\psi$ form a space of dimension $144$ with exactly one $\SU(3)_R$ singlet,
\begin{equation}\label{eq:F7}
F_7=\Delta^2\sum_{i=1}^3\psi^{(i)}\cdot\phi^{(i)},\qquad \Delta=\eps_{mnp}\phi^{(1)}_m\phi^{(2)}_n\phi^{(3)}_p.
\end{equation}
The function $F_7$ is orthogonal, in the Fischer inner product of appendix~\ref{app:conventions}, to all $594$ generators of $S_{B_3}$ at this weight, whose span has dimension $123$. The orthogonality also follows as in section~\ref{sec:exactcert}: the products of $W(B_3)$ invariants of degrees $d$ and $8-d$ do not contain the highest weight $(3,3,2)$, so $S_{B_3}$ contains no singlet. Moreover
\begin{equation}\label{eq:f7pairing}
\langle F_7,F_7\rangle=2160,\qquad \langle F_7,f_7\rangle=-288.
\end{equation}
Hence $f_7\notin S_{B_3}$. If $\is^*\XN_4$ were equal to $\mathcal G+Q\mathcal V$ with $\mathcal G$ a sum of products of $\Qp$-descendants of bosonic invariants, its Coulomb image would lie in $S_{B_3}$, because $Q$-exact operators vanish on the Cartan subalgebra. The restriction $\is^*\XN_4$ is therefore not a graviton combination, and the component of its Coulomb image along $F_7$,
\begin{equation}
\frac{\langle F_7,f_7\rangle}{\langle F_7,F_7\rangle}=-\frac2{15},
\end{equation}
cannot be changed by adding graviton combinations or $Q$-exact operators.

The class counts follow as in section~\ref{sec:singletcohomology} and are collected in table~\ref{tab:seedcomplexes}. The cochain dimensions at $2J=0,1,2$ are $220$, $903$ and $918$. The differential out of $2J=1$ has rank $220$ and is onto, and the rank of the incoming differential is at least $559$, so $h\le124$. The Coulomb images of the graviton products have rank $123$, and adjoining $f_7$ raises it to $124$. Therefore $h=124$, the Coulomb map is injective on the cohomology at these charges, and $\is^*\XN_4$ represents the non-graviton singlet of $\SO(7)$ modulo graviton classes. These numbers agree with those of Gadde, Lee, Raj and Tomar, of Chang and Lin and of the Fortuity collaboration~\cite{Gadde:2025yoa,Chang:2025mqp,FortuitySummary}.

This result is consistent with an observation of the Fortuity collaboration, who found that the $\SO(7)$ singlet restricts to a graviton class on $\mathfrak g_2$-valued letters~\cite{FortuitySummary}. The image $\is(\mathfrak g_2)$ fixes the real octonion unit and hence lies in the stabilizer of a vector, which is conjugate in $\SO(8)$ to $\so(7)\oplus0$. On such letters $\XN_4$ vanishes.

\subsection{The reflection-even partner and the \texorpdfstring{$S_3$}{S3} doublet}
\label{sec:partner}

Since triality mixes the reflection sectors, the image $\tau^*\XN_4$ has a reflection-even part,
\begin{equation}
(\tau^*\XN_4)_{\rm even}=\tfrac12(1+\sigma^*)\tau^*\XN_4,
\end{equation}
which is a $Q$-closed polynomial in traces. We now show that it represents the only class of the reflection-even sector $H^+$ of $\SO(8)$ at $\sect{1}{3,3,3}$ that is not a graviton product. The first step is the cohomology of this sector, which uses neither triality nor the $\SO(7)$ results.

The cohomology was obtained in section~\ref{sec:onset}: by \eqref{eq:h10h8} it has dimension $h^+=145$, against $140$ for $\SO(10)$, the difference being due to the five trace relations of $\Or(8)$ at $2J=0$ (table~\ref{tab:seedcomplexes}). Every class is visible on the Coulomb branch (section~\ref{sec:evenscan}).

Trace gravitons span $140$ of these classes, and products of two Pfaffian-type factors raise the graviton rank to $144$. The remaining class is an $\SU(3)_R$ singlet. To see this, note first that at the four other weights of the singlet column, $(4,3,2)$, $(5,2,2)$, $(4,4,1)$ and $(5,3,1)$, every class is a graviton product, with $h^+=124$, $96$, $82$ and $74$. The singlet multiplicity of $H^+$ is therefore
\begin{equation}
145-2\cdot124+96+82-74=1,
\end{equation}
and every non-singlet component of $H^+$ at weight $(3,3,3)$ lies in the graviton span.

It remains to show that the graviton span contains no singlet, which follows from representation theory alone. As in section~\ref{sec:exactcert} we regard the operators as representations of $\mathrm{GL}(3)$, under which $\psi_m$ transforms in the dual of the fundamental representation multiplied by the determinant, and an $\SU(3)_R$ singlet of weight $(3,3,3)$ is the cube of the determinant. A product of trace gravitons at $2J=1$ has exactly one factor with a fermion. This factor is the component of a multiplet $\tr\Phi^n$, with $n$ even, that contains one fermion and $n-1$ scalars, and it transforms in the irreducible representation of highest weight $(n,1,0)$, that is $[n-1,1]$. The other factors are symmetrized traces of scalars and span a polynomial representation $V$ of degree $8-n$. The cube of the determinant occurs in the product of $(n,1,0)$ with $V$ only if $V$ contains the representation of highest weight $(3,2,3-n)$. Since $V$ is polynomial and $n$ is even, this requires $n=2$, and $V$ must then contain $(3,2,1)$, the representation $[1,1]$. The six scalars are distributed over traces of even length, which gives
\begin{equation}\label{eq:bosonicsix}
[6,0],\qquad [4,0]\otimes[2,0]=[6,0]+[4,1]+[2,2],\qquad \mathrm{Sym}^3[2,0]=[6,0]+[2,2]+[0,0],
\end{equation}
and none of these contains $[1,1]$. The singlet projection of every product of trace gravitons at these charges therefore vanishes identically, at every rank. A similar scarcity of $\SU(3)_R$-singlet gravitons in the BMN sector of $\SU(N)$ at low $N$ was observed by Gaikwad, Kibe, van Leuven and Mathieson~\cite{Gaikwad:2025ugk}.

The representation content also gives the weight multiplicities of the Fock space generated by the trace gravitons, which are $140$, $121$, $94$, $80$ and $72$ at the five weights of \eqref{eq:alternation}. These are the dimensions of the reflection-even cohomology of $\SO(10)$ at $2J=1$ and these weights, where every class is a product of trace gravitons, and their alternating sum $140-2\cdot121+94+80-72$ vanishes: the reflection-even singlet column of $\SO(10)$ has no cohomology at $2J=1$, in agreement with the $\SO(9)$ data of the Fortuity collaboration~\cite{FortuitySummary}.

For Pfaffian pairs the argument is of the same kind. With eight letters at $2J=1$, a product of two Pfaffian-type factors pairs the scalar Pfaffian multiplet $[4,0]$ with a factor $V'$ at $2J=1$ and $|q|=5$. A singlet in $[4,0]\otimes V'$ requires $[0,4]\subset V'$, which is impossible for a polynomial representation of $\mathrm{GL}(3)$ of degree five. The graviton rank is thus exactly $144$, and the remaining class is a singlet. The trace-graviton count $140$ is exact as well, because the trace gravitons of $\SO(8)$ are restrictions of those of $\SO(10)$, whose span has dimension at most $h^+_{\SO(10)}=140$.

This class is fortuitous with respect to the trace covering of Chang and Lin~\cite{Chang:2024zqi}. Let $\mathcal O$ be a representative and $\widetilde{\mathcal O}$ the same polynomial in traces evaluated on $\so(10)$. Restriction maps $Q\widetilde{\mathcal O}$ to $Q\mathcal O=0$, so $Q\widetilde{\mathcal O}$ is a trace relation of $\Or(8)$. If it vanished, the singlet projection of $\widetilde{\mathcal O}$ would be closed on $\so(10)$ and hence exact, because the reflection-even singlet column of $\SO(10)$ has no cohomology at $2J=1$, and by restriction the singlet projection of $\mathcal O$ would be exact on $\so(8)$. The non-singlet part of $\mathcal O$ is exact because the class is a pure singlet, so $\mathcal O$ itself would be exact, a contradiction. For a singlet representative, $Q\widetilde{\mathcal O}$ is therefore a non-zero multiple of the singlet trace relation $\mathcal R$ of $\Or(8)$ at $\sect{0}{3,3,3}$, which is unique up to normalization. Operators with at most eight letters have a unique expression in traces, so the coefficient of $\mathcal R$ does not depend on the representative. At these letter numbers the formal multi-trace complex coincides with the reflection-even complex of $\SO(10)$ (section~\ref{sec:onset}), and the class is not in the image of the covering cohomology.

The identification with the triality image follows from restriction to $\so(7)$. Since $\sigma$ fixes $\so(7)\oplus0$,
\begin{equation}\label{eq:partnerrestriction}
\iv^*(\tau^*\XN_4)_{\rm even}=\iv^*\tau^*\XN_4=\is^*\XN_4.
\end{equation}
The vector restriction maps trace gravitons to trace gravitons of $\SO(7)$ and annihilates products of Pfaffian-type factors. If $(\tau^*\XN_4)_{\rm even}$ were a graviton combination up to a $Q$-exact term, the same would hold for $\is^*\XN_4$, in contradiction with section~\ref{sec:SO7restriction}. Hence $(\tau^*\XN_4)_{\rm even}$ represents the non-graviton class of $H^+$ modulo graviton products.

The same conclusion follows directly on the Coulomb branch of $\SO(8)$, where the partner of the function $F$ of \eqref{eq:F} is
\begin{equation}\label{eq:Edef}
E=F\circ T-\tfrac12F.
\end{equation}
In the notation of \eqref{eq:F} it reads
\begin{equation}\label{eq:Eexplicit}
E=\frac14\sum_{i<j<k}\Delta_{ijk}^2\,\big(s_{ii}+s_{jj}+s_{kk}-5\,s_{ll}\big),
\end{equation}
where $l$ is the block that does not occur in $\{i,j,k\}$. The function $E$ is reflection-even, $W(D_4)$-invariant and an $\SU(3)_R$ singlet, with
\begin{equation}\label{eq:Epairing}
\langle E,E\rangle=3240,\qquad \langle F,E\rangle=0.
\end{equation}
It is orthogonal to all $642$ generators $B\,\Qp^m(A)$ of $S_{\rm even}$, whose span has dimension $144$, as it must be, because $S_{\rm even}$ contains no singlet (section~\ref{sec:exactcert}).

Unlike $F$, the function $E$ is not fixed by its symmetries. The reflection-even invariant functions have dimensions $177$, $149$, $115$, $97$ and $89$ at the five weights of \eqref{eq:alternation}, so they contain $177-2\cdot149+115+97-89=2$ singlets. The singlet orthogonal to $E$ is
\begin{equation}\label{eq:Eperp}
E^\perp=\sum_{i<j<k}\Delta_{ijk}^2\;\sum_{l=1}^4s_{ll}=\tfrac1{16}\,C\Big(\det_{mn}\big(\tr\phi_m\phi_n\big)\sum_p\tr(\phi_p\psi_p)\Big),
\end{equation}
the Coulomb image of a bosonic invariant multiplied by an operator of Konishi type, which is not $Q$-closed. Only $E$ occurs in the Coulomb image of the cohomology. The Coulomb map sends the graviton span of $H^+$ isomorphically onto $S_{\rm even}$, which contains no singlet, so the only singlets in the image are the multiples of the image of the single singlet class of $H^+$, which is the multiple of $E$ determined below. The function $E^\perp$ is not the Coulomb image of any closed operator.

An element $g\in\SO(8)$ with entries in $\{0,\pm1/\sqrt2\}$ makes $\mathrm{Ad}(g)\circ\tau$ preserve the Cartan subalgebra, on which it acts as \eqref{eq:tauCartan}. Gauge invariance then gives $C(\tau^*\XN_4)=C(\XN_4)\circ t_\tau$, and
\begin{equation}
\big\langle E,\,C\big((\tau^*\XN_4)_{\rm even}\big)\big\rangle=-1728.
\end{equation}
With $(\tau^*)^2$ in place of $\tau^*$ the pairing changes sign. As for $F$ in section~\ref{sec:exactproof}, the non-zero pairing excludes any expression of $(\tau^*\XN_4)_{\rm even}$ through products of $\Qp$-descendants of bosonic invariants and $Q$-exact terms. On the Cartan subalgebra of $\so(7)$, where the fourth block vanishes, $F$ restricts to zero, $E$ restricts to $\frac14F_7$ and $S_{\rm even}$ restricts into $S_{B_3}$, which reproduces $f_7\equiv-\frac2{15}F_7$ modulo $S_{B_3}$.

The two non-graviton classes form a doublet of $S_3$. At $\sect{1}{3,3,3}$ the Coulomb map is injective on $H=H^+\oplus H^-$ and maps the graviton span onto $S=S_{\rm odd}+S_{\rm even}$, whose two parts have dimensions $118$ and $144$, equal to the graviton ranks. The space $S$ is stable under $\sigma$ and $T$, so the outer automorphisms act on the two-dimensional quotient $H/G^H\cong C(H)/S$ through $\sigma$ and $T$. Explicitly,
\begin{equation}\label{eq:doubletimages}
C(\XN_4)\equiv\tfrac8{15}F,\qquad C\big((\tau^*\XN_4)_{\rm even}\big)\equiv-\tfrac8{15}E\pmod S,
\end{equation}
and, using $F\circ T=\frac12F+E$ and $E\circ T=\frac34F-\frac12E$, the action in the basis $(F,E)$ is
\begin{equation}
\sigma=\begin{pmatrix}-1&0\\0&1\end{pmatrix},\qquad T=\begin{pmatrix}\frac12&\frac34\\[2pt]1&-\frac12\end{pmatrix}.
\end{equation}
This is the standard two-dimensional representation of $S_3$, with characters $(2,0,-1)$. The orbit sum $\XN_4+\tau^*\XN_4+(\tau^*)^2\XN_4$ has Coulomb image in $S$ and is therefore a graviton combination up to a $Q$-exact term. At weight $(3,3,3)$ the graviton spans of $H^-$ and $H^+$ contain no singlets, so the singlet classes of $H$ are exactly this doublet.

Both members are visible in the index. The reflection-odd index of the singlet column is $-1$ by \eqref{eq:singletindex}. The reflection-even index of the same column is $17-34+29-30+21-6=-3$ for $\SO(8)$, compared with $18-34+29-30+21-6=-2$ for $\SO(9)$ and $\SO(10)$. The difference is the singlet trace relation at $2J=0$.

\subsection{An explicit trace representative}
\label{sec:O8}

The reflection-even member of the doublet is defined by the triality image $(\tau^*\XN_4)_{\rm even}$, which is fixed by the four-term family \eqref{eq:singletfamily} and the automorphism \eqref{eq:spinorembedding}. For explicit computations with the corrections to the supercharge it is convenient to have a polynomial in traces with integer coefficients. We use an operator $\mathcal O_8$ with $438$ terms, each containing seven scalars and one fermion.\footnote{The full expressions of $\mathcal O_8$ and of the operator $\mathcal Z$ of section~\ref{sec:survival} are provided in the ancillary files.} It was obtained by solving the linear conditions for a closed $\SU(3)_R$ singlet in the space of eight-letter multi-traces at these charges, which is $936$-dimensional. The statements below about its closure, its singlet property and its class refer to this particular polynomial. They show how the closure of the class comes about through trace relations, and they enter the explicit corrected operators of section~\ref{sec:survival}; the general results of section~\ref{sec:quantum} do not depend on them.

The image $Q\mathcal O_8$ is a combination of $207$ multi-traces of nine scalars that are linearly independent as formal multi-traces. It is therefore non-zero on $\so(n)$ for every $n\ge9$ (section~\ref{sec:onset}), and its vanishing on $\so(8)$ is a consequence of trace relations of $\Or(8)$. These relations arise from the generalized Kronecker delta with nine upper and nine lower indices, which vanishes identically in eight dimensions. This is the antisymmetrizer construction of orthogonal trace relations~\cite{Procesi:1976,Lehrer:2012}. Kibe, van Leuven and Mathieson used it for $\so(7)$~\cite{Kibe:2026qid}. Let $(M_1,\dots,M_9)=(X,X,X,Y,Y,Y,Z,Z,Z)$ and number the $18$ matrix indices so that $M_k$ carries the row index $2k-1$ and the column index $2k$. Let $a_1,\dots,a_{18}$ denote the values of these indices. For a set $U=\{u_1<\dots<u_9\}$ of nine index positions, with complement $U'=\{u'_1<\dots<u'_9\}$, we define
\begin{equation}\label{eq:Kcontractions}
R_U=\sum_{\pi\in S_9}\mathrm{sgn}(\pi)\,\prod_{i=1}^9\delta_{a_{u_i}a_{u'_{\pi(i)}}}\,\prod_{k=1}^9(M_k)_{a_{2k-1}a_{2k}},
\end{equation}
with all eighteen indices summed. Each term is a product of traces, read with $M^T=-M$ for a matrix traversed from its column to its row index, and $R_U=0$ on $\so(8)$. With the five index sets
\begin{equation}\label{eq:Kindexlists}
\begin{aligned}
U_1&=\{1,4,6,10,11,15,16,17,18\}, & U_2&=\{1,4,7,10,13,14,15,16,18\},\\
U_3&=\{1,2,6,8,9,10,11,12,15\}, & U_4&=\{1,4,7,9,11,12,13,17,18\},\\
U_5&=\{1,2,3,7,9,10,11,14,15\},&&
\end{aligned}
\end{equation}
one finds, as an identity of formal multi-traces,
\begin{equation}\label{eq:closurecert}
Q\mathcal O_8=\tfrac{15}{32}R_{U_1}-\tfrac{15}{32}R_{U_2}-\tfrac{15}8R_{U_3}-\tfrac{15}8R_{U_4}-\tfrac{45}{16}R_{U_5}.
\end{equation}
The five relations are linearly independent. Their number equals the number of trace relations of $\Or(8)$ at this weight, the difference $237-232$ between the formal count and the dimension of the cochain space, so they span all of them. Equation~\eqref{eq:closurecert} shows that $\mathcal O_8$ is $Q$-closed on $\so(8)$ and on no larger orthogonal algebra.

The $\SU(3)_R$ raising operators annihilate $\mathcal O_8$ as a formal multi-trace, so it is a flavour singlet. Its Coulomb image is
\begin{equation}\label{eq:O8Coulomb}
C(\mathcal O_8)=64E,\qquad \langle E,C(\mathcal O_8)\rangle=207360,
\end{equation}
so $\mathcal O_8$ is not cohomologous to a combination of descendant products. Comparing \eqref{eq:O8Coulomb} with \eqref{eq:doubletimages}, and using the injectivity of the Coulomb map on $H^+$ at this weight, we obtain
\begin{equation}\label{eq:O8normalization}
[\mathcal O_8]=-120\,\big[(\tau^*\XN_4)_{\rm even}\big]
\end{equation}
modulo graviton classes. Up to this factor, $\mathcal O_8$ represents the reflection-even member of the doublet. Its vector restriction $\iv^*\mathcal O_8$ is a $Q$-closed polynomial in traces of $\SO(7)$ with Coulomb image
\begin{equation}\label{eq:O8so7}
C(\iv^*\mathcal O_8)=16F_7,
\end{equation}
in agreement with \eqref{eq:partnerrestriction} and with the component $-\frac2{15}=-\frac{16}{120}$ of $f_7$ along $F_7$.

\subsection{Coulomb visibility of the non-graviton singlets}
\label{sec:so7coulomb}

The three non-graviton singlets of this section are visible on the Coulomb branch, and they remain visible after any subtraction of graviton classes. For the $\SO(7)$ singlet, Kibe, van Leuven and Mathieson noted this visibility modulo monotone classes~\cite{Kibe:2026qid}. Here it follows from the $\SO(8)$ doublet, and it comes with an explicit witness: the function $F_7$, which is orthogonal to the Coulomb images of all graviton classes and along which the class has a non-zero component. For $\SO(7)$ the Coulomb image of the cohomology at $\sect{1}{3,3,3}$ has dimension $124$: the graviton images span the $123$-dimensional space $S_{B_3}$, and the image of the singlet class is a non-zero multiple of $F_7$, which is orthogonal to $S_{B_3}$. No representative of the singlet, modified by graviton combinations and $Q$-exact terms, vanishes on the Cartan subalgebra. The explicit operator $\iv^*\mathcal O_8$ shows the same directly, since by \eqref{eq:O8so7} its Coulomb image is $16F_7$. The same holds for the two members of the $\SO(8)$ doublet, whose images are multiples of $F$ and $E$ modulo $S$.

At these charges the statement can also be read modulo monotone classes. The formal multi-trace cohomology at $\sect{1}{3,3,3}$ is that of $\SO(10)$, whose $140$ classes are trace-graviton products, and graviton polynomials restrict to graviton polynomials. The monotone classes of $\SO(7)$ and of the reflection-even sector of $\SO(8)$ at these charges are therefore graviton classes, and the singlets are visible modulo monotone classes as well.

The same conclusion follows, without any of our class counts, from the representative printed by Gadde, Lee, Raj and Tomar and by Choi and Lee~\cite{Gadde:2025yoa,Choi:2025bhi}. We read its letters $X,Y,Z,\psi_m$ as our $\phi_1,\phi_2,\phi_3,\psi_m$. This is the only assignment under which every term has weight $(3,3,3)$, and it is the one used by Chang~\cite{Chang:2026bil}. With the graviton projection indicated in the two original works left out, the Coulomb image $c_7$ of the representative satisfies
\begin{equation}
\langle F_7,c_7\rangle=576,\qquad c_7-\tfrac4{15}F_7\in S_{B_3}.
\end{equation}
A change of $c_7$ within $S_{B_3}$ does not affect its component along $F_7$, so no graviton subtraction makes this representative vanish on the Cartan subalgebra. Its $Q$-closure on $\so(7)$ is the statement of the cited works. Kibe, van Leuven and Mathieson obtain the corresponding statement for a fortuitous operator of $\SO(7)$ outside the BMN sector, which they construct and to which the BMN class is related by a descendant relation~\cite{Kibe:2026qid}.

These results bear on several statements in the literature about the Coulomb branch. Gadde, Lee, Raj and Tomar give a Cartan argument that the graviton spectra of $\SO(2N+1)$ and $\USp(2N)$ agree~\cite{Gadde:2025yoa}. Chang and Lin argue, by a conjugation that relates the two Cartan subalgebras and preserves the multi-trace structure, that the Coulomb-branch cohomologies of $\SO(2N+1)$ and $\USp(2N)$ are isomorphic, and that they embed into the monotone cohomology because fortuitous classes are identified with classes of trace relations, which vanish on the Coulomb branch~\cite{Chang:2024zqi,Chang:2025mqp}. Choi and Lee use the non-Coulomb character of the known fortuitous classes of $\Nfour$ SYM as a premise in a conditional argument~\cite{Choi:2025bhi}. Choi and Kim state that all known non-graviton cohomologies of $\Nfour$ SYM can be chosen to be of non-Coulomb type by subtracting suitable multi-gravitons~\cite{Choi:2025pqr}. At $\sect{1}{3,3,3}$ the Coulomb image of the $\SO(7)$ cohomology has dimension $124$. Chang and Lin found that the $\USp(6)$ cohomology and the $\SO(7)$ monotone cohomology in this sector both have dimension $123$~\cite{Chang:2025mqp}. The Coulomb-branch cohomologies of $\SO(7)$ and $\USp(6)$ are therefore not isomorphic there, and the Coulomb image of the $\SO(7)$ cohomology is larger than its monotone cohomology.

The origin of the difference can be traced through the two steps of the argument. The conjugation identifies the Coulomb images of one polynomial in traces evaluated in the two theories. That polynomial is $Q$-closed in both theories when it is closed as a formal multi-trace, as monotone representatives are. The singlet of $\SO(7)$ is represented by a polynomial in traces, the trace expansion of $\is^*\XN_4$, that is closed on $\so(7)$ without being closed as a formal multi-trace; for the representative $\iv^*\mathcal O_8$ the relations responsible are those of \eqref{eq:closurecert}. If such a polynomial were closed on $\mathfrak{usp}(6)$-valued letters as well, its Coulomb image, which lies outside the span of the graviton images, would give a $\USp(6)$ class beyond the $123$ known ones.

The identification of fortuitous classes with classes of trace relations remains valid. It proceeds through the connecting map $[\mathcal O]\mapsto[Q\widetilde{\mathcal O}]$, with $\widetilde{\mathcal O}$ a formal lift of $\mathcal O$, and it constrains the lift only through its $Q$-image. It therefore does not determine the restriction of $\mathcal O$ itself to the Cartan subalgebra. Kibe, van Leuven and Mathieson also note that the vanishing of a trace relation $Q\widetilde{\mathcal O}$ on the Cartan subalgebra does not imply that of its preimage~\cite{Kibe:2026qid}.

The main results of the cited works are unaffected. The BMN cohomology of $\SO(7)$ and the S-duality mismatch stand as established by Gadde, Lee, Raj and Tomar and by Chang and Lin~\cite{Gadde:2025yoa,Chang:2025mqp}. So does the agreement of the Coulomb images of graviton and monotone classes of $\SO(2N+1)$ and $\USp(2N)$, which is what these Cartan arguments establish. The monotone cohomologies themselves differ: Chang and Lin found an additional bosonic monotone class of $\SO(7)$ of non-Coulomb type, and their conclusion that the Coulomb-branch cohomology does not exhaust the monotone cohomology stands~\cite{Chang:2025mqp,Choi:2025bhi}. At $\sect{1}{3,3,3}$ the two spaces differ in the opposite direction as well. The lifting found by Choi and Lee is also unaffected, and its target is of non-Coulomb type~\cite{Choi:2025bhi}. Sections~\ref{sec:firsttargets} and~\ref{sec:defect} explain why a visible source can have an invisible target. What does not hold in general is the inference that a fortuitous class vanishes on the Cartan subalgebra after subtraction of monotone classes, as Kibe, van Leuven and Mathieson showed with a fortuitous operator of $\SO(7)$~\cite{Kibe:2026qid}. The Pfaffian-pair classes of section~\ref{sec:evenscan}, which are fortuitous with respect to the trace covering and visible modulo trace monotones, show that such visibility can occur for reasons unrelated to black-hole physics.

\subsection{Coverings and the choice of frame}
\label{sec:interpretation}

The two members of the doublet behave alike with respect to the graviton spectrum and differently with respect to coverings. The non-graviton property, with Pfaffian-type gravitons included, is triality covariant at this weight: both members are non-graviton classes, and the graviton span is stable under $\sigma$ and $\tau$, its Coulomb image being the $S_3$-stable space $S$.

With respect to coverings built on the vector representation the two members look completely different. The reflection-odd member is a generalized monotone of the covering with ground state $\Pf(\phi_1)$ (section~\ref{sec:singletgm}), and classically a candidate non-symmetrized excitation of the wrapped brane. The reflection-even member is a polynomial in traces that is fortuitous with respect to the trace covering. It is $Q$-closed only because of a trace relation of $\Or(8)$, in the same way as the $\SO(7)$ singlet~\cite{Gadde:2025yoa,Chang:2025mqp}. The fortuitous classes of $\SU(N)$ are closed by trace relations in the same way~\cite{Chang:2024zqi}. Trace fortuity alone does not distinguish the reflection-even member from products of brane states, because the Pfaffian-pair classes in the columns $[2,2]$ and $[6,0]$ are trace-fortuitous in the same way (section~\ref{sec:evenscan}). The distinction requires the larger class of coverings that includes Pfaffian families.

In the reflection-even sector, the analogue of the space $GM^-_N$ of section~\ref{sec:monotone} is a space $GM^+_N$. It is spanned by trace monotones, by products of an even number of Pfaffian-type generalized monotones, and by the classes of families with an even number $B\ge2$ of $\eps$-tensors that are closed at every rank (appendix~\ref{app:multipf}), together with their products with trace monotones. In all these families every index of every $\eps$-tensor is carried by a word of at least one letter. By the threshold $j\ge2BN$, families with two $\eps$-tensors can contribute from $j=16$ and must be considered at $j=18$.

The class of $(\tau^*\XN_4)_{\rm even}$ does not lie in $GM^+_4$. More generally, it is not cohomologous to the sum of a trace monotone and a $Q$-closed operator of $\SO(8)$ built from $\eps$-contractions in which every index of every $\eps$-tensor is carried by a word. The argument uses the vector restriction. For $\so(7)\oplus0$-valued letters every word has vanishing eighth row and column, while each non-zero term of $\eps_{a_1\cdots a_8}$ has one index equal to $8$, so $\iv^*$ annihilates all such $\eps$-contractions. A trace monotone is represented by a polynomial in traces that is closed as a formal multi-trace and hence on $\so(10)$, where at this weight every class is a trace-graviton class; by block restriction it is a graviton combination up to a $Q$-exact term on $\so(7)$ as well. If $(\tau^*\XN_4)_{\rm even}$ were of the stated form, $\is^*\XN_4=\iv^*(\tau^*\XN_4)_{\rm even}$ would be a graviton combination of $\SO(7)$ up to a $Q$-exact term, which section~\ref{sec:SO7restriction} excludes. The argument has the same structure as the one given by Choi, Choi and Kim for the conifold theory, where a consistent truncation maps every generalized monotone built from their generators to a graviton class while the truncation of their fortuitous operator is not one~\cite{Choi:2026faq}.

This argument does not cover contractions of two $\eps$-tensors in which some indices are contracted directly with Kronecker deltas. These are subdeterminant operators: two $\eps$-tensors contracted with six copies of $X$ and two deltas, for instance, give a multiple of the elementary symmetric function $e_6$ of the eigenvalues of $X$. At each rank they are polynomials in traces, they survive the restriction to $\so(7)\oplus0$, and Choi, Choi and Kim discuss an example of this kind~\cite{Choi:2026faq}. Calling the reflection-even member strongly fortuitous with respect to such families as well would require the analogue of their working assumption that there is no non-factorizable generalized monotone independent of ordinary monotones, which they adopt for the conifold theory at $N=2$. For $\SO(7)$, which has no Pfaffians, we take generalized monotones to be ordinary monotones, with the same proviso about subdeterminant families.

The dependence on the frame is explicit. A convention that also admitted the images of rank-uniform families under all outer automorphisms would make the reflection-even member a generalized monotone, since
\begin{equation}
(\tau^*\XN_4)_{\rm even}=\tfrac12\big(\tau^*+(\tau\sigma)^*\big)\XN_4
\end{equation}
is a combination of such images. Triality exists only at rank four, however, so this image is not itself a member of a rank-uniform family. A related frame dependence appears in the work of Choi, Choi and Kim~\cite{Choi:2026faq}. Fortuitous classes of ABJM theory were constructed by Belin et al.\ and by Behan and Pipolo de Gioia~\cite{Belin:2025hsg,Behan:2025hbx}. Choi, Choi and Kim point out that such classes can be products of baryonic generalized monotones~\cite{Choi:2026faq}. In the conifold theory at $N=2$ an enhanced $\SU(4)$ flavour symmetry mixes mesons with baryons, that is, unwrapped states with wrapped D3-branes, and the cohomology is organized into $\SU(4)$ multiplets~\cite{Choi:2026faq}. Here the mixing is an outer automorphism of the gauge algebra, and the split defined in the vector frame is not $S_3$ covariant.

The relation to $\SO(7)$ should be read in the same way. The known non-graviton singlet of $\SO(7)$ is represented by the restriction, along the spinor embedding, of the rank-four member of a generalized-monotone family of $\SO(8)$, and by the restriction, along the vector embedding, of a class that is strongly fortuitous with respect to $GM^+_4$. This is an algebraic statement and not a relation between bulk objects. Restriction along $\is$ is a chain map of BMN complexes, but no Coulomb-branch vacuum realizes it, since $\is(\so(7))$ has rank three, and the $\SO(7)$ and $\SO(8)$ theories are dual to different orientifolds~\cite{Witten:1998xy}. Whether a class belongs to a rank-uniform family thus depends on the embedding used to compare ranks. Triality preserves the graviton span of $\SO(8)$ at this weight, and the spinor restriction of the reflection-odd member and the vector restriction of the reflection-even member both give the non-graviton singlet of $\SO(7)$. Not every restriction preserves the non-graviton character: the vector restriction annihilates the reflection-odd member, and restricting the $\SO(7)$ singlet further to $\mathfrak g_2$ gives a graviton class~\cite{FortuitySummary}. Section~\ref{sec:quantum} shows that the quantum fate distinguishes the two theories as well: the $\SO(8)$ doublet is protected in perturbation theory, whereas the $\SO(7)$ singlet to which both of its members restrict is lifted at first order.
\section{Quantum fate and the failure of restriction}
\label{sec:quantum}

The classes studied so far are classes of the classical supercharge, and perturbative corrections to the supercharge can lift them. Both members of the $\SO(8)$ doublet restrict to the $\SO(7)$ singlet, and Choi and Lee showed that the first correction lifts that singlet~\cite{Choi:2025bhi}. In this section we show that the doublet itself is not lifted at any order of perturbation theory, and that the same holds for the four-term family at every rank. The two statements are compatible because the classical differential involves only brackets of the letters, whereas the quantum corrections contract pairs of letters through the full gauge algebra of the ambient theory. Restriction to a subalgebra therefore commutes with the former and not with the latter.

The argument rests on two results. The first-order variation of a closed operator vanishes on the Cartan subalgebra, a result of Kibe, van Leuven and Mathieson for the trace sector that we extend to the Pfaffian sector~\cite{Kibe:2026qid}. For BMN classes with $2J=1$ this variation is built from letters of the lowest degree only, and in $\SO(2N)$ the cohomology of such operators is represented faithfully on the Cartan subalgebra, for reflection-odd operators with any number of letters and for reflection-even operators with at most $2N$ letters. Together they imply that the variation is exact. After recalling the first quantum correction we establish these two results in turn, deduce the survival of the classes at first order and construct explicit corrected operators to first order. We then prove protection to all perturbative orders. Finally we identify the term by which restriction to $\so(7)$ fails to commute with the correction and show how it accounts for the lifting of the $\SO(7)$ singlet.

\subsection{The first quantum correction}
\label{sec:Q1}

We write the supercharge as $Q(\eta)=Q_0+\eta Q_1+\eta^2Q_2+\cdots$, where $Q_0$ is the classical differential \eqref{eq:Q} and $\eta$ is proportional to the squared coupling. This is the normalization of Kim, Kim, Lee and Park, which the Fortuity collaboration also uses~\cite{Kim:2026rnx,FortuitySummary}. Nilpotence of $Q(\eta)$ implies
\begin{equation}\label{eq:c8-squares}
Q_0^2=0,\qquad \{Q_0,Q_1\}=0,\qquad Q_1^2=-\{Q_0,Q_2\}.
\end{equation}
The second relation shows that $Q_1$ induces a differential $d_1=[Q_1]$ on the cohomology of $Q_0$. A classical class survives at first order if it defines a non-zero class in $\ker d_1/\operatorname{im}d_1$~\cite{Budzik:2023xbr,Choi:2025bhi,Budzik:2025zvu,Kibe:2026qid}. This requires that the variation $Q_1\mathcal O$ of a representative be $Q_0$-exact and that the class not be the image of another class. If $Q_1\mathcal O=-Q_0\mathcal O'$, the corrected operator $\mathcal O+\eta\mathcal O'$ is closed to this order,
\begin{equation}\label{eq:c8-firstorder}
Q(\eta)(\mathcal O+\eta\mathcal O')=O(\eta^2).
\end{equation}
If instead $\mathcal O=Q_1\mathcal P+Q_0\mathcal W$ with $Q_0\mathcal P=0$, then $Q(\eta)(\mathcal P+\eta\mathcal W)=\eta\,\mathcal O+O(\eta^2)$, and $\mathcal O$ pairs with $\mathcal P$ into a long multiplet.

The correction $Q_1$ acts on the complete set of $\frac1{16}$-BPS letters, of which the BMN letters are a subset. We use the superfield notation of Chang and Yin and of Chang and Lin~\cite{Chang:2013fba,Chang:2022mjp}. All letters are coefficients of a Grassmann-odd superfield,
\begin{equation}\label{eq:superfield}
\Psi=\sum_{k_\pm\ge0}\;\sum_{S\subset\{1,2,3\}}(z^+)^{k_+}(z^-)^{k_-}\,\theta_S\,\Psi_{k_+,k_-;S},\qquad \Psi_{0,0;\varnothing}=0,
\end{equation}
where $z^\pm$ are commuting and $\theta_m$ Grassmann coordinates, and $\theta_S=\prod_{m\in S}\theta_m$, in increasing order, stands to the left of the coefficient. The BMN letters are $\phi_m=\Psi_{0,0;m}$, $\psi_1=\Psi_{0,0;23}$, $\psi_2=-\Psi_{0,0;13}$, $\psi_3=\Psi_{0,0;12}$ and $f=-\Psi_{0,0;123}$, and the gauginos are $\lambda_+=\Psi_{1,0;\varnothing}$ and $\lambda_-=\Psi_{0,1;\varnothing}$. The classical differential is $\{Q_0,\Psi\}=\Psi^2$, which reproduces \eqref{eq:Q} on the BMN letters and expresses every component variation as a graded commutator. Restricting all components to the Cartan subalgebra therefore extends the Coulomb map $C$ to the full complex, with $C\,Q_0=0$ as before.

We grade operators by their letter number $L$, by their total degrees $n_\pm$ in $z^\pm$, which count derivatives and gauginos, and by their total degrees $t_m$ in $\theta_m$. In terms of these, the angular momenta and the charges $q_m$ are
\begin{equation}\label{eq:allorder-charges}
\begin{gathered}
2J_1=-L+2n_++\sum_mt_m,\qquad 2J_2=-L+2n_-+\sum_mt_m,\\
q_m=R_m+\tfrac12(J_1+J_2)=\tfrac12(n_++n_-)+t_m,
\end{gathered}
\end{equation}
where $R_m$ are the Cartan charges of $\SO(6)_R$. BMN operators have $n_+=n_-=0$ and $J_1=J_2=J$. The classical differential raises $L$ by one and preserves $n_\pm$ and $t_m$. The first correction preserves $L$, raises each of $n_\pm$ by one and lowers each $t_m$ by one. Since $n_+=n_-=0$ are the lowest possible values, a non-zero BMN class cannot be a $Q_1$-image modulo a $Q_0$-boundary~\cite{Kim:2026rnx,Kim:2026mch,Kibe:2026qid,FortuitySummary}. This holds for both reflection parities. For BMN classes, survival at first order therefore reduces to the first condition, $d_1[\mathcal O]=0$.

A further grading organizes the arguments below. We call $d=n_++n_-+\sum_mt_m$ the degree of an operator. It is the sum of the degrees of its letters in the superspace coordinates, and every letter has degree at least one because $\Psi_{0,0;\varnothing}=0$. By \eqref{eq:allorder-charges},
\begin{equation}\label{eq:degree}
d-L=J_1+J_2,
\end{equation}
so an operator with $J_1+J_2=0$ consists of letters of degree one only. These letters are the scalars $\phi_m$ and the gauginos $\lambda_\pm$, and all of them are $Q_0$-closed. For the classes at $\sect{1}{3,3,3}$ the variation $Q_1\mathcal O$ is of this kind. It has $n_+=n_-=1$, vanishing $J_1$ and $J_2$ and the charges $q=(3,3,3)$ of the nine-scalar BMN sector, but it contains two gauginos and six scalars, so it has eight letters and lies outside the BMN sector.

Near-BPS spectra and the weak-coupling form of the first non-graviton operator were obtained from the one-loop dilatation operator by Chang, Feng, Lin and Tao and by Budzik, Murali and Vieira~\cite{Chang:2023zqk,Budzik:2023vtr}. The first correction to the supercharge has been obtained in several steps. Choi and Lee inferred the term that contracts a scalar with a fermion from the generalized Konishi anomaly~\cite{Cachazo:2002ry,Choi:2025bhi}. Budzik and Kulp derived the complete pair operation, with a compact superfield expression and component rules that include letters with derivatives~\cite{Budzik:2025zvu}. Their derivation builds on work on semi-chiral operators and higher operations in perturbation theory~\cite{Budzik:2023xbr,Gaiotto:2024gii}. Kim, Kim, Lee and Park wrote out the component correction for $\SU(2)$~\cite{Kim:2026rnx}. On the one-fermion representatives considered below only the scalar--fermion term acts, but the full correction is needed whenever $Q_1$ is used as an operator on general cochains.

To state the scalar--fermion term for $\so(n)$, choose the generators $t_A=M_{ij}=e_{ij}-e_{ji}$ with $1\le i<j\le n$ of section~\ref{sec:homomorphism}. They are orthonormal for $-\frac12\tr$, and $t^A=-t_A$ is the dual basis for $+\frac12\tr$. They obey the completeness relation
\begin{equation}\label{eq:completeness}
\sum_A(t_A)_{ab}(t_A)_{cd}=\delta_{ac}\delta_{bd}-\delta_{ad}\delta_{bc},
\end{equation}
and the pair operation is
\begin{equation}\label{eq:Q1pair}
Q_1(\phi_m\otimes\psi_p)=-\delta_{mp}\sum_A\big([t_A,\lambda_+]\otimes[t^A,\lambda_-]-[t_A,\lambda_-]\otimes[t^A,\lambda_+]\big).
\end{equation}
On a gauge-invariant operator $Q_1$ is the sum over unordered pairs of letters, with the graded permutation signs. The first tensor factor replaces the earlier and the second the later of the two contracted letters; the combination is graded symmetric, so the opposite placement gives the same result. Pairs of scalars give zero. The operation can join letters in different traces, and it is not a derivation on products.

The pair operation depends on the algebra in which the generators are summed. For an antisymmetric $n\times n$ matrix $U$ the relation \eqref{eq:completeness} gives
\begin{equation}\label{eq:c8-adjointcasimir}
\sum_At_AUt_A=-U,\qquad \sum_At_At_A=-(n-1)\,1,\qquad \sum_A[t_A,[t_A,U]]=-2(n-2)\,U,
\end{equation}
and hence
\begin{equation}\label{eq:Q1normalization}
Q_1\tr(\phi_m\psi_p)=4\delta_{mp}(n-2)\tr(\lambda_+\lambda_-).
\end{equation}
The two terms of \eqref{eq:Q1pair} contribute equally, because $\tr(\lambda_-\lambda_+)=-\tr(\lambda_+\lambda_-)$. The factor $n-2$ is the dual Coxeter number of $\so(n)$, and the size of the matrices enters through the closed index loop $\tr1=n$. On polynomials in traces, $Q_1$ is thus given by the same formal rules for every $\so(n)$, and $n$ enters only through the value assigned to a closed loop. These are the trace rules of the Fortuity collaboration, which states them for unitary and odd orthogonal algebras~\cite{FortuitySummary}. The completeness relation \eqref{eq:completeness} holds for every $n$ and extends them to even $n$. The factor $n-2$ agrees with the Konishi anomaly coefficient of Choi and Lee, and the normalization \eqref{eq:Q1normalization} with that of the Fortuity collaboration~\cite{Choi:2025bhi,FortuitySummary}.

The case $n=4$ provides a check that involves no new computation. The Casimir tensor of $\so(4)=\su(2)\oplus\su(2)$ is the sum of those of the two factors with equal coefficients, so the correction acts on each factor separately. The $\SU(2)$ threshold class survives at first order~\cite{Budzik:2023vtr,Kim:2026rnx,FortuitySummary}. It follows that $d_1[\mathcal O^-_{24}]=0$ for the reflection-odd threshold class of $\SO(4)$ of section~\ref{sec:control}, which is a BMN class and therefore survives at first order (appendix~\ref{app:quantumconventions}).

\subsection{Coulomb images of first-order variations}
\label{sec:firsttargets}

The singlets of section~\ref{sec:so7coulomb} retain a non-zero image on the Cartan subalgebra after any subtraction of graviton or monotone classes. Their first-order variations do not. For every polynomial multi-trace operator $\mathcal Y$ of an orthogonal group,
\begin{equation}\label{eq:targettheorem}
Q_0\mathcal Y=0\quad\Longrightarrow\quad C(Q_1\mathcal Y)=0,
\end{equation}
whether or not $C(\mathcal Y)$ vanishes. This result is due to Kibe, van Leuven and Mathieson, who establish it for all classical families~\cite{Kibe:2026qid}. Their argument retains the Coulomb remainder of the source, which need not vanish after the subtraction of monotone classes, as their own fortuitous operator of $\SO(7)$ shows. We recall it for the orthogonal groups and then extend it to the reflection-odd operators of $\SO(2N)$.

The idea is to split a closed source into an expression in gravitons and an exact term, at the cost of inverting differences of scalar eigenvalues, and to clear the denominators before applying $Q_1$. We diagonalize the scalar $\phi_1$, with distinct eigenvalues $\zeta_a$, and invert their differences. The off-diagonal components of the letters then form contractible pairs of $Q_0$. For example, $Q_0\psi_2=[\phi_3,\phi_1]$ pairs an off-diagonal component of $\psi_2$ with $(\zeta_b-\zeta_a)(\phi_3)_{ab}$. For a pair with $Q_0w=w'$ and $Q_0w'=0$, the anticommutator of $w\,\partial_{w'}$ with $Q_0$ counts the pair variables, so a closed component of positive pair degree is exact. Each diagonal letter has a closed replacement $\widetilde\ell_a$, constructed recursively from single-trace gravitons with coefficients that are rational in the eigenvalues, with denominators that are products of eigenvalue differences. For a given polynomial source finitely many substitutions suffice. Eliminating the contractible pairs leaves a closed expression in the diagonal replacements,
\begin{equation}\label{eq:localizedremainder}
\mathcal Y=\mathcal Y_{\rm loc}+Q_0\widetilde{\mathcal H},\qquad \mathcal Y_{\rm loc}=C(\mathcal Y)(\zeta,\widetilde\ell)=\sum_Ir_I(\zeta)\,g_I(\mathcal G_\alpha),
\end{equation}
where the $\mathcal G_\alpha$ are closed single-trace gravitons, the $g_I$ are products of them and the $r_I$ are rational functions. The contracted pairs correspond to the massive off-diagonal modes on the Coulomb branch. In \eqref{eq:localizedremainder}, however, the eigenvalues are variables in a localization of the algebra of operators and not fixed expectation values~\cite{Kibe:2026qid}.

Let $\mathcal D=\prod_{a<b}(\zeta_a-\zeta_b)^2$ be the discriminant of $\phi_1$. Clearing denominators and averaging over the residual gauge symmetry gives an identity of polynomial gauge invariants,
\begin{equation}\label{eq:discriminantgraviton}
\mathcal D^k\,\mathcal Y=\mathcal G+Q_0\mathcal H,
\end{equation}
where $\mathcal G$ is a polynomial in trace gravitons. For the even orthogonal groups the average includes the outer reflection, which excludes a single Pfaffian from the coefficients. Appendix~\ref{app:localization} gives the recursion, the Vandermonde argument for the orthogonal groups and the descent to gauge invariants. The identity extends across the locus of coinciding eigenvalues by polynomiality. The factor $\mathcal D^k$ is essential: the identity does not say that the source itself is a polynomial in gravitons up to an exact term, and for a visible source the Coulomb remainder is retained in $\mathcal G$. Kibe, van Leuven and Mathieson call the corresponding representative with rational coefficients a rational multi-graviton~\cite{Kibe:2026qid}.

The last step uses a property of $Q_1$ that holds only after restriction to the Cartan subalgebra. Consider a product $UV$ of two gauge-invariant operators and a contraction of a letter of $U$ with a letter of $V$. By \eqref{eq:Q1pair} and its analogues for the other pairs of letters, it inserts a commutator $[t_A,u]$ into $U$ and $[t^A,u']$ into $V$, summed over the generators. On the Cartan subalgebra $u$ commutes with the Cartan generators, and for a root vector $t_A=e_\alpha$ the commutator is proportional to $e_\alpha$. An invariant polynomial evaluated on Cartan-valued letters and one root vector vanishes, because it is invariant under the Cartan torus while its argument has the non-zero weight $\alpha$. All contractions between different invariant factors therefore vanish on the Cartan subalgebra, and
\begin{equation}\label{eq:cartanproduct}
C\big(Q_1(UV)\big)=C(Q_1U)\,C(V)+(-1)^{|U|}C(U)\,C(Q_1V).
\end{equation}
For products of traces this is the restricted product rule of Kibe, van Leuven and Mathieson~\cite{Kibe:2026qid}. The argument applies equally to factors built with $\eps$-tensors.

A single-trace graviton has a $Q_0$-exact image under $Q_1$, as used in the trace sector by Kibe, van Leuven and Mathieson on the basis of the closure analysis of Budzik and Kulp~\cite{Budzik:2025zvu,Kibe:2026qid}. Hence $C(Q_1\mathcal G)=0$, and $Q_1\mathcal D^k=0$ because $\mathcal D$ contains only $\phi_1$. Applying $Q_1$ and then $C$ to \eqref{eq:discriminantgraviton} gives
\begin{equation}\label{eq:cartantarget}
C(\mathcal D)^k\,C(Q_1\mathcal Y)=C(Q_1\mathcal G)-C(Q_0Q_1\mathcal H)=0.
\end{equation}
Multiplication by the bosonic discriminant is injective on the algebra of polynomials on the Cartan subalgebra, including its Grassmann variables, which gives \eqref{eq:targettheorem}. When $C(\mathcal Y)$ vanishes the remainder is zero, and \eqref{eq:discriminantgraviton} reduces to the statement of Kibe, van Leuven and Mathieson that a sufficiently high power of the discriminant makes the class exact~\cite{Kibe:2026qid}. For a visible source the same injectivity shows instead that $[\mathcal D^k\mathcal Y]\ne0$.

The result extends to the Pfaffian sector. Let $\mathcal Y$ be a $Q_0$-closed reflection-odd operator of $\SO(2N)$. The product $\Pf(\phi_1)\,\mathcal Y$ is $Q_0$-closed and reflection-even, hence a polynomial in traces, and \eqref{eq:targettheorem} applies to it. Since $Q_1\Pf(\phi_1)=0$, the product rule \eqref{eq:cartanproduct} gives
\begin{equation}\label{eq:oddtarget}
0=C\big(Q_1(\Pf(\phi_1)\,\mathcal Y)\big)=C\big(\Pf(\phi_1)\big)\,C(Q_1\mathcal Y).
\end{equation}
The Coulomb image of $\Pf(\phi_1)$ is the product of the block values of $\phi_1$, which is not a zero divisor. Hence \eqref{eq:targettheorem} holds for every $Q_0$-closed gauge-invariant operator of $\SO(2N)$, of either reflection parity.

\subsection{Operators of lowest degree and their cohomology}
\label{sec:lowestdegree}

The first-order variations of the classes of interest, and the obstructions that can arise at higher orders, consist of letters of degree one. The cohomology of such operators can be described in closed form, and this description is the second ingredient of the argument.

Consider the gauge-invariant operators of $\SO(2N)$ with $L$ letters and degree $d=L$, at fixed gradings $n_\pm$ and $t_m$. Their letters are scalars and gauginos, so all of them are $Q_0$-closed. A $Q_0$-preimage has $L-1$ letters and the same degree, and therefore consists of one letter of degree two and $L-2$ letters of degree one. The letters of degree two are the fermions $\psi_m$, the first derivatives $\Psi_{1,0;m}$ and $\Psi_{0,1;m}$ of the scalars, and the letters $\Psi_{2,0;\varnothing}$, $\Psi_{0,2;\varnothing}$ and $\Psi_{1,1;\varnothing}$. Expanding $\{Q_0,\Psi\}=\Psi^2$ gives, in addition to $Q_0\psi_m$ in \eqref{eq:Q},
\begin{equation}\label{eq:preimage-components}
Q_0\Psi_{1,0;m}=[\lambda_+,\phi_m],\qquad Q_0\Psi_{0,1;m}=[\lambda_-,\phi_m],\qquad Q_0\Psi_{1,1;\varnothing}=\{\lambda_+,\lambda_-\},
\end{equation}
together with $Q_0\Psi_{2,0;\varnothing}=\lambda_+^2$ and $Q_0\Psi_{0,2;\varnothing}=\lambda_-^2$.

Let $R$ be the graded-commutative polynomial algebra in the matrix components of $\phi_m$ and $\lambda_\pm$, and let $I\subset R$ be the ideal generated by the components of $[\phi_m,\phi_n]$, $[\lambda_\pm,\phi_m]$, $\lambda_\pm^2$ and $\{\lambda_+,\lambda_-\}$. Since the letters of degree one are closed, the boundaries are the invariant elements of $I$, and the cohomology is the corresponding graded piece of the invariants $(R/I)^{\SO(2N)}$. Passing to invariants loses no preimages, because the average over the compact gauge group of a preimage of an invariant boundary is an invariant preimage. The quotient $(R/I)^{\SO(2N)}$ is a generalization, with the scalars included, of the classical chiral ring of gaugino bilinears studied by Witten~\cite{Witten:2003ye}. Etingof and Kac proved the conjecture of Cachazo, Douglas, Seiberg and Witten on this ring for $G_2$, and Etingof gave an algebraic proof for $\SO(n)$~\cite{Cachazo:2002ry,Etingof:2003dd,Etingof:2003fy}. Kibe, van Leuven and Mathieson used the gaugino chiral ring to identify the target of a first-order lifting in $\SO(7)$~\cite{Kibe:2026qid}.

In $R/I$ the matrices supercommute, and the gauginos square to zero. Transposing a word of length $\ell$ reverses the order of its letters and multiplies each by $-1$, as in \eqref{eq:slotid}, and restoring the order with the help of supercommutativity cancels the sign of the reversal. Hence $W^T=(-1)^\ell W$: words of even length are symmetric and words of odd length antisymmetric.

Consequently a trace of odd length vanishes, and so does an $\eps$-contraction with a slot of even length. A non-vanishing reflection-even invariant therefore has even $L$, and a non-vanishing reflection-odd invariant has $L\equiv N$ modulo two. The cohomology of operators with $d=L$ vanishes unless the parity \eqref{eq:parity} is $P_N=+1$, exactly as for the restriction to the Cartan subalgebra in section~\ref{sec:gravitons}.

The reflection-even part, $(R/I)^{\Or(2N)}$, is the image of $R^{\Or(2N)}$, which is generated by traces~\cite{Procesi:1976}, with the gauginos treated by the polarization of section~\ref{sec:invariants}. The reflection-even part is therefore spanned by the products of traces of words of even length, where each word is determined by its letter content and contains each gaugino at most once. A product with two equal Grassmann-odd factors vanishes. The remaining products are labelled by the multiset of their words, and for $L\le2N$ they are linearly independent and have linearly independent Coulomb images. To see this, recall that on the Cartan subalgebra a trace is a sum over the $N$ blocks of the product of the block values of its letters. In the image of a product of $k$ traces, the monomials that involve exactly $k$ blocks are those in which the traces occupy distinct blocks, and each such monomial records the words: the letters carried by a block are those of one word. Products of $k$ traces with different multisets of words thus contribute different monomials on $k$ blocks, and products of fewer traces contribute none. Since $k\le L/2\le N$, an induction that descends in the number of traces shows that a vanishing combination of Coulomb images has vanishing coefficients. The Coulomb map is therefore injective on the reflection-even cohomology of operators with $d=L\le2N$.

The reflection-odd part is spanned by products of one $\eps$-word with traces. The covariance of the $\eps$-tensor under $\mathrm{GL}(2N)$ gives, for a matrix $W_0$ of Grassmann parity $p$ and slots of parities $|W_s|$,
\begin{equation}\label{eq:absorbtrace}
\begin{gathered}
\tr(W_0)\,\eps(W_1,\dots,W_N)=\sum_{s=1}^N(-1)^{p\sum_{i<s}|W_i|}\,\eps\big(W_1,\dots,W_{s-1},\widehat W_s,W_{s+1},\dots,W_N\big),\\
\widehat W_s=W_0^TW_s+(-1)^{p|W_s|}\,W_sW_0.
\end{gathered}
\end{equation}
This follows by applying $\eps(g^TW_1g,\dots,g^TW_Ng)=\det g\;\eps(W_1,\dots,W_N)$ to $g=1+\vartheta W_0$, with a parameter $\vartheta$ of parity $p$ and $\vartheta^2=0$. The second sign arises when $\vartheta$ is moved to the left of $W_s$, and the first when it is then moved to the front of the contraction, past the slots $W_1,\dots,W_{s-1}$. In $R/I$ a word $W_0$ of even length is symmetric and supercommutes with $W_s$, so $\widehat W_s=2W_0W_s$, and the right-hand side is twice the signed sum of the $\eps$-words in which $W_0$ is absorbed into one slot. The reflection-odd cohomology is therefore spanned by $\eps$-words without trace factors, whose $N$ slots are words of odd length determined by their letter content. An $\eps$-word with two equal Grassmann-odd slots vanishes. By \eqref{eq:epsCartan} the Coulomb image of each of the others is, up to a non-zero factor, the sum over all assignments of its $N$ slots to the $N$ blocks, and every monomial in it records the multiset of slot words. Different multisets give different monomials. These $\eps$-words therefore form a basis, and the Coulomb map is injective on the reflection-odd cohomology of operators with $d=L$, for every $N$ and every $L$.

For $\SO(8)$ and the gradings $n_+=n_-=1$, $t=(2,2,2)$ of the variations of the classes at $\sect{1}{3,3,3}$, the reflection-even basis consists of $1$, $26$, $57$ and $23$ products of one, two, three and four traces. The cohomology thus has dimension $107$, and Weyl alternation over the neighbouring gradings $t=(3,2,1)$, $(4,1,1)$, $(3,3,0)$ and $(4,2,0)$, where the dimensions are $85$, $62$, $47$ and $43$, shows that it contains three flavour singlets. The reflection-odd basis of $\eps$-words has $92$ elements at these gradings and $73$, $50$, $39$ and $34$ at the neighbouring ones, so that the reflection-odd cohomology contains one flavour singlet. After reduction of the trace words by graded cyclicity and reversal, the space of eight-letter multi-traces at these gradings has dimension $848$, and the image of $Q_0$ from the seven-letter operators has dimension $741=848-107$. It is annihilated by $107$ independent integer linear functionals, and the matrix of $Q_0$ has a non-zero minor of size $741$ modulo a prime.

\subsection{First-order survival}
\label{sec:survival}

Let $\mathcal Y$ be a $Q_0$-closed BMN operator of $\SO(2N)$ with one fermion, $2J=1$, and $L$ letters. Its variation $Q_1\mathcal Y$ has $J_1+J_2=0$ and hence, by \eqref{eq:degree}, degree $d=L$. It is $Q_0$-closed, and its Coulomb image vanishes by section~\ref{sec:firsttargets}. If $\mathcal Y$ is reflection-odd, or reflection-even with $L\le2N$, the Coulomb map is injective on the cohomology in which the class of $Q_1\mathcal Y$ lies (section~\ref{sec:lowestdegree}). The class therefore vanishes: $Q_1\mathcal Y$ is $Q_0$-exact. Since no BMN class is a $Q_1$-image, we conclude that every reflection-odd BMN class of $\SO(2N)$ with $2J=1$ survives at first order, at every rank, and so does every reflection-even BMN class with $2J=1$ and at most $2N$ letters.

For $\SO(8)$ at the charges $\sect{1}{3,3,3}$, where $L=8$, this covers the whole cohomology in both reflection sectors: all $145$ reflection-even classes and all $119$ reflection-odd classes survive at first order (table~\ref{tab:seedcomplexes}). In particular both members of the doublet survive. The bound $L\le2N$ is saturated. Eight is the largest number of letters for which $\Or(8)$ has no trace relations, and the reflection-even member is closed only because of the relations among nine letters at $2J=0$, while its variation still lies in a sector that the four blocks of the Cartan subalgebra resolve completely. For the four-term family the result holds at every rank: the classes of $\XN_N$ survive at first order for all $N\ge4$. Survival does not follow from an absence of possible targets. The target cohomology has dimension $107$ in the reflection-even case, and what vanishes is the map $d_1$.

The argument shows that a preimage exists, and for the two singlets we also construct it. We begin with the reflection-even member, represented by the operator $\mathcal O_8$ of section~\ref{sec:O8}. By \eqref{eq:Q1pair} its variation $v=Q_1\mathcal O_8$ is a polynomial in traces of eight letters, with two gauginos and six scalars, $n_+=n_-=1$ and $t_m=2$. It is non-zero as a formal multi-trace, and hence non-zero on $\so(8)$, where formal multi-traces of eight letters are linearly independent. A preimage consists of terms of three kinds: a fermion $\psi_m$ with both gauginos and four scalars; a derivative of a scalar, $\Psi_{1,0;m}$ or $\Psi_{0,1;m}$, with one gaugino and five scalars; and the letter $\Psi_{1,1;\varnothing}$ with six scalars. Solving the linear system defined by \eqref{eq:Q} and \eqref{eq:preimage-components} for $2v$ gives a polynomial in traces $\mathcal Z$ with $675$ terms and integer coefficients, such that
\begin{equation}\label{eq:oneloop}
2\,Q_1\mathcal O_8=Q_0\mathcal Z.
\end{equation}
Of the terms of $\mathcal Z$, $231$ contain a fermion with both gauginos, $365$ a derivative of a scalar with one gaugino, and $79$ the letter $\Psi_{1,1;\varnothing}$. The identity \eqref{eq:oneloop} holds over the integers and uses only graded cyclicity, reversal of words and the value $\tr1=8$ of a closed loop.

The corrected representative is
\begin{equation}\label{eq:O8correction}
\mathcal O_8(\eta)=\mathcal O_8-\tfrac\eta2\mathcal Z+O(\eta^2),\qquad Q(\eta)\,\mathcal O_8(\eta)=O(\eta^2).
\end{equation}
The correction necessarily leaves the BMN sector. Adding a $Q_0$-closed operator to $\mathcal Z$ gives another solution, so $\mathcal O_8(\eta)$ represents the corrected class and is not a distinguished BPS operator. The chosen $\mathcal Z$ is not a flavour singlet, but its average over $\SU(3)_R$ is a singlet with the same $Q_0$-image. A change of representative $\mathcal O_8\mapsto\mathcal O_8+Q_0B$ changes the variation by $-Q_0Q_1B$, by \eqref{eq:c8-squares}, so exactness depends only on the classical class.

For the reflection-odd member we use triality. Every automorphism of $\so(8)$, outer ones included, preserves the bracket and the invariant form, and hence the Casimir tensor contracted by every pair rule. It therefore commutes with the full $Q_1$, also on operators built with $\eps$-tensors. We define
\begin{equation}\label{eq:Omega}
\Omega=\tfrac12(1-\sigma^*)\tau^*\mathcal O_8,\qquad \mathcal Z_{\rm odd}=\tfrac12(1-\sigma^*)\tau^*\mathcal Z.
\end{equation}
Equivariance gives $Q_0\Omega=0$ and $2Q_1\Omega=Q_0\mathcal Z_{\rm odd}$. The Coulomb image of $\Omega$ follows from $C(\tau^*\mathcal O_8)=48F-32E$ and is $C(\Omega)=48F$, so $\Omega$ represents a non-zero multiple of the reflection-odd singlet class. To obtain a corrected operator whose leading term is the singlet projection $\XN_4^{(0)}$ of the covering family itself, we need the boundary that relates the two representatives. It is given by the identity
\begin{equation}\label{eq:directidentity}
\Omega-90\,\XN_4^{(0)}=Q_0\Lambda,
\end{equation}
in which $\Lambda$ is a reflection-odd combination of $\eps$-words with two fermions, written out in appendix~\ref{app:directidentity}. The normalization agrees with the Coulomb images, $C(\Omega)=48F=90\,C(\XN_4^{(0)})$. Subtracting $Q(\eta)\Lambda$ from the corrected $\Omega$ changes the leading term from $\Omega$ to $90\,\XN_4^{(0)}$ and the first-order term accordingly, and it preserves closure because $Q(\eta)^2=0$. Division by $90$ gives
\begin{equation}\label{eq:X4correction}
\XN_4^{(0)}(\eta)=\XN_4^{(0)}-\frac\eta{90}\Big(\tfrac12\mathcal Z_{\rm odd}+Q_1\Lambda\Big)+O(\eta^2),\qquad Q(\eta)\,\XN_4^{(0)}(\eta)=O(\eta^2).
\end{equation}
Here $Q_1\Lambda$ is the full correction, including its action on pairs of fermions. The relations \eqref{eq:c8-squares} and the two identities \eqref{eq:oneloop} and \eqref{eq:directidentity} fix this deformation without an evaluation of $Q_1\Lambda$.

In the reflection-even sector $d_1$ vanishes on the monotone classes of the trace covering, as Kibe, van Leuven and Mathieson established~\cite{Kibe:2026qid}; directly, the $Q_1$-images of a set of $324$ trace-graviton polynomials at this weight are annihilated by the $107$ functionals that characterize the boundaries. For the products of two Pfaffian-type factors, which are fortuitous with respect to the trace covering, triality gives an independent argument. Let ${\Pf}_{\rm tr}=-\frac1{16}\tr X^4+\frac1{64}(\tr X^2)^2$. By \eqref{eq:Pftriality} and $\tau^3=1$ on invariants,
\begin{equation}\label{eq:Pfdoublet}
\tau^*{\Pf}=-\tfrac12\,{\Pf}+{\Pf}_{\rm tr},\qquad \tau^*{\Pf}_{\rm tr}=-\tfrac34\,{\Pf}-\tfrac12\,{\Pf}_{\rm tr},
\end{equation}
and consequently, in the tensor square of this two-dimensional representation,
\begin{equation}\label{eq:Pftensor}
\begin{aligned}
{\Pf}\otimes{\Pf}={}&\tfrac89\Big(\tau^*{\Pf}_{\rm tr}\otimes\tau^*{\Pf}_{\rm tr}+(\tau^*)^2{\Pf}_{\rm tr}\otimes(\tau^*)^2{\Pf}_{\rm tr}\Big)\\
&-\tfrac49\,{\Pf}_{\rm tr}\otimes{\Pf}_{\rm tr}.
\end{aligned}
\end{equation}
The $30$ products ${\Pf}\cdot\Qp^m\Pf$ at weight $(3,3,3)$ are thereby expressed through triality images of products of trace gravitons, on which $d_1$ vanishes. Since $Q_1$ commutes with triality, which acts here on whole products, $d_1$ vanishes on the Pfaffian pairs as well, and no Leibniz rule for $Q_1$ is needed.

\subsection{Protection to all perturbative orders}
\label{sec:allorders}

The doublet of $\SO(8)$ and the classes of the four-term family at every rank admit a non-trivial deformation to every order in $\eta$. The grading of the perturbative complex leaves only two further orders at which an obstruction could arise, and the obstructions are again operators of lowest degree. For the reflection-odd classes the description of section~\ref{sec:lowestdegree} shows that both vanish in cohomology. For the reflection-even member of the doublet the same description excludes the obstruction at second order, and triality excludes the one at third order. A projection back to the BMN sector then shows that the deformed classes are not exact.

We use the homogeneous perturbative complex of local operators, in which $Q_r$ changes the letter number by $1-r$~\cite{Budzik:2023xbr,Budzik:2025zvu,Kim:2026rnx}. In the cubic holomorphic description of Budzik and Kulp the operation of order $r$ has $r+2$ entries, the operator and $r+1$ insertions of the interaction, and is computed by graphs with $r+2$ vertices~\cite{Budzik:2025zvu}. These graphs are Laman graphs, whose number of edges is twice the number of vertices minus three~\cite{Budzik:2025zvu,Gaiotto:2024gii,Budzik:2022mpd}. They therefore have $2r+1$ propagators, and hence $\Delta L=3(r+1)-2(2r+1)=1-r$. This is a grading of the perturbative construction and not a conserved letter number at finite coupling. Every $Q_r$ lowers $J_1$ and $J_2$ by $\frac12$ and preserves the charges $q_m$. Together with $\Delta L=1-r$, the relations \eqref{eq:allorder-charges} fix the change of the gradings,
\begin{equation}\label{eq:allorder-grading}
\Delta_{Q_r}(n_+,n_-;t_1,t_2,t_3)=(r,r;-r,-r,-r).
\end{equation}
Each order thus trades one unit of every $\theta$-degree for one unit of each $z$-degree. A finite list of obstruction sectors of this kind also enters the all-orders argument for $\SU(2)$ reported by the Fortuity collaboration~\cite{FortuitySummary}.

Consider a class of $\SO(2N)$ at the charges $\sect{1}{N-1,3,3}$ of the family, which for $N=4$ are the charges of the doublet. A representative $\mathcal O_0$ has $L=N+4$ letters and gradings $(0,0;N-1,3,3)$. We write the corrected operator as $\mathcal O(\eta)=\sum_{r\ge0}\eta^r\mathcal O_r$. The equation at order $r$ is
\begin{equation}\label{eq:allorder-obstruction}
Q_0\mathcal O_r=-\mathcal B_r,\qquad \mathcal B_r=\sum_{s=1}^rQ_s\mathcal O_{r-s}.
\end{equation}
The correction $\mathcal O_r$ has $N+4-r$ letters and the obstruction $\mathcal B_r$ has $N+5-r$, and both have gradings $(r,r;N-1-r,3-r,3-r)$. Nilpotence makes $\mathcal B_r$ closed once the equations of lower order hold. The sum includes the action of the lower $Q_s$ on the corrections already added to the seed, so the vanishing of $Q_r\mathcal O_0$ alone would not settle the equation; all terms have the same gradings, however. For $r\ge4$ the $\theta$-degrees $3-r$ are negative and there is no obstruction. For $r\le3$ the obstruction has $J_1+J_2=0$, so by \eqref{eq:degree} it is an operator of lowest degree, and its class lies in the cohomology described in section~\ref{sec:lowestdegree}. The first-order obstruction vanishes by section~\ref{sec:survival}. Table~\ref{tab:allorder-targets} lists the sectors of the remaining two for $N=4$.

\begin{table}[t]
\centering
\begin{tabular}{cclcc}
\toprule
$r$ & letters & letter content & reflection-even & reflection-odd\\
\midrule
$2$ & $7$ & $\lambda_+^2\lambda_-^2\phi_1\phi_2\phi_3$ & $0$ & $0$\\
$3$ & $6$ & $\lambda_+^3\lambda_-^3$ & $\operatorname{span}_{\mathbb C}\{[\Xi^3]\}$ & $0$\\
\bottomrule
\end{tabular}
\caption{The sectors in which an obstruction to the deformation of the $\SO(8)$ doublet could arise at orders two and three, and their cohomology in the two reflection sectors, with $\Xi=\tr(\lambda_+\lambda_-)$. The letter content gives multiplicities and not an ordering inside traces. For the reflection-even sector, in a basis of reduced trace words the obstruction sectors have dimensions $200$ and $7$, and the classical differential into them is a $200\times714$ matrix of rank $200$ and a $7\times15$ matrix of rank $6$. The cohomology therefore has dimensions $200-200=0$ and $7-6=1$, in agreement with the entries.}
\label{tab:allorder-targets}
\end{table}

For the reflection-odd classes both obstructions vanish, at every rank. At order two the obstruction is reflection-odd with $N+3$ letters, so $P_N=-1$ and the selection rule of section~\ref{sec:lowestdegree} leaves no cohomology. At order three it has $N+2$ letters, with content $\lambda_+^3\lambda_-^3\phi_1^{N-4}$. A basis element of the reflection-odd cohomology is an $\eps$-word with $N$ slots of odd length, hence with one slot of three letters and $N-1$ slots of one letter. The single-letter slots can contain $\lambda_+$ and $\lambda_-$ at most once each, since equal Grassmann-odd slots vanish, and the three-letter slot contains each gaugino at most once. At most four of the six gauginos can be accommodated, so the cohomology vanishes. Every reflection-odd class at the charges $\sect{1}{N-1,3,3}$ therefore admits corrections $\mathcal O_1$, $\mathcal O_2$ and $\mathcal O_3$ with
\begin{equation}\label{eq:allorder-protected}
Q(\eta)\big(\mathcal O_0+\eta\,\mathcal O_1+\eta^2\mathcal O_2+\eta^3\mathcal O_3\big)=0,
\end{equation}
as an identity in the perturbative complex, in this scheme. This applies to the classes of $\XN_N$ for all $N\ge4$, and for $N=4$ to the reflection-odd member of the doublet. Since $Q(\eta)$ commutes with $\SU(3)_R$, the same holds for the $\SU(3)_R$ multiplets that these classes generate.

For the reflection-even member of the doublet the second-order obstruction has seven letters and vanishes in cohomology by the same selection rule. At order three only the gauginos occur. Every word of three or more letters vanishes in $R/I$, the only trace is $\Xi=\tr(\lambda_+\lambda_-)$, and the cohomology is the line spanned by $[\Xi^3]$, which is non-zero because on the Cartan subalgebra $\Xi=-2\sum_{i=1}^4a_ib_i$ with independent Grassmann variables $a_i,b_i$.

To exclude an obstruction along this line we use triality. The two singlets span the standard representation of $S_3$ (section~\ref{sec:partner}), and since the graviton spans contain no singlets at this weight, this doublet is a subspace of the cohomology itself. Every colour tensor of the perturbative expansion is built from the Lie bracket and the invariant quadratic form~\cite{Budzik:2025zvu}. Each $Q_r$ therefore commutes with triality on local operators. Representatives and successive preimages can then be chosen equivariantly in the full perturbative complex, by averaging a linear choice over $S_3$ and over $\SU(3)_R$; the explicit $\mathcal Z$ of \eqref{eq:oneloop} need not have this property. With such a choice the class $[\mathcal B_r]$ defines an $S_3$-equivariant map from the doublet to the cohomology of the obstruction sector. This map vanishes on the reflection-odd member, as just shown. Its kernel is an invariant subspace of an irreducible representation, so it is the whole doublet. Consistently, $\Xi$ is the invariant form applied to the two gauginos and is fixed by triality, so the line $[\Xi^3]$ could not be the reflection-even vector of a standard representation. The identity \eqref{eq:allorder-protected} thus holds for both members of the doublet.

The deformed classes are not exact, and they are not the target of a lifting at any order. The projection $P_{\rm BMN}$ onto $n_+=n_-=0$ satisfies $P_{\rm BMN}\,Q(\eta)=Q_0\,P_{\rm BMN}$ by \eqref{eq:allorder-grading}, which extends to all orders the argument that excluded incoming pairing at first order~\cite{Kim:2026rnx}. It is linear over power series in $\eta$ and maps the corrected operator $\mathcal O(\eta)$ to its classical seed $\mathcal O_0$, since every correction $\mathcal O_r$ with $r\ge1$ has $n_\pm=r$. An incoming pairing at order $r\ge0$ would be a relation $\eta^r\mathcal O(\eta)=Q(\eta)\mathcal P(\eta)$ with $\mathcal P(\eta)=\sum_{k\ge0}\eta^k\mathcal P_k$; the case $r=0$ is exactness of the corrected class. Applying $P_{\rm BMN}$ and taking the coefficient of $\eta^r$ gives $\mathcal O_0=Q_0P_{\rm BMN}\mathcal P_r$, which the classical class excludes. The doublet of $\SO(8)$ and the classes of $\XN_N$ for all $N\ge4$ are therefore protected to all orders of perturbation theory.

The same grading protects the Pfaffian-pair classes of section~\ref{sec:evenscan}, which lie in the columns $[2,2]$ and $[6,0]$ with highest weights $(5,3,1)$ and $(7,1,1)$ at $2J=1$. At these weights the obstruction of order $r$ has $\theta$-degree $1-r$ in the third direction, so only the first-order obstruction can occur, and it vanishes by section~\ref{sec:survival}, since these classes have eight letters. Each $Q_r$ commutes with $\SU(3)_R$, so the lowering operators carry the corrected highest-weight operators to corrections of the whole multiplets, and the projection argument above shows that the corrected classes are not exact. The Pfaffian-pair classes are therefore protected to all orders as well.

The conclusion about cohomology is scheme independent, although the higher corrections and the termination of the representative at order three depend on the scheme. Different schemes are related by quasi-isomorphisms of the associated $L_\infty$-conformal algebras, which induce isomorphisms of the perturbative cohomology~\cite{Budzik:2025zvu}. Such quasi-isomorphisms preserve closure and non-exactness of classes, but a transported representative need not terminate. Survival beyond perturbation theory remains open.

\subsection{The restriction defect and the lifting of the \texorpdfstring{$\SO(7)$}{SO(7)} singlet}
\label{sec:defect}

Each member of the protected doublet of $\SO(8)$ has a non-zero classical image in $\SO(7)$, and that image is lifted. To understand the difference one must distinguish the restriction of a corrected operator from the correction of a restricted operator. For the vector restriction these involve $\iv^*Q_1^{(8)}$ and $Q_1^{(7)}\iv^*$, where the superscript indicates the algebra. After the letters have been restricted, $\iv^*Q_1^{(8)}$ still sums over all $28$ generators of $\so(8)$, whereas $Q_1^{(7)}\iv^*$ sums over the $21$ generators of $\so(7)$. The difference is the contribution of the seven complementary generators, and on polynomials in traces it amounts precisely to the value of the closed loop, eight or seven. The Fortuity collaboration identified the analogous contractions outside the block, for restrictions that separate branes, with the exchange of W-bosons~\cite{FortuitySummary}. The embeddings considered here are not brane separations, and we make no such identification.

Classical restriction still gives a one-way test of lifting~\cite{FortuitySummary}. If $\iv^*Q_1^{(8)}\mathcal O$ is not exact, then $Q_1^{(8)}\mathcal O$ is not exact either. The converse fails, because restriction need not be injective on cohomology, and the separate computation of $Q_1^{(7)}\iv^*\mathcal O$ implies nothing about the ambient theory. For a closed operator $\mathcal O$ in the class of the reflection-even member the two induced operations disagree,
\begin{equation}\label{eq:restrictionQ1}
\iv^*\,d_1^{(8)}[\mathcal O]=0,\qquad d_1^{(7)}[\iv^*\mathcal O]\ne0.
\end{equation}
The first statement is the result of section~\ref{sec:survival}. The second follows from the lifting of the $\SO(7)$ singlet found by Choi and Lee~\cite{Choi:2025bhi}. Indeed, the restriction $\iv^*\mathcal O$ represents that singlet modulo graviton classes (section~\ref{sec:partner}), and $d_1$ annihilates the graviton classes~\cite{Kibe:2026qid}. Equivalently, the restriction defect
\begin{equation}\label{eq:restrictiondefect}
\delta_{\rm Res}=Q_1^{(7)}\iv^*-\iv^*Q_1^{(8)},\qquad [\delta_{\rm Res}\mathcal O]=[Q_1^{(7)}\iv^*\mathcal O]\ne0,
\end{equation}
is non-trivial in cohomology, because $\iv^*Q_1^{(8)}\mathcal O$ is exact.

For the explicit representative $\mathcal O_8$ both statements can be seen directly. With the value $\tr1=7$ the rules that give \eqref{eq:oneloop} compute the correction of $\so(7)$, and the residual $2Q_1^{(7)}\mathcal O_8-Q_0\mathcal Z$ is a non-zero combination of $120$ multi-traces. Let $B_n$ denote the space of reflection-even $Q_0$-boundaries of $\so(n)$ in the sector of the variation, and let $v_7=Q_1^{(7)}\iv^*\mathcal O_8$. Then
\begin{equation}\label{eq:boundarydimensions}
\dim B_8=\dim B_7=741,\qquad \dim(B_7+\mathbb Q\,v_7)=742.
\end{equation}
The value $\dim B_8=741$ was derived in section~\ref{sec:lowestdegree}, and it bounds $\dim B_7$ from above, since evaluation on $\so(7)$ cannot raise a rank. Evaluation at integer points of $\so(7)$, modulo a prime, gives the lower bounds $741$ for the boundary space and $742$ after $v_7$ is adjoined. If $v_7$ were an element of $B_7$, adjoining it could not raise a dimension of at most $741$ to $742$. Since $-1\in\Or(7)$ acts trivially on the adjoint representation, every invariant of $\SO(7)$ is $\Or(7)$-invariant and hence a polynomial in traces~\cite{Procesi:1976,Aslaksen:1995}, so $B_7$ is the entire boundary space, and the restricted class is lifted. The dimension $741$ of $B_7$ was found by Chang and Lin and used by Choi and Lee, and the Fortuity collaboration also found the lifting~\cite{Chang:2025mqp,Choi:2025bhi,FortuitySummary}. Restricting the corrected operator \eqref{eq:O8correction} gives
\begin{equation}\label{eq:restrictedcorrection}
Q^{(7)}(\eta)\,\iv^*\big(\mathcal O_8-\tfrac\eta2\mathcal Z\big)=\eta\,\delta_{\rm Res}\mathcal O_8+O(\eta^2),
\end{equation}
and no first-order correction in the restricted theory removes the right-hand side.

The argument of section~\ref{sec:survival} shows where the two theories differ. The variation $v_7$ vanishes on the Cartan subalgebra of $\so(7)$, as \eqref{eq:targettheorem} requires, but there the Coulomb map is not injective on the cohomology of the sector. The injectivity argument of section~\ref{sec:lowestdegree} applies to the odd orthogonal groups, all of whose invariants are generated by traces, but only when the number of letters is at most twice the rank, and here there are eight letters at rank three. The Cartan subalgebra has three blocks, which do not separate the products of four traces from the others: on three blocks the Coulomb images of the $107$ products of section~\ref{sec:lowestdegree} span a space of dimension $84$, the number of products of at most three traces. In addition, eight letters exceed the range in which $\Or(7)$ has no trace relations.

The flavour singlets of this cohomology make this explicit. We write them in the notation of Choi and Lee, with traces written out~\cite{Choi:2025bhi}. With $\lambda^+=\lambda_-$, $\lambda^-=-\lambda_+$ and a sum over $\alpha\in\{+,-\}$, they are, up to an overall constant,
\begin{equation}\label{eq:ChoiLeetargets}
\begin{aligned}
M_1&=\eps_{mnp}\eps_{qrs}\tr(\phi_m\phi_q)\tr(\phi_n\phi_r)\tr(\phi_p\phi_s)\tr(\lambda_\alpha\lambda^\alpha),\\
M_2&=\eps_{mnp}\eps_{qrs}\tr(\phi_m\lambda_\alpha)\tr(\phi_q\lambda^\alpha)\tr(\phi_n\phi_r)\tr(\phi_p\phi_s),\\
M_3&=\eps_{mnp}\eps_{qrs}\tr(\lambda_\alpha\lambda^\alpha\phi_{(m}\phi_{q)})\tr(\phi_n\phi_r)\tr(\phi_p\phi_s),
\end{aligned}
\end{equation}
with $\phi_{(m}\phi_{q)}=\frac12(\phi_m\phi_q+\phi_q\phi_m)$. In $\SO(8)$ these are the three singlets of the $107$-dimensional cohomology, and their Coulomb images are linearly independent. On the Cartan subalgebra of $\so(7)$ they become proportional, $C(M_1)=3\,C(M_2)=6\,C(M_3)$, and Choi and Lee find that the variation of the $\SO(7)$ singlet is proportional, modulo $Q_0$-exact terms, to the combination $M_1-M_2-4M_3$~\cite{Choi:2025bhi}. For this combination
\begin{equation}\label{eq:c8-ChoiLeeCartan}
C(M_1-M_2-4M_3)=\big(1-\tfrac13-\tfrac46\big)C(M_1)=0.
\end{equation}
The source is visible after every allowed subtraction, with Coulomb image proportional to
\begin{equation}\label{eq:F7Konishi}
F_7=\tfrac1{16}\,C\Big(\det_{mn}\big(\tr\phi_m\phi_n\big)\sum_p\tr(\phi_p\psi_p)\Big),
\end{equation}
while its non-zero lifting target is invisible. The polynomial inside the parentheses is not closed, and although its last factor is of Konishi type, with $Q_1\sum_p\tr(\phi_p\psi_p)=60\,\tr(\lambda_+\lambda_-)$ in $\SO(7)$ by \eqref{eq:Q1normalization}, contractions that join different traces also contribute to the variation of the closed operator. In $\SO(8)$ no invisible combination of this kind exists in the sector, and the variation is exact.

The same defect accounts for the spinor restriction of the reflection-odd member. By \eqref{eq:partnerrestriction}, applied to the singlet projection, the spinor restriction of the reflection-odd member is the vector restriction of the reflection-even member,
\begin{equation}\label{eq:c8-spinorimage}
\is^*\XN_4^{(0)}=\iv^*\big(\tau^*\XN_4^{(0)}\big)_{\rm even},\qquad \is^*[\XN_4^{(0)}]=-\tfrac1{120}\,[\iv^*\mathcal O_8],
\end{equation}
where the second relation expresses the first through the explicit representative, by the singlet projection of \eqref{eq:O8normalization}. Since $\is^*=\iv^*\tau^*$ and $Q_1^{(8)}$ commutes with triality, the class $[\XN_4^{(0)}]$ survives in $\SO(8)$ while its spinor image is lifted in $\SO(7)$. The classes of $\SO(9)$ and $\SO(11)$ that the Fortuity collaboration identifies as brane fusions of the $\SO(7)$ singlet, which restrict to it under block embeddings, are lifted as well~\cite{FortuitySummary}. Their restriction maps separate branes and differ from the embeddings used here.
\section{Conclusions}
\label{sec:conclusions}

In this paper we have studied the classical $\frac1{16}$-BPS cohomology of $\Nfour$ super-Yang--Mills theory with an even orthogonal gauge group in the BMN sector, for the operators that carry the charge of a D3-brane wrapped on the non-contractible three-cycle of the dual geometry. For gauge group $\SO(8)$ we determined this sector through level eighteen. All but one of its $106$ $\SU(3)_R$ multiplets are products of trace gravitons and of supersymmetry multiplets of Pfaffians whose entries are replaced by symmetrized words. The bosonic operators of this kind are counted by the partition function of Arai, Fujiwara and Imamura, and the first excited one matches a fluctuation mode of the wrapped brane in the analysis of Arai and Imamura~\cite{Arai:2018utu,Arai:2019xmp}. The exception is a flavour singlet with one fermion. Its image on the Coulomb branch is not generated by descendants of bosonic invariants, so no enlargement of the list of graviton-like generators whose supersymmetry multiplets have bosonic bottom components can account for it.

The covering space that we constructed for the Pfaffian sector, following Choi, Choi and Kim~\cite{Choi:2026faq}, explains this class. The singlet is represented by the singlet projection of the rank-four member of a family of operators that is closed at every rank and defines a non-graviton class for every rank from four on. It is a generalized monotone, and we propose to interpret it as an excitation of the wrapped brane by words that are not symmetrized. Independently of this family, the covering reproduces the whole cohomology at fixed excitation charges above the Pfaffian once the rank is large enough. At the excitation charges of the singlet the covering cohomology has dimension $180$. It coincides with the cohomology of the Pfaffian sector from rank eight on and maps onto the $119$-dimensional cohomology at rank four. As a control, the covering with a single Pfaffian does not reach the threshold non-graviton class of $\SU(2)$ of Chang and Lin and of Choi, Kim, Lee and Park~\cite{Chang:2022mjp,Choi:2022caq}, which appears in the Pfaffian sector of $\SO(4)$.

The large-$N$ Pfaffian sector of the full superconformal index was obtained from the fluctuations of the wrapped brane~\cite{Arai:2019xmp,Fujiwara:2023bdc}. It is therefore natural to expect that the covering cohomology describes the $\frac1{16}$-BPS open-string states of the brane in the BMN sector. This would be in the spirit of the open-string interpretation of fermion bilinears by Gaiotto and Lee and of the relation between trace relations and open-string vacua studied by Lee~\cite{Gaiotto:2021xce,Lee:2023iil}. The BMN index, however, is not a limit of the full index, a signed index does not identify the states of the cohomology one by one, and a derivation from the worldvolume theory of the brane remains to be given.

Triality of the gauge algebra of $\SO(8)$ places the singlet in a doublet with a second class, which lies in the trace sector and is closed only because of trace relations of the finite-rank theory. At the charges of the singlet the non-graviton property is preserved by triality, whereas membership in a rank-uniform family is not. The doublet contains a generalized monotone and a class that is fortuitous with respect to the trace covering and to the Pfaffian coverings in which every index of an $\eps$-tensor is carried by a word. Choi, Choi and Kim noted a related ambiguity in the definition of fortuity~\cite{Choi:2026faq}.

Both members of the doublet restrict to the non-graviton singlet of $\SO(7)$ found by Gadde, Lee, Raj and Tomar, one along the spinor and the other along the vector embedding~\cite{Gadde:2025yoa}. Kibe, van Leuven and Mathieson noted that this class is visible on the Coulomb branch modulo monotone classes~\cite{Kibe:2026qid}. The doublet supplies an explicit witness for this visibility, a function that is orthogonal to the images of all graviton classes and on which the class has a non-zero component. Fortuitous classes can therefore be visible on the Coulomb branch, although the operators with which the known lifted classes pair, the images of the first correction, are not~\cite{Choi:2025bhi,Kibe:2026qid}. Kibe, van Leuven and Mathieson suggest a generalized-monotone interpretation of Coulomb-visible fortuitous classes~\cite{Kibe:2026qid}. The $\SO(7)$ singlet supports this suggestion in a specific sense, since it is the restriction, along the spinor embedding, of a generalized monotone of $\SO(8)$.

The doublet is protected to all orders of perturbation theory, and so are the classes of the four-term family at every rank. The reason is structural. The variation of a BMN class with $2J=1$, built from scalars and one fermion, consists of scalars and gauginos. The cohomology of such operators is detected on the Cartan subalgebra of $\SO(2N)$, for reflection-odd operators with any number of letters and for reflection-even ones with at most $2N$ letters, and first-order variations vanish there by the result of Kibe, van Leuven and Mathieson~\cite{Kibe:2026qid} and its extension to the Pfaffian sector in section~\ref{sec:firsttargets}. At higher orders the possible obstructions lie in sectors of the same kind, whose cohomology vanishes or is excluded by triality. This distinguishes the doublet from the $\SO(7)$ singlet. That class has no counterpart in the classical cohomology of the S-dual $\USp(6)$ theory~\cite{Chang:2025mqp,Chang:2026bil}. S-duality therefore requires it to be lifted, without fixing the order at which this happens, and the calculation of Choi and Lee shows that the lifting occurs already at first order~\cite{Choi:2025bhi}. No such constraint exists for $\SO(8)$, whose gauge algebra is self-dual. In $\SO(7)$ the Cartan subalgebra is too small to detect the variation. The difference between the two theories is carried by the contractions through the seven generators of the gauge algebra of $\SO(8)$ that lie outside that of $\SO(7)$.

Protection in the trace sector also tests the relation between fortuity and parity. On the basis of their $\SU(N)$ data and the lifting of the $\SO(7)$ class, Kim, Lee, Lee and Song conjectured that, at the full quantum level, fortuitous classes have charge-conjugation parity opposite to that of the gravitons with the same charges~\cite{Kim:2026mch}. For $\SO(2N)$ the analogous parity combines the letter number $L$ with the reflection $\sigma$, $P_N=(-1)^L\sigma^N$, and equals $+1$ on graviton products. For $\SO(8)$, $P_4=(-1)^L$, so the analogue would forbid fortuitous classes with even letter number. At level eighteen the two trace-fortuitous Pfaffian-pair multiplets and the reflection-even singlet all have one fermion and eight letters and survive to all perturbative orders. The pairs are products of brane states, whereas the singlet is a non-graviton class. Thus the trace-fortuity analogue fails in perturbation theory for both. Pfaffian coverings admit the pairs as generalized monotones. If the triality image of a rank-uniform family is also admitted, the singlet is not fortuitous either, and an analogue formulated with strong fortuity is not contradicted. Trace fortuity alone thus need not indicate a black-hole microstate at small rank. Kim, Kim, Lee and Park take the lifting of a fortuitous and a monotone class in a pair as a sign that the two notions may become ambiguous at the quantum level~\cite{Kim:2026rnx}.

The protection of the doublet and of the family is perturbative. It relies on the grading of the perturbative complex, in which the classical supercharge adds one letter and the correction of each further order adds one letter fewer, and the fate of these classes at finite coupling is open. The exclusion of rank-uniform families for the reflection-even member does not cover subdeterminant families with direct contractions between two $\eps$-tensors, and for the control class of $\SO(4)$ the exclusion of families with three Pfaffians rests on an assumption of the kind adopted by Choi, Choi and Kim~\cite{Choi:2026faq}.

Several directions for future work suggest themselves. A strongly fortuitous class of the Pfaffian sector would be a candidate black-hole microstate with odd wrapping charge. Since this is a discrete gauge charge, a horizon can carry it~\cite{Krauss:1988zc,Coleman:1991ku}. Such a state need not be a black hole bound to a separate brane. This would distinguish it from the dressed black holes of Choi et al.~\cite{Choi:2024xnv,Choi:2025lck}. It would also differ from the black holes surrounded by wrapped branes on the conifold proposed by Choi, Choi and Kim and studied by Amariti, Salto and Toldo~\cite{Choi:2026faq,Amariti:2026lfc}. We found no such class for $\SO(8)$ through level eighteen.

The parity discussed above indicates where to look. In the sectors with odd parity that we examined modulo a prime, the covering cohomology vanishes. If this holds in general, a flavour multiplet whose contribution to the reflection-odd index has the sign opposite to that of the graviton products, below the level at which families with three Pfaffians begin, would establish a strongly fortuitous class at charges of order $N$. No such multiplet occurs for $\SO(8)$ through level forty. Above the onset of the three-Pfaffian families two known classes are natural targets. One is the non-graviton class of $\SU(4)$ at level thirty-eight, in the seventh symmetric power of the fundamental flavour representation, which Kim, Lee, Lee and Song construct as a brane-fused operator, one whose restriction to $\SU(3)\times\mathrm U(1)$ contains the black-hole class of $\SU(3)$~\cite{Kim:2026mch}. It has fourteen letters and is odd under charge conjugation. Under the isomorphism between the gauge algebras of $\SU(4)$ and $\SO(6)$ it therefore lies in the Pfaffian sector of $\SO(6)$ and has odd parity, whereas the threshold class of $\SU(4)$ at level thirty-six is even under charge conjugation and lies in the trace sector. The other is the threshold class of $\SU(2)$ in the Pfaffian sector of $\SO(4)$, which the single-Pfaffian covering does not reach.

It would also be interesting to classify the fermionic excitations of the wrapped brane. In the sector built from scalars alone, every class of the Pfaffian sector is a combination of excited Pfaffians at every rank. This agrees with the higher polarizations of Wallach and Hunziker~\cite{Wallach:1993,Hunziker:1997}. It also agrees with the restriction theorem of Song, Xia and Xu~\cite{Song:2023chev}. With one fermion this fails, and the four-term family is the first example of what replaces it. On the quantum side, the argument for first-order survival applies to every BMN class with $2J=1$ in the Pfaffian sector, and a natural next step is to extend it to classes with $2J>1$, which contain more fermions or field strengths and whose variations contain letters of higher degree. For restrictions that separate branes, the Fortuity collaboration identifies the contractions through the complementary generators with the exchange of W-bosons~\cite{FortuitySummary}. One would like to know whether these contractions have a physical interpretation for the embeddings used here as well. Indices twisted by triality would provide further tests for $\SO(8)$, in analogy with the reflection-twisted index of section~\ref{sec:index}. Indices with outer-automorphism twists, including the twist of $D_4$ of order three, were studied for theories of class S by Mekareeya, Song and Tachikawa~\cite{Mekareeya:2012tn}. More generally, automorphisms of the operator algebra that do not preserve a chosen covering provide a test for any proposed definition of fortuity.

\acknowledgments

We would like to thank Timm Wrase and Min-Xin Huang for helpful discussions and suggestions. The work of Z.M. is supported in part by the NSF grant PHY-2609419. Z.M. acknowledges the support of the Dr.\ Hyo Sang Lee Graduate Fellowship from the College of Arts and Sciences at Lehigh University. The work of Y.T. is supported by the National Natural Science Foundation of China Grant No.~12247103.

\paragraph*{AI assistance.}
The authors used several large language models (Codex Astra, Claude Opus) to assist with writing, editing, and checking calculations in this paper. The authors have carefully reviewed all such contributions and take full responsibility for the content, accuracy, and conclusions of the work. 

\appendix
\section{Conventions and the Fischer inner product}
\label{app:conventions}

The Grassmann order of an operator is the order in which its letters are written: for an $\eps$-word, slot by slot and then trace by trace, and within a slot or a trace from left to right. Canonical forms use the identities \eqref{eq:slotid}, graded cyclicity of traces and graded commutativity of slots and traces; a necklace that equals its own reflection with sign $-1$ vanishes. The supercharges $Q$ and $\Qp^m$ act as left derivations and acquire the sign $(-1)^k$ when they pass $k$ fermions. The $\SU(3)_R$ generators $E_{mn}$ with $m\neq n$ act as even derivations with $E_{mn}\phi_n=\phi_m$, $E_{mn}\psi_m=-\psi_n$ and $E_{mn}f=0$, and they commute with $Q$.

The Fischer inner product of polynomials in the bosonic block variables $\phi^{(i)}_m$ of section~\ref{sec:exactcert} is $\langle\phi^\alpha,\phi^\beta\rangle=\delta_{\alpha\beta}\,\alpha!$ in multi-index notation. Each fermionic block value $\psi^{(i)}_m$ appears at most linearly in the functions to which the form is applied. The fermions can therefore be treated as commuting variables of degree at most one and taken orthonormal,
\begin{equation}\label{eq:Fischer}
\langle\psi^{(i)}_m\phi^\alpha,\psi^{(j)}_n\phi^\beta\rangle=\delta_{ij}\delta_{mn}\delta_{\alpha\beta}\,\alpha!.
\end{equation}
The adjoint of $E_{mn}$ with respect to this form is $E_{nm}$, so the form is invariant under the flavour symmetry and the projection onto $\SU(3)_R$ singlets is orthogonal. The form is positive definite on real polynomials. A non-zero function $F$ that is orthogonal to a subspace $S$ therefore does not lie in $S$, and if $G-cF\in S$ for a number $c$, then $c=\langle F,G\rangle/\langle F,F\rangle$.
\section{Exact counting of gauge invariants}
\label{app:dims}

This appendix completes section~\ref{sec:dims}: it gives the character and the two measures that enter \eqref{eq:dimprojection}, and describes how the integrals are evaluated exactly and tested.

For an element $g$ of $\Or(2N)$ let $\mathrm{Ad}(g)$ denote its action on antisymmetric matrices. With a fugacity $x_\ell$ for each of the seven letters, the character of the free Fock space is
\begin{equation}\label{eq:Fockcharacter}
\chi_{\rm F}(g;x)=\prod_{\ell\in\{\phi_m,f\}}\frac1{\det\big(1-x_\ell\,\mathrm{Ad}(g)\big)}\prod_{\ell\in\{\psi_m\}}\det\big(1+x_\ell\,\mathrm{Ad}(g)\big),
\end{equation}
and $\chi_s(g)$ is the sum of the coefficients of the monomials $\prod_\ell x_\ell^{n_\ell}$ whose letter numbers $n_\ell$ give the charges of the sector $s$. The fermionic letters enter with a plus sign, because the character counts operators and is not graded by fermion number. The determinants follow from the power sums
\begin{equation}\label{eq:adjointpowers}
\tr\,\mathrm{Ad}(g^k)=\tfrac12\Big[\big(\tr g^k\big)^2-\tr g^{2k}\Big].
\end{equation}

On the identity component the vector representation has eigenvalues $z_1^{\pm1},\dots,z_N^{\pm1}$, and the Weyl integration formula for a class function $\chi$ reads
\begin{equation}\label{eq:Weyl}
\int_{\SO(2N)}\!d\mu\,\chi=\frac1{2^{N-1}N!}\oint\prod_{i=1}^{N}\frac{dz_i}{2\pi\mathrm iz_i}\prod_{i<j}|1-z_iz_j|^2\,|1-z_i/z_j|^2\,\chi\big(z^{\pm1}\big),
\end{equation}
with $|1-w|^2=(1-w)(1-w^{-1})$. On the component $\Or(2N)^-$ the eigenvalues are $z_1^{\pm1},\dots,z_{N-1}^{\pm1},+1,-1$, and the integral reduces to a torus of rank $N-1$,
\begin{multline}\label{eq:twistedWeyl}
\int_{\Or(2N)^-}\!d\mu\,\chi=\frac{1}{2^{N-1}(N-1)!}\oint\prod_{i=1}^{N-1}\frac{dz_i}{2\pi\mathrm iz_i}\\
\times\prod_{i<j<N}|1-z_iz_j|^2\,|1-z_i/z_j|^2\prod_{i<N}|1-z_i^2|^2\,\chi\big(z^{\pm1},+1,-1\big),
\end{multline}
where the density is the Weyl density of type $C_{N-1}$, as given by Wendt~\cite{Wendt:2001}. Both measures have unit mass. On the reflection component the power sums in \eqref{eq:adjointpowers} are $\tr g^k=\sum_{i<N}(z_i^k+z_i^{-k})+1+(-1)^k$.

Both integrals are evaluated exactly. The integrands are Laurent polynomials, so an average over a grid of $M$-th roots of unity reproduces the torus average once $M$ exceeds their exponent span. We compute these averages in $\Fp$ for primes $p\equiv1\pmod M$ and reconstruct the integers by the Chinese remainder theorem with symmetric residues. The product of the primes exceeds $2D_{\rm F}+1$, where $D_{\rm F}$ is the largest Fock-space dimension among the sectors considered and bounds the absolute value of every integral.

The simplest test is the half-BPS sector of a single scalar, for which
\begin{equation}\label{eq:Molien}
\begin{aligned}
\int_{\SO(2N)}\frac{d\mu(g)}{\det\big(1-t\,\mathrm{Ad}(g)\big)}&=\frac1{(1-t^N)\prod_{k=1}^{N-1}(1-t^{2k})},\\
\int_{\Or(2N)^-}\frac{d\mu(g)}{\det\big(1-t\,\mathrm{Ad}(g)\big)}&=\frac1{(1+t^N)\prod_{k=1}^{N-1}(1-t^{2k})}.
\end{aligned}
\end{equation}
The first expression is the Molien series of the invariants of $\SO(2N)$, which are generated freely by the traces $\tr\phi^{2k}$ with $k<N$ and the Pfaffian, and the second is the trace of $\sigma$ on them, in which the Pfaffian is counted with a minus sign. The second identity tests the twisted density.
\section{Properties of the Pfaffian covering}
\label{app:covering}

This appendix supplies the arguments behind four statements of sections~\ref{sec:monotone} and~\ref{sec:covering}: the faithfulness of the covering algebra at large rank, the restriction formula that relates neighbouring ranks, the stability of covering-closed families under flavour rotations and $\Qp$, and the letter-number threshold and parity of families with several Pfaffians.

\subsection{Faithfulness of the covering algebra}
\label{app:faithful}

Section~\ref{sec:algebra} states that an element $c\in\mathcal C$ which is homogeneous in the excitation charges, contains no factor $\tr1$, and vanishes at $u=\phi_1^{-1}$ on $\so(2N)$-valued letters for a single $N\ge d+r$ is zero in $\mathcal C$. Here $d$ is the largest number of letters other than $\phi_1$ and $u$ in a monomial of $c$, and $r$ is the number of distinct non-zero exponents of pure traces of $\phi_1$ in $c$. The argument reduces $c$ to a multilinear bosonic element, chooses a basis of $\mathcal C$, and then isolates each basis element through one monomial in the matrix entries.

The identities of $\mathcal C$ and the evaluation preserve the number of each letter other than $\phi_1$ and $u$, and rescaling these letters independently shows that each component of fixed letter content vanishes separately. We may therefore assume that $c$ has fixed content and polarize it fully in the letters other than $\phi_1$ and $u$. Polarization replaces them by distinct letters $B_1,\dots,B_d$, each appearing once. It is injective, since restitution recovers a non-zero multiple of $c$, and it commutes with evaluation. A fermionic $B_k$ can be written as $\theta_k\widetilde B_k$ with an odd parameter $\theta_k$ and a bosonic matrix $\widetilde B_k$. This maps graded multilinear monomials injectively to $\pm\theta_1\cdots\theta_{d_F}$ times bosonic monomials of the same shape, where $d_F$ is the number of fermionic letters. It therefore suffices to treat bosonic multilinear $c$.

Write $A=\phi_1$ and $u=A^{-1}$. After cancellation, a multilinear monomial of $\mathcal C$ without the factor $\tr A^0=\tr1$ is a product of pure traces $\tr A^m$ with $m\neq0$ even and of cycles $\tr(A^{n_1}B_{k_1}A^{n_2}B_{k_2}\cdots A^{n_\ell}B_{k_\ell})$ with $n_g\in\mathbb Z$, in which every $B_k$ appears in exactly one cycle. Cycles are defined up to rotation and reflection. Reflection multiplies a cycle by $(-1)^{\sum_gn_g+\ell}$, so a cycle that equals its own reflection with sign $-1$ vanishes. The products of pure traces and of the remaining cycles, up to these identifications, form a basis.

To evaluate the basis elements, take $A=\bigoplus_{j=1}^Nx_j\mathbb J$ with independent variables $x_j$, and let the $B_k$ be generic antisymmetric matrices; since $c$ is polynomial in their entries, it then vanishes for all complex antisymmetric $B_k$. The eigenvectors $v_{j,\pm}=(e_{2j-1}\pm\mathrm ie_{2j})/\sqrt2$ of $A$ have eigenvalues $\alpha_{j,\pm}=\pm\mathrm ix_j$. Let $P$ be the matrix with these columns, so that $P^{-1}AP=\mathrm{diag}(\alpha_\nu)$, where $\nu=(j,\pm)$ labels the eigenvectors. The matrix $P$ is not orthogonal: the transformed bilinear form $K=P^TP$ pairs each index $\nu=(j,\pm)$ with its partner $\bar\nu=(j,\mp)$, so that $K_{\nu\nu'}=\delta_{\nu'\bar\nu}$ and $K^2=1$. For an antisymmetric $B$ the matrix $B'=P^{-1}BP$ is therefore not antisymmetric but satisfies $B'^TK+KB'=0$. The lowered matrix $b=KB'=P^TBP$ is antisymmetric in the ordinary sense, and $B\mapsto b$ is a linear isomorphism of the complex antisymmetric matrices. The entries $b_{\nu\nu'}$ with $\nu<\nu'$ are thus independent linear coordinates, and $B'_{\nu\nu'}=(Kb)_{\nu\nu'}=b_{\bar\nu\nu'}$. A cycle evaluates to
\begin{equation}\label{eq:cycleevaluation}
\tr\big(A^{n_1}B_{k_1}\cdots A^{n_\ell}B_{k_\ell}\big)=\sum_{\nu_1,\dots,\nu_\ell}\prod_{g=1}^\ell\alpha_{\nu_g}^{n_g}\,b^{(k_g)}_{\bar\nu_g\,\nu_{g+1}},\qquad\nu_{\ell+1}=\nu_1,
\end{equation}
a sum over assignments of basis indices $\nu_g$ to the $\ell$ gaps between consecutive letters, in which each $B_{k_g}$ contributes the coordinate that connects the gaps on its two sides.

Each basis monomial $T$ now receives a test monomial. Assign to the $d$ gaps of $T$ indices $\nu_g=(j_g,s_g)$ with distinct blocks $j_g\le d$, which is possible since $N\ge d$, and let $\mu_T$ be the resulting product of coordinates $b^{(k)}_{\{\bar\nu_g,\nu_{g+1}\}}$, one for each $B_k$. Every block $j_g$ occurs exactly twice among these index pairs: in the pairs of the two letters adjacent to gap $g$, or twice in the pair of the single letter of a cycle of length one. Hence any assignment of any basis monomial that produces $\mu_T$ puts its $d$ gaps into the blocks $j_1,\dots,j_d$, one gap per block, with the same adjacency of letters and gaps as in $T$. Its cycles are those of $T$ as cyclic sequences of letters, up to reflection, and only the exponents $n_g$ and the pure traces can differ.

Within a cycle of length $\ell\ge3$ the block and the sign of each gap are then determined by $\mu_T$. A cycle of length $\ell\le2$ has a second assignment that produces the same coordinates, obtained by reflecting the cycle: for $\ell=2$ it exchanges the blocks of the two gaps and reverses their signs, and for $\ell=1$ it flips the sign of the gap. Relative to the first assignment, it contributes with the reflection sign $(-1)^{\sum_gn_g+\ell}$ and with the exponents of the gaps reflected. For $\ell=1$ the two contributions add to $(1-(-1)^{n_1})(\mathrm ix_{j_1})^{n_1}$ times the coordinate, which is non-zero for every non-vanishing $\tr(A^{n_1}B)$. For $\ell=2$ they contribute to the same monomial in the $x_j$ only if $n_1=n_2$, and then the reflection sign is $+1$. The coefficient of $\mu_T$ in the evaluation of a basis monomial $T'$ with the adjacency of $T$ is therefore $\xi_{T'}\,p_{T'}(x)$. Here $p_{T'}$ is the product of the evaluated pure traces of $T'$, and $\xi_{T'}$ is a non-zero Laurent polynomial in the used variables $x_{j_1},\dots,x_{j_d}$ whose monomials are $\prod_gx_{j_g}^{n_g}$ for the exponents of $T'$ and their reflections. Basis monomials with distinct cycle parts give disjoint sets of such monomials.

It remains to separate the pure traces. A factor $\tr A^m$ evaluates to $\sum_\nu\alpha_\nu^m=2\mathrm i^m\sum_jx_j^m$. Split the sum into the $d$ blocks used above and the $N-d\ge r$ unused ones. Power sums of the unused variables with $r$ distinct non-zero exponents are algebraically independent over the field of rational functions of the used variables, because their Jacobian is a non-vanishing generalized alternant. The full power sums differ from them by elements of that field and are therefore algebraically independent as well. If $c=\sum_{T'}c_{T'}T'$ vanishes on evaluation, the coefficient of $\mu_T$ gives $\sum_{T'}c_{T'}\,\xi_{T'}\,p_{T'}(x)=0$, where the sum runs over the basis monomials with the adjacency of $T$. Distinct multisets of pure-trace exponents give products of power sums that are linearly independent over that field, so for each multiset the sum of $c_{T'}\xi_{T'}$ over the $T'$ with these pure traces vanishes. These $T'$ have distinct cycle parts, hence $\xi_{T'}$ with disjoint monomials, and every $c_{T'}$ vanishes. This holds for every basis monomial $T$, so $c=0$.

\subsection{The restriction formula}
\label{app:rankrestriction}

We derive the restriction formula behind \eqref{eq:rho} for a general block-diagonal configuration. Restrict each letter to $\so(2N)\oplus\so(2)$, denote its $\so(2N)$ block by a prime, and let its $\so(2)$ block be a multiple of $\mathbb J$ with a coefficient of the same Grassmann parity as the letter. For a word $W$ let $y_W$ be the ordered product of the $\so(2)$ coefficients of its letters. Write the slot list of $X_{N+1}(\vec W,M)$ as $(W_1,\dots,W_{N+1})=(\vec W,\phi_1^{\times(N+1-k)})$, with slot lengths $\ell_s$, and let $M_i$ be the necklaces of $M$, with lengths $\ell(M_i)$. Then
\begin{multline}\label{eq:blockrestriction}
\mathrm{res}\,X_{N+1}(\vec W,M)=\frac{1}{(N+1-k)!}\sum_{s:\,\ell_s\ \mathrm{odd}}2(-1)^{\frac{\ell_s-1}{2}}\kappa_s\,y_{W_s}\,\eps_N(W'_t:t\neq s)\\
\times\prod_i\Big(\tr M'_i+[\ell(M_i)\ \mathrm{even}]\,2(-1)^{\ell(M_i)/2}\,y_{M_i}\Big),
\end{multline}
where $\mathrm{res}$ denotes the restriction, $\eps_N$ is the rank-$N$ contraction, $[\cdot]$ equals $1$ if its condition holds and $0$ otherwise, and $\kappa_s$ is the Koszul sign of moving $y_{W_s}$ to the front. Block-diagonality forces the index pair of each slot into one block, and exactly one slot $s$ uses the $\so(2)$ block. Its two index orders give $y_{W_s}\big((\mathbb J^{\ell_s})_{12}-(\mathbb J^{\ell_s})_{21}\big)$, which is $2(-1)^{(\ell_s-1)/2}y_{W_s}$ for odd $\ell_s$ and zero for even $\ell_s$. Traces split as $\tr_{2N+2}=\tr_{2N}+\tr_2$.

In the restriction of section~\ref{sec:branesep} only $\phi_1$ has an $\so(2)$ component, equal to $x$, so $y_W$ is $x^\ell$ for $W=\phi_1^\ell$ and zero otherwise, and \eqref{eq:blockrestriction} reduces to the three contributions described there. Two steps of that section remain to be justified, one for the trace gravitons and one for the $\eps$-factors. A component of a trace-graviton multiplet restricts to the same component at rank $N$ plus constant multiples of powers of $x$, and the components of $\tr\Phi^{2N+2}$ at rank $N$ are polynomials in those of $\tr\Phi^2,\dots,\tr\Phi^{2N}$ by Newton's identities, polarized and acted on by $\Qp$. For an excited Pfaffian, $\rho$ picks the slots equal to the single letter $\phi_1$ and returns an excited Pfaffian at rank $N$. Every $\eps$-factor starts at order $x$. In a product with one $\eps$-factor the term linear in $x$ is therefore the linear term of that factor multiplied by the $x^0$ terms of the trace gravitons, and a product with three or more $\eps$-factors starts at $x^3$ and is annihilated by $\rho$. Hence $\rho$ maps graviton products to graviton products.

\subsection{Flavour rotations and \texorpdfstring{$\Qp$}{Q+} on families}
\label{app:gl3}

The flavour generators $E_{mn}$ of appendix~\ref{app:conventions} commute with $Q$. For every combination $a$ of formal monomials there are rank-independent combinations $b_0$ and $b_1$ such that
\begin{equation}\label{eq:gl3family}
E_{mn}X_N(a)=X_N(b_0)+N\,X_N(b_1)\qquad\text{for all }N,
\end{equation}
with $b_1\neq0$ only for $E_{1n}$, which turns a slot equal to the single letter $\phi_n$ into a filler. This is a direct computation. A slot $\phi_n$ mapped to $\phi_1$ becomes a filler, and $\eps(\vec W\setminus s,\phi_1^{\times(N-k+1)})\,\tr M/(N-k)!=(N-k+1)X_N(\vec W\setminus s,M)$, where $\vec W\setminus s$ is the slot list without that slot; all other terms are rank independent. Similarly $\Qp^mX_N(a)=X_N(a')$ for a rank-independent $a'$, the extra term coming from $\Qp^m\phi_1=\eps_{m1p}\psi_p$ acting on the fillers.

If $a$ is covering-closed, so are $b_0$, $b_1$ and $a'$. Since $E_{mn}$ commutes with $Q$, equation~\eqref{eq:gl3family} gives $QX_N(b_0)+N\,QX_N(b_1)=0$ for all $N$. Applying $\rho$ to this identity at rank $N+1$ and using \eqref{eq:rho} gives $QX_N(b_0)+(N+1)\,QX_N(b_1)=0$. Subtracting the two identities gives $QX_N(b_1)=0$ and then $QX_N(b_0)=0$ for all $N$, which is covering closure by the large-rank identification of section~\ref{sec:algebra}. For $\Qp^m$ the statement follows because $\{Q,\Qp^m\}$ acts on letters as $\mathrm{ad}(\phi_m)$, which vanishes on invariants.

The classes of covering-closed families therefore form a subspace of $H^-$ at each rank that is stable under $\mathfrak{gl}(3)$ and $\Qp$. This subspace contains every graviton product with one Pfaffian-type factor: bosonic families are closed because $Q\phi_m=0$, and the graviton factors are obtained from bosonic ones by acting with $\Qp$ and multiplying by trace gravitons.

\subsection{Families with several Pfaffians: thresholds and parity}
\label{app:multipf}

A $B$-Pfaffian family is specified by a set of non-filler matrix words, by the tensors and index positions to which their indices are attached, and by a list of necklaces $M$. Let $2k_b$ be the number of indices of the $b$-th tensor carried by non-filler words. The member at rank $N$ contracts $B$ tensors $\eps_{a_1\cdots a_{2N}}$ with these words and with $N-k_b$ filler letters $\phi_1$ on the $b$-th tensor, each filler occupying an index pair of a single tensor. It is multiplied by $\tr M$ and normalized by $\prod_b1/(N-k_b)!$, and it vanishes if $k_b>N$ for some $b$. For $B=1$ this is \eqref{eq:family}. Every index is thus carried by a word of length at least one.

Direct contractions between two $\eps$-tensors, that is, Kronecker deltas, are excluded by convention. They do not reduce to $(B-2)$-Pfaffian families. A family with a fixed number of deltas and rank-dependent fillers, such as $\eps\eps\,\phi_1^{\times(2N-2)}\delta\delta\propto\det\phi_1\,\tr\phi_1^{-2}$, is not a rank-independent formal multi-trace, although at each rank this bosonic example is a polynomial in trace gravitons. The analogous example of Choi, Choi and Kim, discussed in section~\ref{sec:interpretation}, is likewise a polynomial in trace gravitons at each rank~\cite{Choi:2026faq}.

The $2BN$ indices are absorbed two per word, so a member has at least $BN$ letters, and since every letter has $|q|\ge1$, its level obeys $j=2|q|\ge2BN$. This gives the thresholds quoted in section~\ref{sec:monotone}. For a single-Pfaffian family, $L=N-k+\sum_s\ell_s+\ell(M)$ and $\sigma=-1$, so
\begin{equation}\label{eq:familyparity}
P_N=(-1)^{\#\{s:\,\ell_s\ \mathrm{even}\}+\ell(M)},
\end{equation}
independently of $N$. Families for which the number of even-length slots plus the total trace length is odd therefore have $P_N=-1$ (section~\ref{sec:monotone}).
\section{Quantum corrections and explicit identities}
\label{app:quantum}

This appendix supplements section~\ref{sec:quantum}. It derives the trace rules for the pair operation of orthogonal algebras, gives the localization argument behind the vanishing of the Coulomb images of first-order variations, and states the boundary that relates the two representatives of the reflection-odd singlet.

\subsection{Trace rules for the pair operation of orthogonal algebras}
\label{app:quantumconventions}

We derive the two statements used in section~\ref{sec:Q1}: the reduction of the pair operation to rules on traces, and the splitting of the correction for $\so(4)$.

Every term of $Q_1$ replaces two letters $\ell,\ell'$ of an operator by expressions of the form $[t_A,u]\otimes[t^A,u']$, summed over the generators, where $u,u'$ are letters, possibly with derivatives, determined by the component rules of Budzik and Kulp~\cite{Budzik:2025zvu}. Equation~\eqref{eq:Q1pair} is the instance in which a scalar and a fermion are replaced by the two gauginos. For $\so(n)$, the sum over generators inside traces is carried out with the completeness relation \eqref{eq:completeness}. For arbitrary $n\times n$ matrices $U$ and $V$ it gives
\begin{equation}\label{eq:fusionfission}
\begin{gathered}
\sum_A\tr(t_AU)\tr(t_AV)=\tr(UV^T)-\tr(UV),\\
\sum_A\tr(t_AUt_AV)=\tr(UV^T)-\tr U\,\tr V.
\end{gathered}
\end{equation}
The first identity joins two traces and the second splits one, and the transposed words are brought back to standard form with the reversal identity of section~\ref{sec:invariants}. When $U$ or $V$ is the empty word, the second identity produces the closed loop $\tr1=n$. Applying \eqref{eq:fusionfission} to all pairs of letters gives $Q_1$ on polynomials in traces as a formal operation that depends on $n$ only through the value of the closed loop, as stated in section~\ref{sec:Q1}. Neither identity distinguishes even from odd $n$.

For $n=4$ the correction splits. With $t_{ij}=e_{ij}-e_{ji}$, so that the $t_{ij}$ with $i<j$ are the generators $t_A$ of section~\ref{sec:Q1}, the generators of the two factors of $\so(4)=\su(2)_+\oplus\su(2)_-$ are
\begin{equation}\label{eq:so4generators}
J^\pm_i=\tfrac12\Big(\tfrac12\eps_{ijk}\,t_{jk}\pm t_{i4}\Big),\qquad i=1,2,3,
\end{equation}
with $[J^+_i,J^-_j]=0$. Since $t_{jk}=\eps_{ijk}(J^+_i+J^-_i)$ and $t_{i4}=J^+_i-J^-_i$, the Casimir tensor is
\begin{equation}\label{eq:so4Casimir}
\sum_At_A\otimes t_A=2\sum_{i=1}^3\big(J^+_i\otimes J^+_i+J^-_i\otimes J^-_i\big).
\end{equation}
The two factors enter with equal coefficients and there is no mixed term. If every letter is decomposed as $\ell=\ell^++\ell^-$, a contraction $[J^+_i,u]\otimes[J^+_i,u']$ sees only the components in $\su(2)_+$, and a pair with one letter in each factor is not contracted at all. The correction of $\SO(4)$ is therefore the sum of the corrections of the two $\SU(2)$ factors, with equal couplings. This holds for the complete pair operation, since all its terms contract the same Casimir tensor.

On the cohomology $H_{\SO(4)}=H_{\SU(2)}\otimes H_{\SU(2)}$ the induced differential is then $d_1\otimes1+1\otimes d_1$, with the usual sign for odd classes. The reflection-odd class $\mathcal O^-_{24}$ of section~\ref{sec:control} is the graded-antisymmetric combination of the threshold class of $\SU(2)$ in one factor with the identity in the other. The identity is annihilated by $d_1$, and so is the threshold class~\cite{Budzik:2023vtr,Kim:2026rnx,FortuitySummary}. Hence $d_1[\mathcal O^-_{24}]=0$.

\subsection{Localization on the eigenvalues of a scalar}
\label{app:localization}

Here we supply the steps behind equations~\eqref{eq:localizedremainder} and~\eqref{eq:discriminantgraviton}. The construction is that of Kibe, van Leuven and Mathieson~\cite{Kibe:2026qid}. We write it in the notation of section~\ref{sec:Q1} for the algebra $\so(n)$ of rank $r$, and denote by $x_i$ the block values of $\phi_1$ on the Cartan subalgebra, so that its eigenvalues are $\pm\mathrm ix_i$, together with one further eigenvalue $0$ for odd $n$.

Let $\Phi=\Psi_{0,0;1}$ be the undifferentiated scalar $\phi_1$. Where its eigenvalues $\zeta_a$ are distinct we diagonalize it by a complex change of basis, in which the invariant bilinear form pairs each eigenvalue with its negative. For odd $n$ the eigenvalue $0$ is paired with itself, and the corresponding diagonal entry of every letter vanishes by antisymmetry; its differences with the other eigenvalues are $\pm\mathrm ix_i$. The eigenvalues are $Q_0$-closed, and so are the inverses of their differences, which we adjoin to the algebra of operators. In the equation $\{Q_0,\Psi\}=\Psi^2$, the terms linear in $\Phi$ give, for every $(k_+,k_-;S)\neq(0,0;\varnothing)$ with $1\notin S$ and for $a\ne b$,
\begin{equation}\label{eq:offdiagonalpairs}
Q_0\big(\Psi_{k_+,k_-;S\cup\{1\}}\big)_{ab}=\pm(\zeta_a-\zeta_b)\big(\widehat\Psi_{k_+,k_-;S}\big)_{ab},
\end{equation}
where $\widehat\Psi_{k_+,k_-;S}$ equals $\Psi_{k_+,k_-;S}$ plus terms that are at least bilinear in the letters, divided by $\zeta_a-\zeta_b$. The four cases of \eqref{eq:offdiagonalpairs} pair the off-diagonal components of the derivatives of $\phi_1$ with those of the gauginos and their derivatives, of $\psi_3$ with $\phi_2$, of $\psi_2$ with $\phi_3$, and of $f$ with $\psi_1$, together with their derivatives, so that every off-diagonal component of every letter belongs to exactly one pair. Since $Q_0^2=0$ and the eigenvalue differences are closed and invertible, the hatted components are $Q_0$-closed. The change of variables from the original off-diagonal components to the hatted ones is invertible in finitely many steps at each order in $z^\pm$, because the corrections to a component involve only components that are recovered earlier.

The diagonal components must be replaced by closed variables as well, because their variations contain off-diagonal components. We grade the letters by the weight $W=n_++n_-+t_2+t_3$, which $Q_0$ preserves and which is positive for every letter except $\Phi$. For a letter $Y$ of positive weight and an integer $s\ge1$, let $\mathcal G_{2s,Y}$ be a closed single-trace graviton in which $\tr(\Phi^{2s-1}Y)$ enters with unit coefficient. Such a graviton is obtained from the superform $\tr(\mathrm d\Psi)^{2s}$, where $\mathrm d$ is the exterior derivative in the coordinates $z^\pm$ and $\theta_m$~\cite{Chang:2013fba,Chang:2025mqp,Budzik:2023xbr}. Since $\{Q_0,\Psi\}=\Psi^2$ gives $Q_0\,\mathrm d\Psi=\pm[\mathrm d\Psi,\Psi]$, the superform is closed, and so is each of its coefficients in the coordinates and their differentials. Take the coefficient of $(\mathrm d\theta_1)^{2s-1}$ times a monomial in the coordinates and differentials that multiplies $Y$ in $\mathrm d\Psi$. In its terms with $2s-1$ factors $\mathrm d\theta_1\,\Phi$, where $\Phi$ is the lowest component of $\partial_{\theta_1}\Psi$, the remaining factor is $Y$, because a letter is determined by its gradings, and the $2s$ cyclic placements of this factor add up to a non-zero multiple of $\tr(\Phi^{2s-1}Y)$. The other terms contain fewer copies of $\Phi$ and therefore at least two letters of positive weight, each of smaller weight than $Y$.

On the Cartan subalgebra, with $y_i$ the block values of $Y$, the trace formula of section~\ref{sec:exactproof} gives
\begin{equation}\label{eq:gravitonmoments}
C(\mathcal G_{2s,Y})=2(-1)^s\sum_{i=1}^rx_i^{2s-1}y_i+R_{2s,Y},\qquad s=1,\dots,r,
\end{equation}
where $R_{2s,Y}$ depends only on the eigenvalues and on diagonal letters of weight smaller than $W(Y)$. The matrix of the linear system \eqref{eq:gravitonmoments} for the $y_i$ has the determinant
\begin{equation}\label{eq:orthogonalVandermonde}
\det\big(x_i^{2s-1}\big)_{i,s=1,\dots,r}=\prod_{i=1}^rx_i\prod_{i<j}\big(x_j^2-x_i^2\big),
\end{equation}
whose factors are, up to constants, differences of the eigenvalues $\pm\mathrm ix_i$ of $\Phi$. Proceeding in increasing weight, we define closed variables $\widetilde y_i$ by solving
\begin{equation}\label{eq:closeddiagonal}
2(-1)^s\sum_{i=1}^rx_i^{2s-1}\,\widetilde y_i=\mathcal G_{2s,Y}-\widetilde R_{2s,Y},\qquad s=1,\dots,r,
\end{equation}
where $\widetilde R_{2s,Y}$ is obtained from $R_{2s,Y}$ by replacing every diagonal letter of smaller weight by its closed counterpart. The right-hand side is closed, and on the Cartan subalgebra the subtraction removes exactly $R_{2s,Y}$, so that
\begin{equation}
Q_0\,\widetilde y_i=0,\qquad C(\widetilde y_i)=y_i.
\end{equation}
The recursion terminates for any given operator, because the weight decreases at each step and is bounded below. The closed diagonal variables are combinations of single-trace gravitons whose coefficients are rational functions of the eigenvalues with denominators that divide a power of \eqref{eq:orthogonalVandermonde}.

In the new variables $Q_0$ acts only on the off-diagonal pairs, which are contractible by the homotopy of section~\ref{sec:firsttargets}. A closed operator $\mathcal Y$ therefore differs by an exact term from its part of degree zero in the pairs, which is obtained from its Coulomb image by replacing each diagonal entry by its closed counterpart. This is \eqref{eq:localizedremainder}. Both terms on its right-hand side can have poles where eigenvalues coincide, although $\mathcal Y$ is polynomial.

The descent to gauge invariants uses the residual symmetry. The operator $\mathcal Y$ is invariant and $Q_0$ commutes with the normalizer of the Cartan subalgebra, so both terms of \eqref{eq:localizedremainder} can be averaged over the Weyl group and, for even $n$, over the reflection $\sigma$, under which a polynomial in traces is invariant. The products of gravitons $g_I$ are gauge invariant, so the average acts on their coefficients $r_I(\zeta)$ only. These become rational functions invariant under permutations and arbitrary sign changes of the $x_i$, with denominators that divide a power of the discriminant $\mathcal D$. After multiplication by $\mathcal D^k$ with $k$ large enough they are invariant polynomials, hence polynomials in the traces of even powers of $\Phi$, which are themselves gravitons. Without the average over $\sigma$ the coefficients for even $n$ could also contain the Pfaffian of $\Phi$. The same power of the discriminant clears the denominators of the exact term, and the result is the polynomial identity \eqref{eq:discriminantgraviton}. Since both sides are polynomials in the matrix entries of the letters, the identity extends from the dense set of $\Phi$ with distinct eigenvalues to all configurations.

\subsection{An explicit boundary relating the reflection-odd representatives}
\label{app:directidentity}

Section~\ref{sec:survival} uses the identity \eqref{eq:directidentity}, $\Omega-90\,\XN_4^{(0)}=Q_0\Lambda$, which relates the reflection-odd part of the triality image of $\mathcal O_8$ to the singlet projection of the covering family. Here we state the three objects precisely and explain how the identity is established.

The singlet projection is $\XN_4^{(0)}=\Pi_0\XN_4$, where $\XN_4$ is the rank-four member of \eqref{eq:singletfamily} with its factorial normalization. At weight $(3,3,3)$ the projector can be written in terms of the $\SU(3)_R$ generators $E_{mn}$ of appendix~\ref{app:conventions} as
\begin{equation}\label{eq:singletprojector}
\Pi_0=\prod_{c\in\{6,12,16,24,30,36,40,54,72\}}\Big(1-\frac{C_0}{c}\Big),\qquad C_0=\sum_{m\ne n}E_{nm}E_{mn}.
\end{equation}
On weight zero, $C_0$ is twice the quadratic Casimir operator of $\SU(3)_R$, and the product runs over its non-zero eigenvalues on the ninth tensor power of the fundamental representation. This power contains all operators with seven scalars and one fermion, because $\psi_m$ transforms in the antifundamental representation, which is the antisymmetric square of the fundamental one. Expanded in $\eps$-words, $\XN_4^{(0)}$ has $588$ terms. The operator $\Omega$ is defined in \eqref{eq:Omega}, with $\tau$ as in \eqref{eq:spinorembedding} and with the representative $\mathcal O_8$ of section~\ref{sec:O8}.

The preimage $\Lambda$ is a reflection-odd flavour singlet at $\sect{2}{3,3,3}$, with two fermions and five scalars. In terms of the unnormalized contraction \eqref{eq:epsword} it is
\begin{equation}\label{eq:Lambda}
\begin{aligned}
64\,\Lambda=\sum_{m,n}\eps_{pqr}\Big[&-61\,\eps(\psi_m,\phi_m,\phi_n,\phi_p)\tr(\psi_n\phi_q\phi_r)\\
&+204\,\eps(\psi_m,\psi_n,\phi_m,\phi_p\phi_q)\tr(\phi_n\phi_r)\\
&-168\,\eps(\psi_m,\phi_m,\phi_p,\psi_n\phi_q)\tr(\phi_n\phi_r)\\
&+720\,\eps(\psi_m,\phi_m,\phi_p,\phi_n\phi_q)\tr(\psi_n\phi_r)\\
&-1038\,\eps(\psi_m,\psi_n,\phi_m,\phi_n\phi_p\phi_q\phi_r)-152\,\eps(\psi_m,\psi_n,\phi_m,\phi_p\phi_n\phi_q\phi_r)\\
&-846\,\eps(\psi_m,\phi_m,\phi_n,\psi_n\phi_p\phi_q\phi_r)+444\,\eps(\psi_m,\phi_m,\phi_p,\psi_n\phi_n\phi_q\phi_r)\\
&-268\,\eps(\psi_m,\phi_m,\phi_n,\phi_p\psi_n\phi_q\phi_r)+1276\,\eps(\psi_m,\phi_m,\phi_p,\phi_n\psi_n\phi_q\phi_r)\\
&+4\,\eps(\psi_m,\phi_m,\phi_p,\phi_q\psi_n\phi_n\phi_r)+44\,\eps(\psi_m,\phi_n,\phi_p,\phi_q\psi_n\phi_m\phi_r)\Big],
\end{aligned}
\end{equation}
with a sum over $p,q,r$ understood and the Grassmann order of appendix~\ref{app:conventions}. After the flavour sums are carried out and the words are brought to canonical form, $\Lambda$ has $393$ terms. It is determined only up to the addition of $Q_0$-closed operators.

The three operators $\Omega$, $\XN_4^{(0)}$ and $Q_0\Lambda$ belong to the space of reflection-odd flavour singlets of $\SO(8)$ with seven scalars and one fermion. This space has dimension $23$, the multiplicity of the singlet at $2J=1$ that enters \eqref{eq:singletindex}. Its elements are linear in the fermion and can be evaluated on matrices by replacing each $\psi_m$ with a commuting antisymmetric matrix; for $Q_0\Lambda$ the supercharge is applied before this replacement. An element of the space is zero if it vanishes at $23$ points at which a basis has a non-singular matrix of values. The residual $\Omega-90\,\XN_4^{(0)}-Q_0\Lambda$ vanishes in exact rational arithmetic at $23$ such points with integer entries, which establishes \eqref{eq:directidentity}.

That a relation of this form exists follows from the results of sections~\ref{sec:singlet}, \ref{sec:partner} and~\ref{sec:survival}: the two operators are closed singlets with the same Coulomb image $48F$, the Coulomb map is injective on the cohomology at this weight, and so their difference is exact.

\bibliographystyle{JHEP}
\bibliography{refs}
\end{document}